\documentclass{elsart}

\usepackage[square,comma]{natbib}
\usepackage{graphicx}
\usepackage{pxfonts}
\usepackage{lineno}

\usepackage{amssymb}

\journal{}

\begin{document}

\thispagestyle{empty}
\begin{Large}
\textbf{DEUTSCHES ELEKTRONEN-SYNCHROTRON}

\textbf{\large{Ein Forschungszentrum der Helmholtz-Gemeinschaft}\\}
\end{Large}

DESY 13-013

January 2013

\begin{eqnarray}
\nonumber &&\cr \nonumber && \cr \nonumber &&\cr
\end{eqnarray}
\begin{eqnarray}
\nonumber
\end{eqnarray}
\begin{center}
\begin{Large}
\textbf{Wake monochromator in asymmetric and symmetric Bragg and
Laue geometry for self-seeding the European X-ray FEL}

\end{Large}
\begin{eqnarray}
\nonumber &&\cr \nonumber && \cr
\end{eqnarray}

\begin{large}
Gianluca Geloni,
\end{large}
\textsl{\\European XFEL GmbH, Hamburg}
\begin{large}

Vitali Kocharyan, Evgeni Saldin, Svitozar Serkez and Martin Tolkiehn
\end{large}
\textsl{\\Deutsches Elektronen-Synchrotron DESY, Hamburg}
\begin{eqnarray}
\nonumber
\end{eqnarray}
\begin{eqnarray}
\nonumber
\end{eqnarray}
ISSN 0418-9833
\begin{eqnarray}
\nonumber
\end{eqnarray}
\begin{large}
\textbf{NOTKESTRASSE 85 - 22607 HAMBURG}
\end{large}
\end{center}
\clearpage
\newpage

\begin{frontmatter}




\title{Wake monochromator in asymmetric and symmetric Bragg and Laue geometry for self-seeding the European X-ray FEL}


\author[XFEL]{Gianluca Geloni\thanksref{corr},}
\thanks[corr]{Corresponding Author. E-mail address: gianluca.geloni@xfel.eu}
\author[DESY]{Vitali Kocharyan}
\author[DESY]{Evgeni Saldin}
\author[DESY]{Svitozar Serkez}
\author[DESY]{and Martin Tolkiehn}

\address[XFEL]{European XFEL GmbH, Hamburg, Germany}
\address[DESY]{Deutsches Elektronen-Synchrotron (DESY), Hamburg,
Germany}

\begin{abstract}
We discuss the use of self-seeding schemes with wake monochromators
to produce TW power, fully coherent pulses for applications at the
dedicated bio-imaging bealine at the European X-ray FEL, a concept
for an upgrade of the facility beyond the baseline previously
proposed by the authors. We exploit the asymmetric and symmetric
Bragg and Laue reflections (sigma polarization) in diamond crystal.
Optimization of the bio-imaging beamline is performed with extensive
start-to-end simulations, which also take into account effects such
as the spatio-temporal coupling caused by the wake monochromator.
The spatial shift is maximal in the range for small Bragg angles. A
geometry with Bragg angles close to $\pi/2$ would be a more
advantageous option from this viewpoint, albeit with decrease of the
spectral tunability. We show that it will be possible to cover the
photon energy range from 3 keV to 13 keV by using four different
planes of the same crystal with one rotational degree of freedom.
\end{abstract}

%
%
%
\end{frontmatter}



\section{\label{sec:intro} Introduction}

One of the highest priority for experiments at any advanced XFEL
facility is to establish a beamline for studying biological objects
at the mesoscale, including large macromolecules, macromolecular
complexes, and cell organelles. This requires 2-6 keV photon energy
range and TW peak-power pulses \cite{HAJD}-\cite{BERG}. However,
higher photon energies (up to 13 keV) are needed to reach the
K-edges of commonly used elements, such as Se, for anomalous
experimental phasing. Studies at intermediate resolutions need
access to the water window at 0.5 keV. The pulse duration should be
adjustable from 2 fs to 10 fs.

A basic concept and design of an undulator system for a dedicated
bio-imaging beamline at the European XFEL was proposed in
\cite{OURCC}, and optimized in \cite{OURCD}. All the requirements in
terms of photon beam characteristics can be satisfied by the use of
a very efficient combination of self-seeding, fresh bunch and
undulator tapering techniques \cite{SELF}-\cite{LAST}. In
particular, a combination of self-seeding and undulator tapering
techniques would allow to meet the design TW output power. The
bio-imaging beamline would be equipped with two different
self-seeding setups. For soft X-ray self-seeding, the monochromator
consists of a grating \cite{SELF}. Starting around the energy of 3
keV it is possible to use a single crystal (wake) monochromator
instead of a grating \cite{OURCD}.

In \cite{OURCD} we demonstrated that it will be possible to cover
the photon energy range between 3 keV and 13 keV using the C(111),
C(220) and C(004) symmetric Bragg reflections. In this scenario,
three different crystals would enable self-seeding for the different
spectral range. In particular, we proposed to exploit the C(111)
symmetric Bragg reflection in the photon energy range between 3 keV
and 5 keV.

While developing a design for the bio-imaging beamline, the authors
first priority was to have it satisfying all requirements. Having
achieved that goal, the next step is to optimize the design, making
it as simple as possible. The design presented here aims for
experimental simplification and cost reduction of the self-seeding
setups. In order to improve the original design, here we propose to
exploit asymmetric Bragg  and Laue C(111), C(113) and C(333)
reflections together with the symmetric Bragg reflection C(004). The
novel design of the self-seeding setup combines a wide photon energy
range with a much needed experimental simplicity. Only one diamond
crystal is needed, and only one rotational degree of freedom is
required.

While, in this article, we consider applications for the bio-imaging
beamline in particular, the present study can also be applied to
other beamlines, for example the SASE1-SASE2 lines at the European
XFEL as well.

\section{\label{sec:diffra} Dynamical diffraction theory and
asymmetric-cut crystals}

\begin{figure}[tb]
\begin{center}
\includegraphics[width=0.75\textwidth]{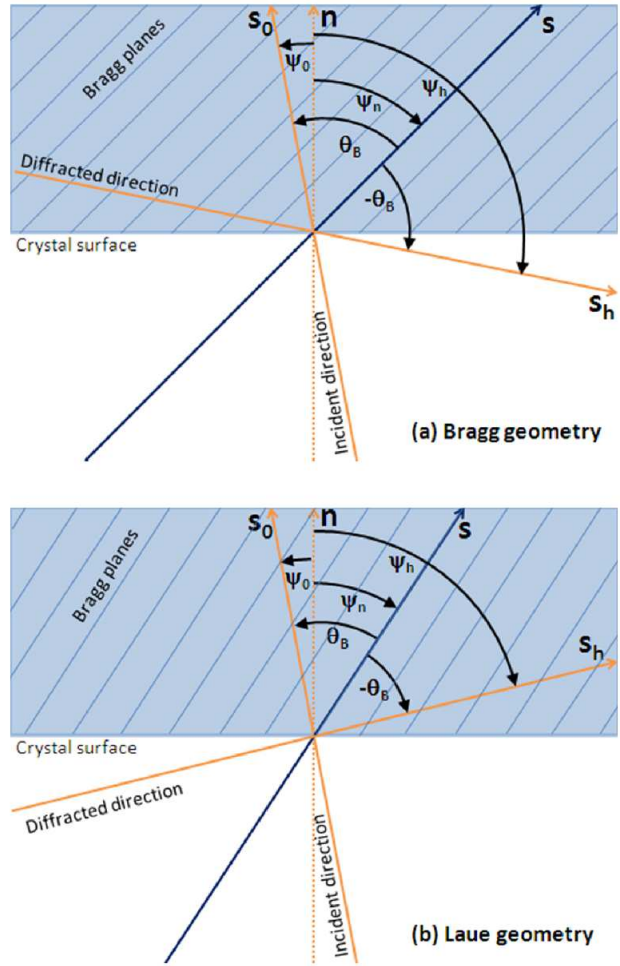}
\end{center}
\caption{Sketch of (a) Bragg and (b) Laue scattering geometry.}
\label{geoscatt}
\end{figure}

We begin our analysis of asymmetric-cut crystals by specifying the
scattering geometry under study. With reference to Fig.
\ref{geoscatt}, and following the notation in \cite{AUTH} we
identify the crystal surface with the unit vector normal to the
crystal surface $\vec{n}$, directed inside the crystal. The unit
vector $\vec{s}$ indicates the trace of the Bragg planes. The
direction of the incident beam is specified by the unit vector
$\vec{s}_0$, while that of the diffracted beam is given by
$\vec{s}_h$. Always following \cite{AUTH}, we call $\psi_n$ the
angle between $\vec{n}$ and $\vec{s}$, $\psi_0$ the angle between
$\vec{n}$ and $\vec{s}_0$, and $\psi_h$ the angle between $\vec{n}$
and $\vec{s}$. The signs of these angles are univocally fixed by
requiring that the angle between $\vec{s}$ and $\vec{s}_0$ is
positive (and equal to the Bragg angle $\theta_B$), while the angle
between $\vec{s}$ and $\vec{s}_h$ is negative (and equal to
-$\theta_B$). Fig. \ref{geoscatt} shows two generic examples for
both Bragg and Laue geometries.

Let us consider an electromagnetic plane wave in the x-ray frequency
range incident on an infinite, perfect crystal. Within the
kinematical approximation, according to the Bragg law, constructive
interference of waves scattered from the crystal occurs if the
angles between the reflecting lattice planes and both incident and
reflected beams are equal to Bragg angle $\theta_B$, Fig.
\ref{geoscatt}. The wavelength $\lambda$ and the Bragg angle are
related by the well known equation

\begin{eqnarray}
\lambda =  2 d \sin(\theta_B)~, \label{bragg}
\end{eqnarray}
where $d$ is the distance between Bragg planes, and from now on we
assume reflection into the first order. This equation shows that for
a given wavelength of the x-ray beam diffraction is possible only at
certain angles determined by $d$. It is important to remember the
following geometrical relationships:

\begin{itemize}

\item{In the kinematical approximation, the Bragg reflection is a mirror
reflection i.e. the angle between the incident X-ray beam and the
normal to the Bragg planes is equal to that between the normal and
the reflected beam.}

\item{The incident beam and the forward diffracted (i.e. the transmitted) beam
have the same direction.}

\end{itemize}

Diffraction from an asymmetric-cut crystal is equivalent to
diffraction from a blazed grating. This equivalence has already been
noted in \cite{BAJT}. The Bragg-reflected light from an
asymmetric-cut crystal behaves as the light diffracted from a
grating with period $D = d/\cos(\psi_n)$ and is dispersed in
accordance with the grating equation

\begin{eqnarray}
\lambda = D [\sin(\psi_0) - \sin(\psi_h)]~.\label{grat}
\end{eqnarray}
When the Bragg condition Eq. (\ref{bragg}) and the mirror reflection
condition are simultaneously met one obtains

\begin{eqnarray}
\psi_0 = \psi_n + \theta_B \label{psi0B}
\end{eqnarray}
and

\begin{eqnarray}
\psi_h = \psi_n - \theta_B \label{psihB}
\end{eqnarray}
By direct substitution in Eq. (\ref{grat}) and using simple
trigonometric relations, it is easy to check that Eq. (\ref{psi0B})
and Eq. (\ref{psihB}) satisfy the grating equation.

We now turn our attention beyond the kinematical approximation to
the dynamical theory of diffraction by a crystal. With reference to
Fig. \ref{geoscatt}, let us define the angle between the crystal
surface and the input beam as $\theta_i = \pi/2-\psi_0$. Similarly,
we define the diffracted angle $\theta_d = \pi/2 - \psi_h$, with
$\theta_d$ a function of the frequency, according to Eq.
(\ref{grat}). We also define the angle between the Bragg planes and
the crystal surface as $\alpha = \pi/2-\psi_n$. All these new
quantities follow the previously defined convention as concerns
their signs. Note that Eq. (\ref{psi0B}) and Eq. (\ref{psihB}) can
be rewritten in terms of incident and diffracted angles as $\theta_i
= \alpha - \theta_B$ and $\theta_d = \alpha + \theta_B$.

It is useful to describe the modification of the incident beam by
means of a transfer function. The reflectivity curve - the
reflectance - can be expressed in the frame of dynamical theory as

\begin{eqnarray}
R(\theta_i, \theta_d, \omega) = R[\Delta \omega + \omega_B \Delta
\theta_i \cot(\theta_B), \Delta \theta_d] ~, \label{reflectance}
\end{eqnarray}
where $\Delta \omega =  (\omega - \omega_B)$,  $\Delta \theta_i =
\theta_i - \theta_B + \alpha$, and $\Delta \theta_d = \theta_d -
\theta_B - \alpha$ are, respectively, the deviations of frequency,
incident angle and diffracted (i.e. reflected) angle from the
resonance (Bragg) frequency and angles, respectively. The frequency
$\omega_B$ and the angle $\theta_B$ are given by the Bragg law
$\omega_B \sin(\theta_B) = \pi c/d$.

Consider now a perfectly collimated, white beam incident on the
crystal. Within the kinematic approximation, the transfer function
as a function of the first argument is a Dirac $\delta$-function,
$\delta[\Delta \omega + \omega_B \Delta \theta_i \cot(\theta_B)]$
which is another representation of the Bragg law in differential
form:

\begin{eqnarray}
\frac{d \lambda}{d \theta_i} = \lambda \cot(\theta_B)~.
\label{bragdif}
\end{eqnarray}
In addition to this, within the kinematic approximation, reflection
from a lattice plane is always a mirror reflection, and

\begin{eqnarray}
\frac{d \theta_d}{d \theta_i} = 1 ~. \label{mirror2}
\end{eqnarray}
Moving to the framework of the dynamical theory, and in contrast to
what has just been said, the reflectivity width is finite. This
means that there is a reflected beam even when the incident angle
and the wavelength of the incoming beam are not exactly related by
the Bragg equation. It is interesting to note that incident beam and
transmitted beam continue to have the same direction also in the
framework of dynamical theory. However, in the framework of the
dynamical theory and in the case of asymmetric-cut crystal, the
reflection from lattice planes are not a mirror reflection anymore:
the grating equation Eq. (\ref{grat}),  holds, and the reflection
from the crystal surface is always a grating reflection.

The diffraction from the asymmetric-cut crystal in Bragg and Laue
geometry can be described an asymmetry parameter, $b$, defined by

\begin{eqnarray}
b = \frac{\vec{n}\cdot \vec{s_0}}{\vec{n}\cdot\vec{s_h}} =
\frac{\cos(\psi_0)}{\cos(\psi_h)}~.\label{bas}
\end{eqnarray}
In literature on X-ray crystal diffraction, the asymmetry parameter
is sometimes defined as inverse of $b$, and indicated with the
letter $\gamma$, for example in \cite{AUTH}. Here we follow the same
convention used in \cite{BAJT}. Note that $|b| =
\sigma_\mathrm{in}/\sigma_\mathrm{out}$ is the ratio of the widths
of incident and diffracted beams. From Eq. (\ref{grat}) follows that

\begin{eqnarray}
b = \frac{d \theta_d}{d\theta_i} \label{bratio} ~. \end{eqnarray}
As has been pointed out elsewhere, this is a consequence of
Liouville theorem \cite{BAJT}. In fact, a quasi-monochromatic,
collimated beam of finite width that is reduced in width must be
dispersed in angle to preserve the space-angle phase volume.

For the Bragg geometry, where the diffracted light exists on the
same side of the surface as the incident beam, we have $b < 0$,
whereas $b > 0$ describes Laue geometry. In particular, $b = -1$
corresponds the symmetric Bragg diffraction, and $b = 1$ corresponds
to the symmetric Laue diffraction.

The dispersion of the diffracted light from the asymmetric-cut
crystal follows directly from the grating equation, yielding

\begin{eqnarray}
\lambda \frac{d \theta_d}{d \lambda} =  (1+b) \tan(\theta_B)~.
\label{disp} \end{eqnarray}

As expected, there is no dispersion for symmetric Bragg geometry $(b
= -1)$ because there is no grating structure in that case.

\begin{figure}[tb]
\begin{center}
\includegraphics[width=0.5\textwidth]{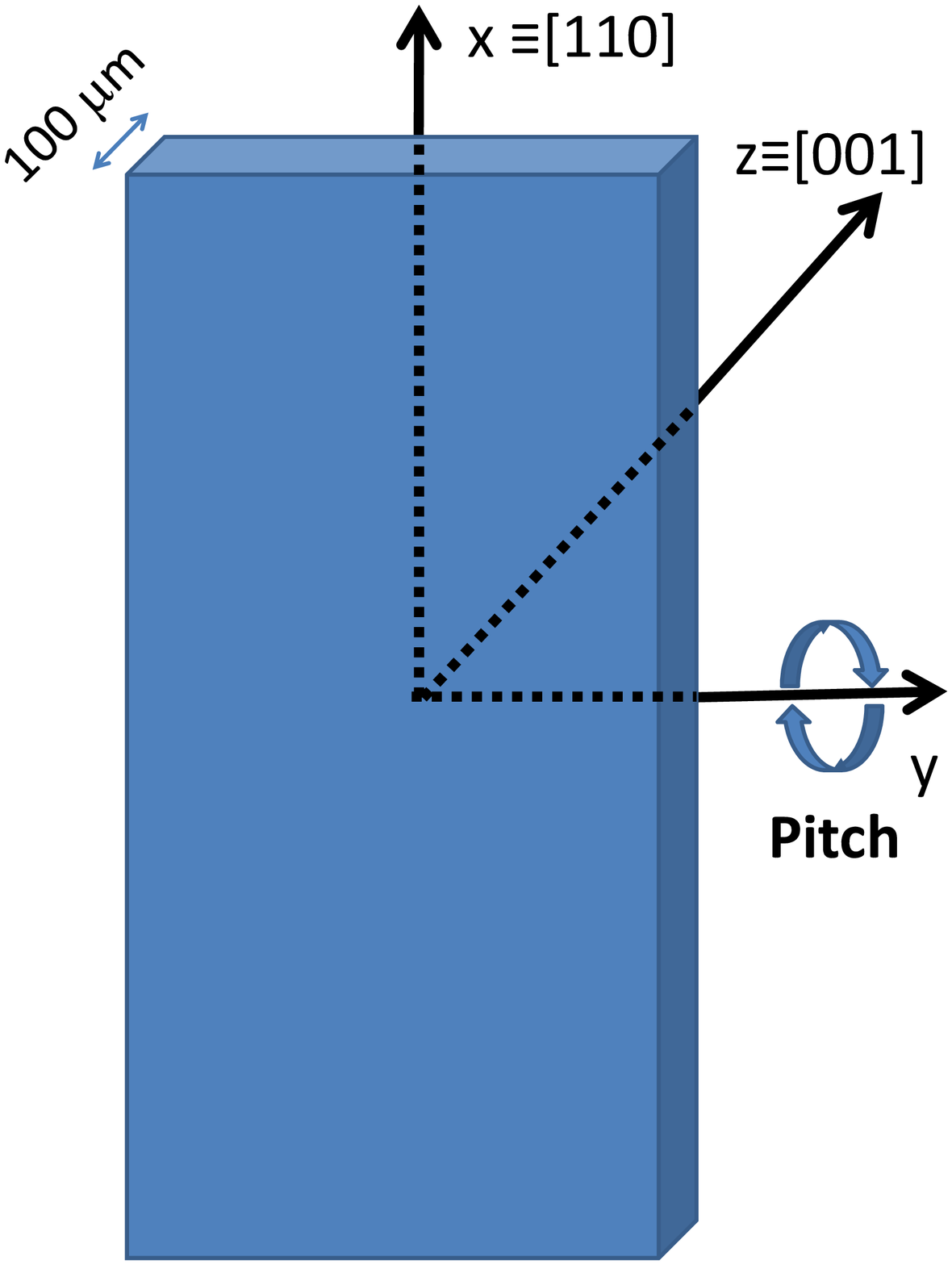}
\end{center}
\caption{Drawing of the orientation of the diamond crystal
considered in this article. } \label{LCLScry}
\end{figure}

In the following, we will consider one single diamond crystal, as
the $100~\mu$m thick crystal proposed in \cite{OURY5b}, and
currently used into the LCLS self-seeding setup \cite{EMNAT}. We
will define a cartesian reference system $\{x,y,z\}$ linked with the
crystal. The direction $z$ corresponds to the direction identified
by the Miller indexes $0$, $0$, $1$, which also coincides with the
direction of $\vec{n}$, while $x$ and $y$ are specified as in Fig.
\ref{LCLScry}, so that the direction $[110]$ corresponds to the $x$
direction.

\begin{table}
\caption{Useful Diamond Reflections}
\begin{small}\begin{tabular}{ l c c}
\hline hkl & Min. Energy (keV) &  $\Delta E$ (meV) \\ \hline
111      & 3.01034                  & 192.0     \\
311      & 5.76401                   & 56.0   \\
400      & 6.95161                   & 60.6    \\
333      & 9.03035                   & 27.3   \\
444      & 12.0404                 & 25.0 \\
\hline
\end{tabular}\end{small}
\label{reflT}
\end{table}

The crystal can rotate freely around the y axis (pitch angle) as
indicated in the figure. In this way, we can exploit several
symmetric and asymmetric reflection: we will be able to cover the
entire energy range between $3$ keV and $13$ keV by changing the
pitch angle of the crystal in Fig. \ref{LCLScry}. In the low energy
range between $3$ keV and $5$ keV we will use the C(111) asymmetric
reflection (in Bragg or Laue geometry, depending on the energy). At
higher energies between $5$ keV and $7$ keV we propose to go to the
C(113) asymmetric reflection in Bragg geometry. In the range between
$7$ keV and $9$ keV we can use instead the C(004) symmetric
reflection, in Bragg geometry. Finally, for energies larger than $9$
keV we propose the use of the C(333) asymmetric reflection, in Bragg
geometry\footnote{It should be remarked that above $12$ keV the
C(444) asymmetric Bragg reflection turns out to be a valid
alternative, though results are not explicitly analyzed in this
article. The Bragg angle is, in this case, very near to $\pi/2$,
$\theta_B = 84.4$ deg, while $\psi_n = 35.3$ deg. The relative
Darwin width amount to about $2\cdot10^{-6}$, with a very high
transmissivity (about $97\%$). Our calculations show that one can
reach pulse power in excess of $600$ GW at $12$ keV with a FWHM
relative bandwidth of about $5 \cdot 10^{-5}$.}. The list of
reflections considered in this article is summarized in Table
\ref{reflT}, extracted from \cite{SHVI}.

For self-seeding implementation, we are interested in the forward
diffracted, i.e. in the transmitted beam for each of these
reflections. From this viewpoint, the crystal can be characterized
as a filter with given complex transmissivity.  We will now consider
the three transmissivity separately, showing their amplitude and
phase, together with the crystal geometry for fixed energy points.

\begin{figure}[tb]
\begin{center}
\includegraphics[width=0.5\textwidth]{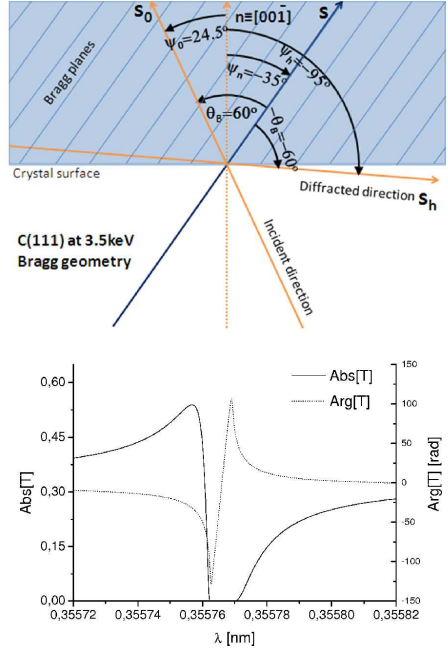}
\end{center}
\caption{Scattering geometry, modulus and phase of the transmittance
for the C(111) asymmetric Bragg reflection from the $100~\mu$m thick
perfect diamond crystal in Fig. \ref{LCLScry} at $3.5$ keV.}
\label{C111B}
\end{figure}

\begin{figure}[tb]
\begin{center}
\includegraphics[width=0.5\textwidth]{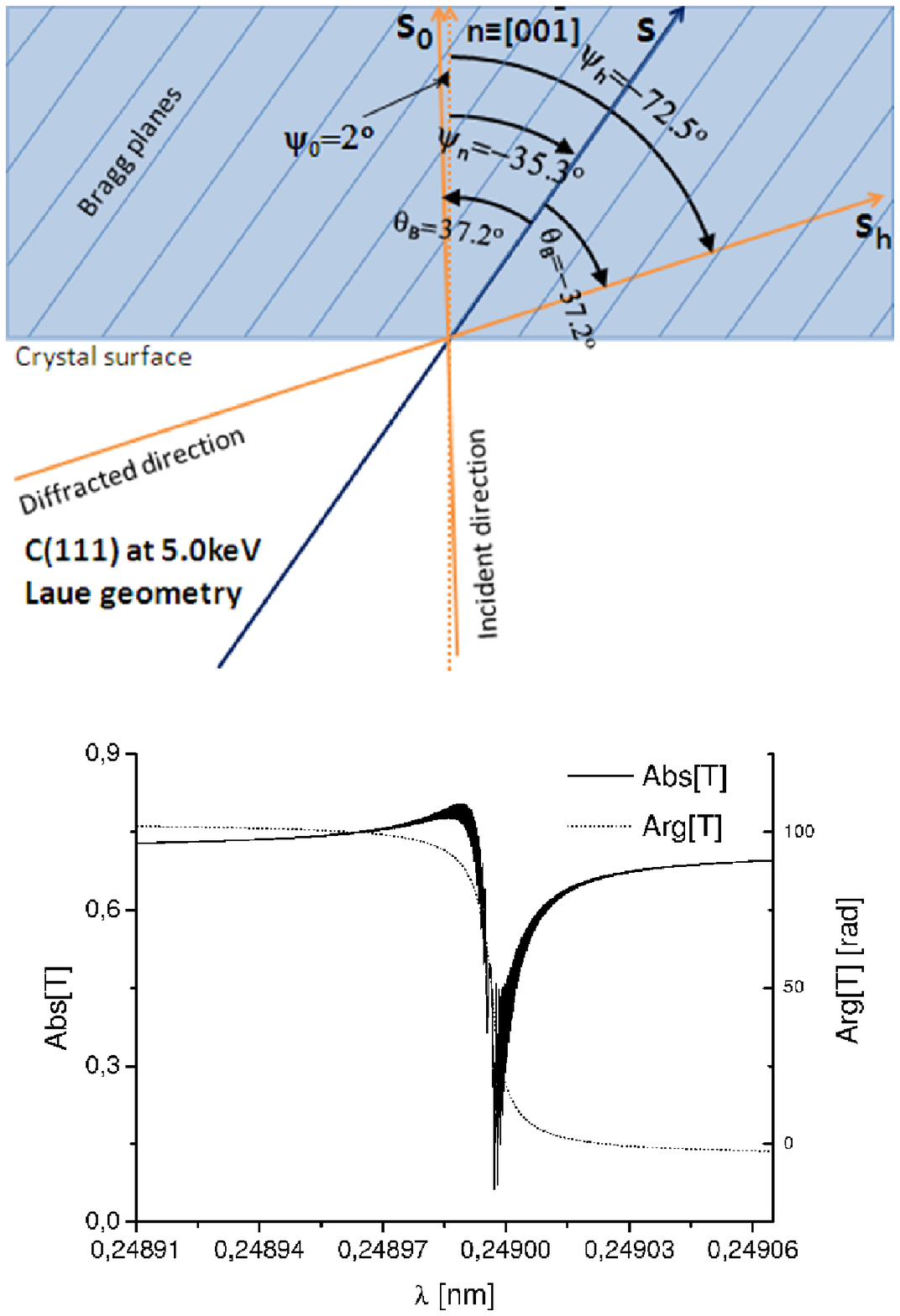}
\end{center}
\caption{Scattering geometry, modulus and phase of the transmittance
for the asymmetric C(111) Laue reflection from the $100~\mu$m thick
perfect diamond crystal in Fig. \ref{LCLScry} at $5$ keV.}
\label{C111L}
\end{figure}

\begin{figure}[tb]
\begin{center}
\includegraphics[width=0.5\textwidth]{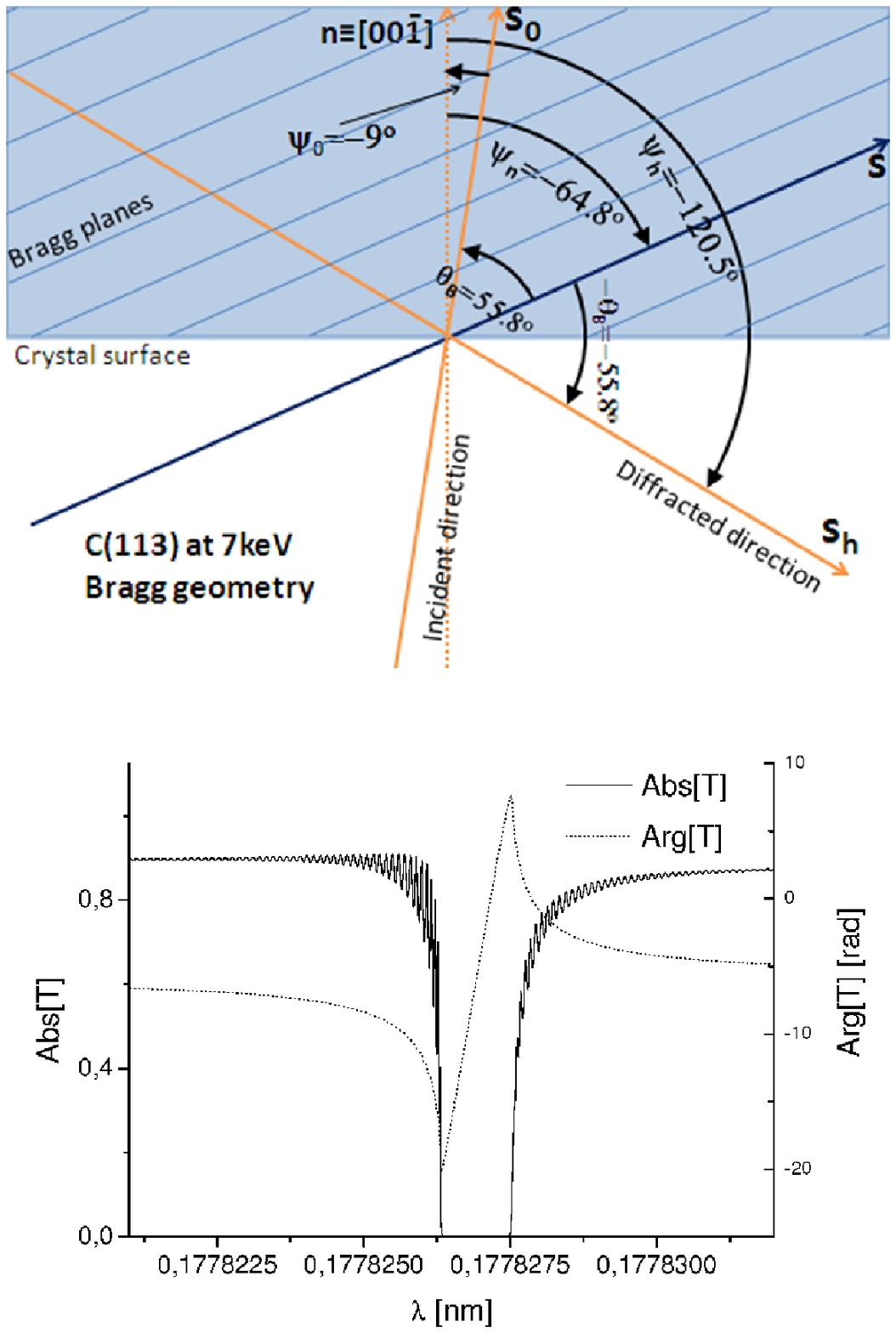}
\end{center}
\caption{Scattering geometry, modulus and phase of the transmittance
for the asymmetric C(113) Bragg reflection from the $100~\mu$m thick
perfect diamond crystal in Fig. \ref{LCLScry} at $7$ keV.}
\label{C113}
\end{figure}

\begin{figure}[tb]
\begin{center}
\includegraphics[width=0.5\textwidth]{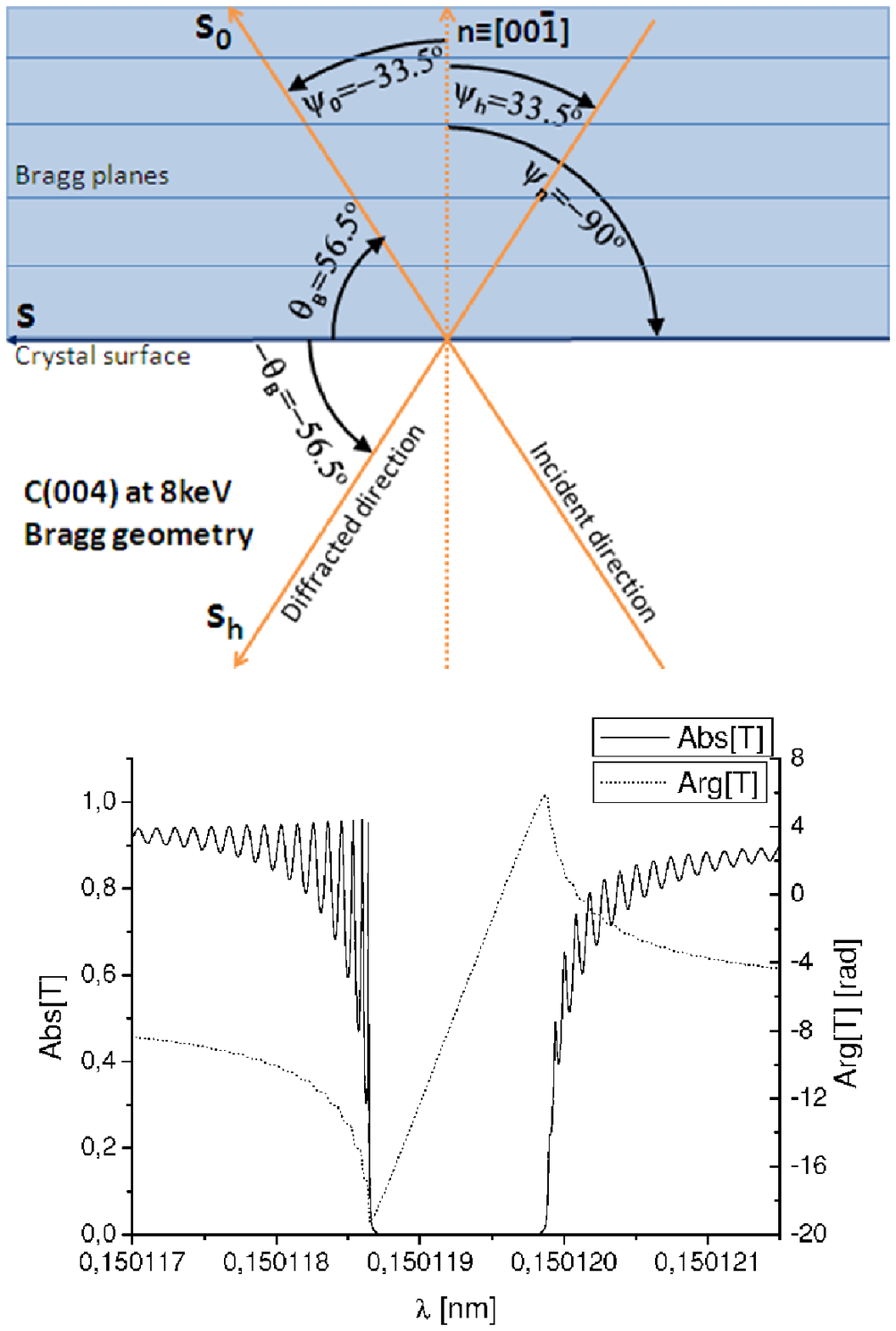}
\end{center}
\caption{Scattering geometry, modulus and phase of the transmittance
for the symmetric C(004) Bragg reflection from the $100~\mu$m thick
perfect diamond crystal in Fig. \ref{LCLScry} at $8$ keV.}
\label{C004}
\end{figure}

\begin{figure}[tb]
\begin{center}
\includegraphics[width=0.5\textwidth]{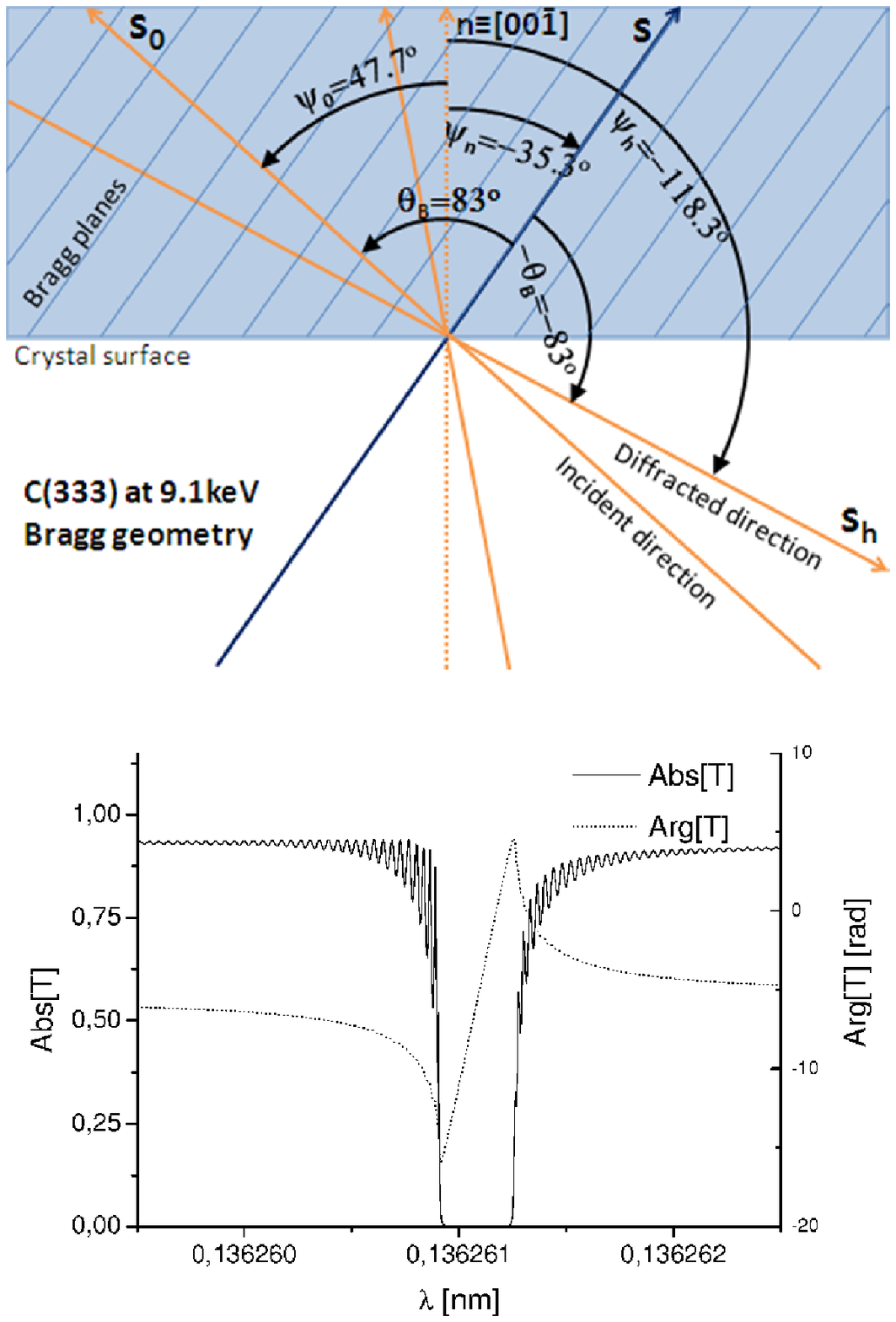}
\end{center}
\caption{Scattering geometry, modulus and phase of the transmittance
for the asymmetric C(333) Bragg reflection from the $100~\mu$m thick
perfect diamond crystal in Fig. \ref{LCLScry} at $9$ keV.}
\label{C333}
\end{figure}

In Fig. \ref{C111B} we show the amplitude and the phase for the
C(111) asymmetric Bragg reflection at $3.5$ keV. The following Fig.
\ref{C111L} shows the same reflection at a higher energy, $5$ keV.
In this case, one has Laue geometry. In Fig. \ref{C113} we plot the
same quantities for the C(113) asymmetric Bragg geometry at $7$ keV.
In Fig. \ref{C004} we consider the symmetric Bragg reflection C(004)
at 8 keV and, finally, in Fig. \ref{C333} we plot amplitude and
phase for the asymmetric Bragg reflection C(333) at $9$ keV. By
inspecting the plots, the reader can easily see the difference
between Bragg and Laue geometry, as well as that between symmetric
and asymmetric reflections.

Amplitude and phase in the plots are related by Kramers-Kroning
relations.  Let us regard our crystal as a filter with transmission
$T(\omega) = |T(\omega)| \exp[i\Phi(\omega)]$. According to
Titchmarsch theorem (see \cite{LUCC} and references therein for a
recent review on the subject) causality\footnote{Causality simply
requires that the filter can respond to a physical input after the
time of that input and never before.} and square-integrability of
the inverse Fourier transform of $T(\omega) = |T(\omega)| \exp[i
\Phi(\omega)]$, which will be indicated with $\mathcal{T}(t)$, is
equivalent\footnote{Here we are tacitly assuming, as required by
Titchmarsh theorem, that $T(\omega)$ is a square integrable
function.} to the existence of an analytic continuation of
$T(\omega)$ to $\Omega = \omega + i \omega'$ on the upper complex
$\Omega$-plane (i.e. for $\omega'>0$). The same theorem also shows
that the two previous statements are equivalent to the fact that
real and imaginary part of $T(\omega)$ are connected by Hilbert
transformation. Since $\mathcal{T}(t)$ must be real (thus implying
that $T^*(\omega)=T(-\omega)$), from the Hilbert transformation
follows the well-known Kramers-Kroninig relations \cite{KRAM,KRON},
linking real and imaginary part of $T(\omega)$.

A similar reasoning can be done for the modulus $|T(\omega)|$ and
the phase $\Phi(\omega)$, see \cite{TOLL}. In fact, one can write

\begin{eqnarray}
\mathrm{ln}[T(\omega)] = \mathrm{ln}[|T(\omega)|] + i
\Phi(\omega)~,\label{ln}
\end{eqnarray}
that play a similar role to real and imaginary part of the
refractive index of a given medium.

Note that $T^*(\omega)=T(-\omega)$ implies that
$|T(\omega)|=|T(-\omega)|$ and that $\Phi(\omega) = -
\Phi(-\omega)$. Therefore, using Eq. (\ref{ln}) one also has that
$\mathrm{ln}[T(\omega)]^*=\mathrm{ln}[T(-\omega)]$. Application of
Titchmarsh theorem shows that the analyticity of
$\mathrm{ln}[|T(\Omega)|]$ on the upper complex $\Omega$-plane
implies that $\Phi(\omega)$ can be obtained from $|T(\omega)|$. As
is well known however, in applying such procedure one tacitly
assumes that $\mathrm{ln}[T(\Omega)]$ is analytical on the upper
complex $\Omega$-plane. While causality implies this fact for
$T(\Omega)$, it does not imply it automatically for
$\mathrm{ln}[|T(\Omega)|]$. In fact, such function is singular where
$T(\Omega)=0$. If $T(\Omega)$ has zeros on the upper complex plane,
these zeros would contribute adding extra terms to the total phase.
It can be shown that $T(\Omega)$ does not have zeros on the upper
complex plane, and that Titchmarsch theorem actually applies.

The simplest check of the correctness of the calculations by means
of dynamical theory consists, therefore, in verifying that amplitude
and phase of the transmissivity are actually related by
Kramers-Kroning relations.

\section{\label{shift} Spatiotemporal shift and its influence on input coupling
factor}

A wake monochromator introduces spatiotemporal deformations of the
seeded X-ray pulse, which can be problematic for seeding. The
spatiotemporal coupling in the electric field relevant to
self-seeding schemes with crystal monochromators has been analyzed
in the framework of classical dynamical theory of X-ray diffraction
\cite{SHVI}.

This analysis shows that a crystal in Bragg or Laue geometry
transforms the incident electric field $E(x,t)$ in the $\{x,t\}$
domain into $E(x-at,t)$, where $a=\mathrm{const}$. The physical
meaning of this distortion is that the beam spot size is independent
of time, but the beam central position changes as the pulse evolves
in time. First we will show in a simple manner that, based on the
only use of the Bragg law, we can directly arrive to an explanation
of spatiotemporal coupling phenomena in the dynamical theory of
diffraction \cite{TILT}.

The transmissivity curve - the transmittance - in  asymmetric and
symmetric Bragg or Laue geometry can be expressed in the framework
of dynamical theory with the help of Eq. (\ref{reflectance}).
Transmitted (i.e. forward diffracted) beam and incident beam have
the same direction. Therefore we set $\theta_i=\theta_d$. Using the
symbol $T$ for transmittance, Eq. (\ref{reflectance}) yields

\begin{eqnarray}
T(\Delta \omega, \Delta \theta_i) = T(\Delta \omega + \omega_B
\Delta \theta \cot(\theta_B))~, \label{T}
\end{eqnarray}
where $\Delta \omega = (\omega - \omega_B)$ and $\Delta \theta =
(\theta_i - \theta_B)$ are the deviations of frequency and incident
angle of the incoming beam from Bragg frequency and Bragg angle
respectively. The frequency $\omega_B$ and the angle $\theta_B$ are
obviously related by the Bragg law: $\omega_B \sin(\theta_B) = \pi
c/d$.

It should be realized that the crystal does not introduce an angular
dispersion similar to the grating. However, a more detailed analysis
based on the expression for the transmissivity, Eq. (\ref{T}), shows
that a less well-known spatiotemporal coupling exists. Let us
discuss this fact. It is evident by inspection of Eq. (\ref{T}),
that the transmissivity is invariant under angle and frequency
transformations obeying

\begin{eqnarray}
\Delta \omega + \omega_B \Delta \theta \cot(\theta_B) =
\mathrm{const}~. \label{transf}
\end{eqnarray}
This corresponds to the coupling in the Fourier domain. In general,
one would indeed expect the transformation to be symmetric in both
the Fourier domain $\{k_x,\omega\}$ (with $k_x = \omega_B \Delta
\theta/c$) and in the space-time domain $\{x,t\}$, due to the
symmetry of the transfer function. However, it is reasonable to
expect the influence of a nonsymmetric input beam distribution. The
field transformation for the XFEL pulse after the crystal in the
$\{x,t\}$ domain is given by

\begin{eqnarray}
E_{\mathrm{out}} (x,t) =   FT\{T[\Delta \omega, (k_x)_\mathrm{in},
(k_x)_\mathrm{out}]E_{\mathrm{in}}(\Delta \omega, k_x)\} ~,
\label{Eoutxt}
\end{eqnarray}
where $FT$ indicates a Fourier transform from the frequency domain
$\{k_x,\omega\}$ to the space-time domain $\{x,t\}$.

In the self-seeding case, the incoming XFEL beam is well collimated,
meaning that its angular spread is a few times smaller than the
angular width of the transfer function. Yet, the spectral bandwidth
of the incoming beam is much wider than the bandwidth of the
transfer function.  Under  the limit of a wide spectral bandwidth
for the incoming beam (with respect to the bandwidth of the transfer
function), and applying the shift theorem twice we obtain

\begin{eqnarray}
&& E_\mathrm{out}(x,t) = \eta(t) \int d k_x \exp(i \cot \theta_B k_x
c t) \exp(-i k_x x) E(\omega_B,k_x) \cr && = \eta(t) a(x - c \cot
\theta_B t) ~, \label{eoutdel}
\end{eqnarray}
where

\begin{eqnarray}
\eta(t) = \frac{1}{2\pi} \int d \omega \exp( -i \omega t) T(\omega)
\label{tempFT}
\end{eqnarray}
is the inverse temporal Fourier transform of the transmittance curve
(impulse response). Eq. (\ref{eoutdel}) is universal i.e. valid for
both symmetric and asymmetric Bragg and Laue geometries.

The spatial shift given by Eq. (\ref{eoutdel}) is proportional to
$\cot(\theta_B)$, and is maximal in the range for small values of
$\theta_B$. Therefore, the spatiotemporal coupling is an issue, and
efforts are necessarily required to avoid distortions. A geometry
with $\theta_B$ close to $\pi/2$ would be a more advantageous option
from this view point, albeit with decrease in the spectral
tunability \cite{SHVI}.

\subsection{Influence of the offset on the coupling factor}

The influence of the spatiotemporal distortion on the operation of
the self-seeding setup can be quantified by studying the input
coupling factor between the seed beam and the ground eigenmode of
the FEL amplifier. When the undulator is sufficiently long, the
output energy in the FEL pulse grows exponentially with the
undulator length, and the eenrgy gain, $G =
E_\mathrm{out}/E_\mathrm{ext}$, can be written as

\begin{eqnarray}
G = A \exp[z/L_g] ~, \label{powergain}
\end{eqnarray}
where $A$ is the input coupling factor. In the linear regime the
energy gain does not depend on the input energy $E_\mathrm{ext}$,
and for the case including spatial shift the input coupling factor
$A$ allows for a simple and convenient measure of the departure from
the ideal situation. In our case this measure is simply defined as

\begin{eqnarray}
Q \equiv \frac{E_\mathrm{nonideal}}{E_\mathrm{ideal}} =
\frac{A_\mathrm{nonideal}}{A_\mathrm{ideal}}~. \label{measure}
\end{eqnarray}
We will consider a particular example for $\lambda = 0.15$ nm,
corresponding to about $8$ keV. We simulated the SASE process in the
first $11$ cells, followed by the self-seeding setup exploiting the
C(004) reflection. The seed beam is then superimposed again onto the
electron bunch (the modulation having been washed out by the
chicane). We calculate the output power and spectrum after $6$
cells, which is still in the linear regime, and we scan over the
transverse offset.

\begin{figure}[tb]
\begin{center}
\includegraphics[width=0.75\textwidth]{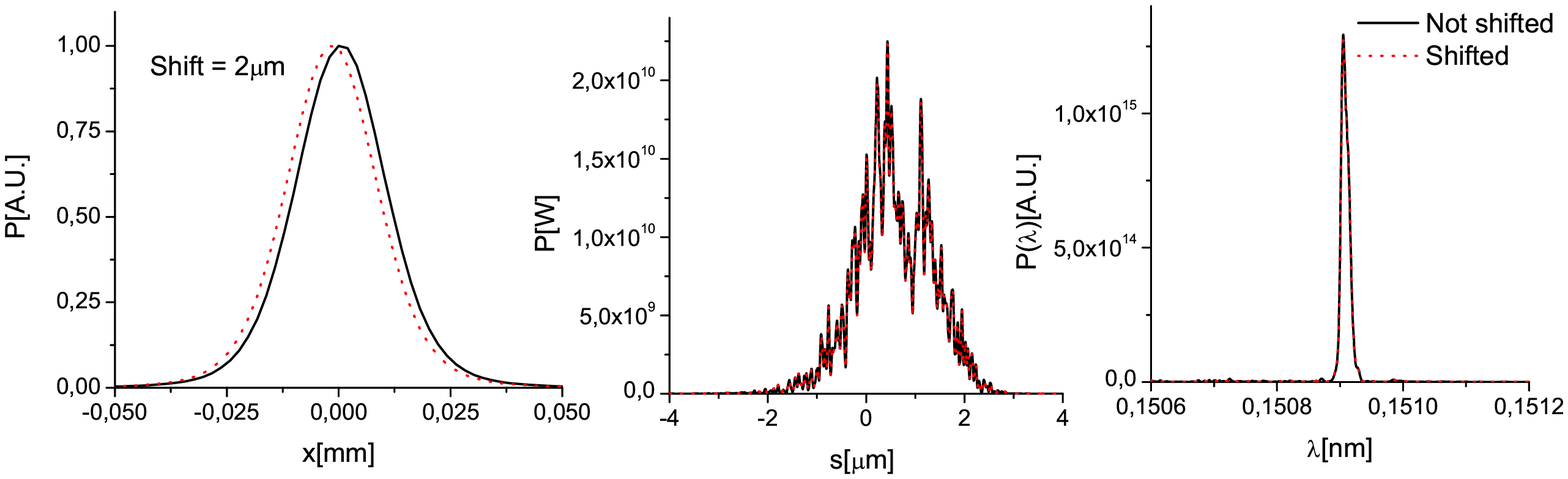}
\includegraphics[width=0.75\textwidth]{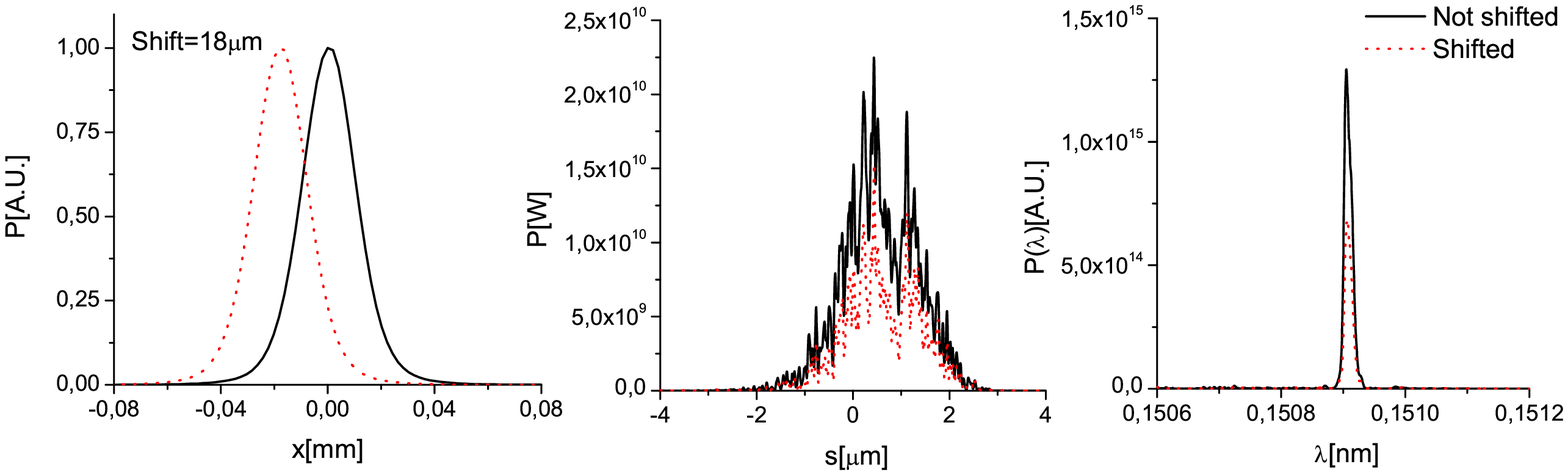}
\includegraphics[width=0.75\textwidth]{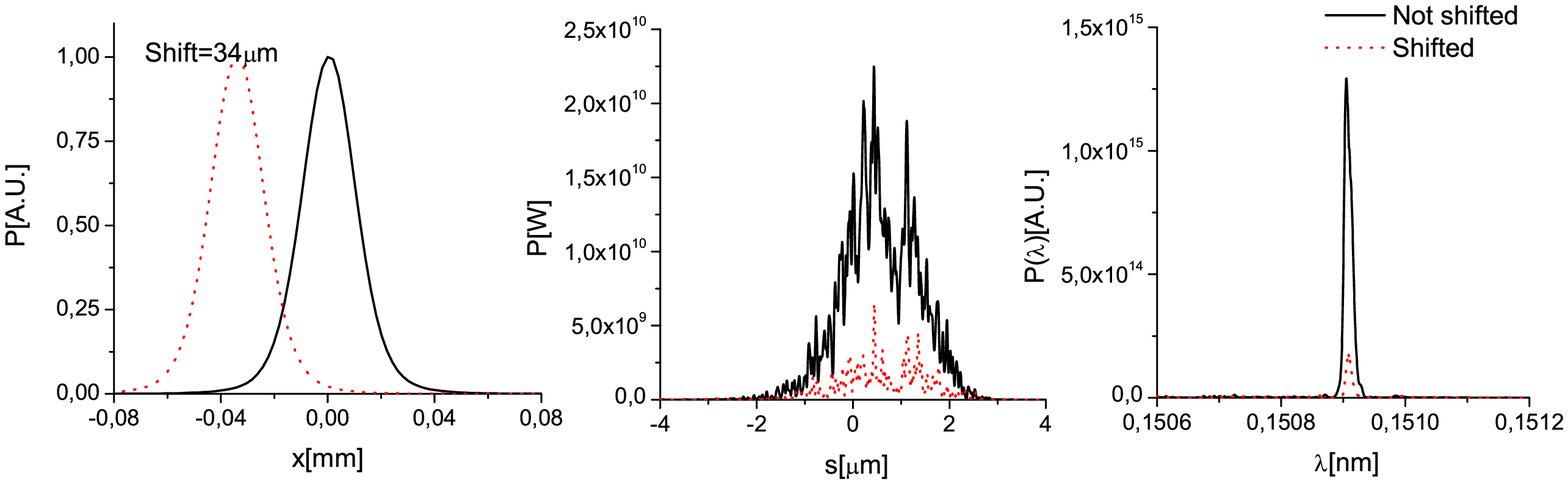}
\end{center}
\caption{Comparison between the self-seeded output with (red, dotted
lines) and without (black, solid lines) transverse offset, at
different offset values. The left plots show the photon beam
transverse profiles, the middle plots show the power, the right
plots show the spectra. All plots refer to a single simulation
shot.} \label{shift}
\end{figure}
A few examples of the output obtained with this procedure are shown
in Fig. \ref{shift}, which compares the result for zero offset
(black solid lines) with those for $2~\mu$m, $18~\mu$m, and
$34~\mu$m (red dotted lines). All results refer to a single
simulation shot. The plots on the left show the transverse profile
of the photon beam after $6$ undulator cells following the
self-seeding setup. The figures in the middle show the power, and
those on the left the spectra. Qualitatively, the effect due to the
transverse offset is easy to see by inspection.

\begin{figure}[tb]
\begin{center}
\includegraphics[width=0.75\textwidth]{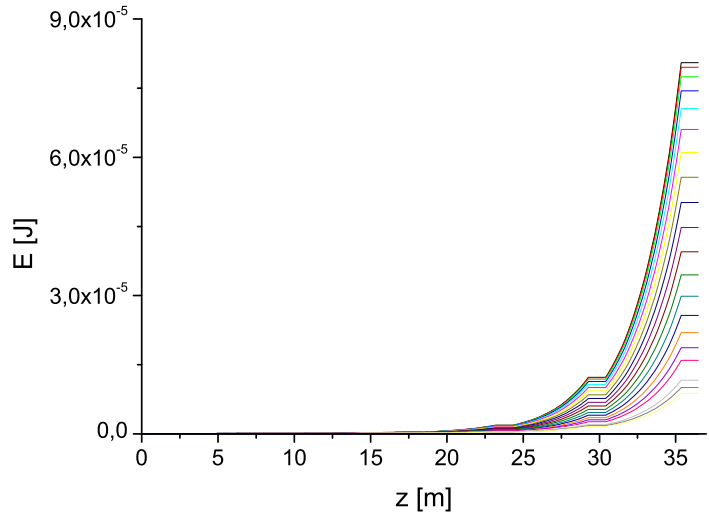}
\end{center}
\caption{Collected energies as a function of the position inside the
radiator, for increasing offset from $0~\mu$m to $40~\mu$m with
steps of $2~\mu$m.} \label{Enez}
\end{figure}

\begin{figure}[tb]
\begin{center}
\includegraphics[width=0.75\textwidth]{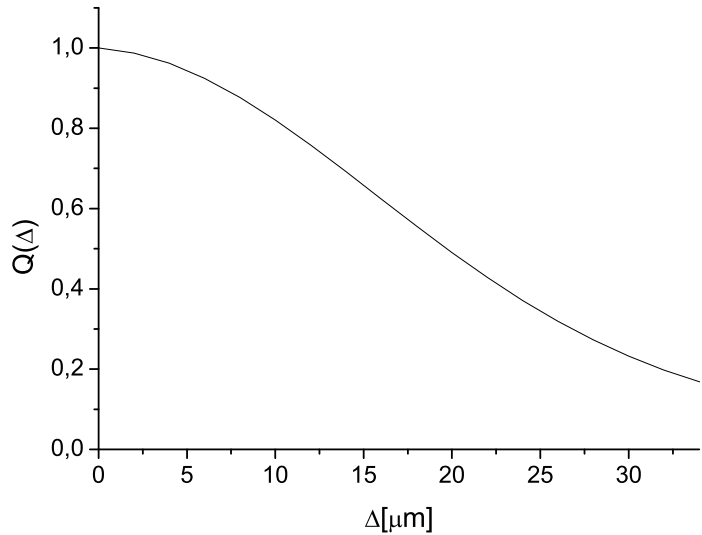}
\end{center}
\caption{Measure of the dependence of the input coupling factor on
the offset $Q(\Delta)$. } \label{QD}
\end{figure}
The energy as a function of the position into the radiator is shown
in Fig. \ref{Enez}. The function $Q(\Delta)$ instead, is shown in
Fig. \ref{QD}. Note that the input coupling factor does not drop
abruptly to zero when $\Delta$ is of order of the rms transverse
size of the photon beam. This is because the photon beam intensity
does not drop off as a Gaussian function. As a result, on can
qualitatively go to larger deviations of $\theta_B$, still obtaining
a reasonable coupling factor.

We hope that these considerations will be useful for FEL physicists
in the design stage of wake monochromator setups.

\section{\label{studies} FEL studies}

\begin{figure}[tb]
\begin{center}
\includegraphics[width=1.0\textwidth]{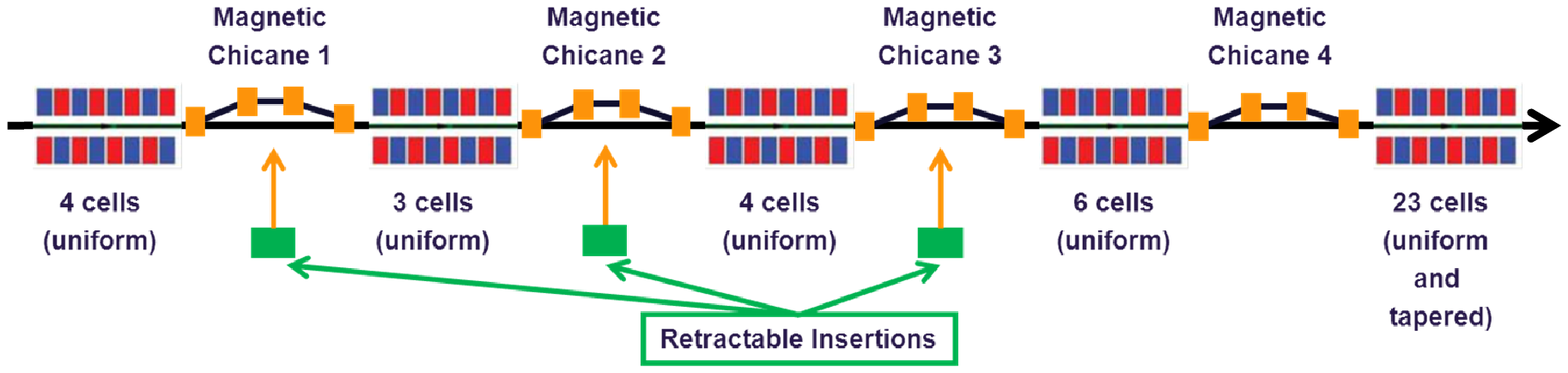}
\end{center}
\caption{Schematic setup of the bio-imaging beamline discussed in
\cite{OURCD}.} \label{setup}
\end{figure}

\begin{figure}[tb]
\includegraphics[width=0.5\textwidth]{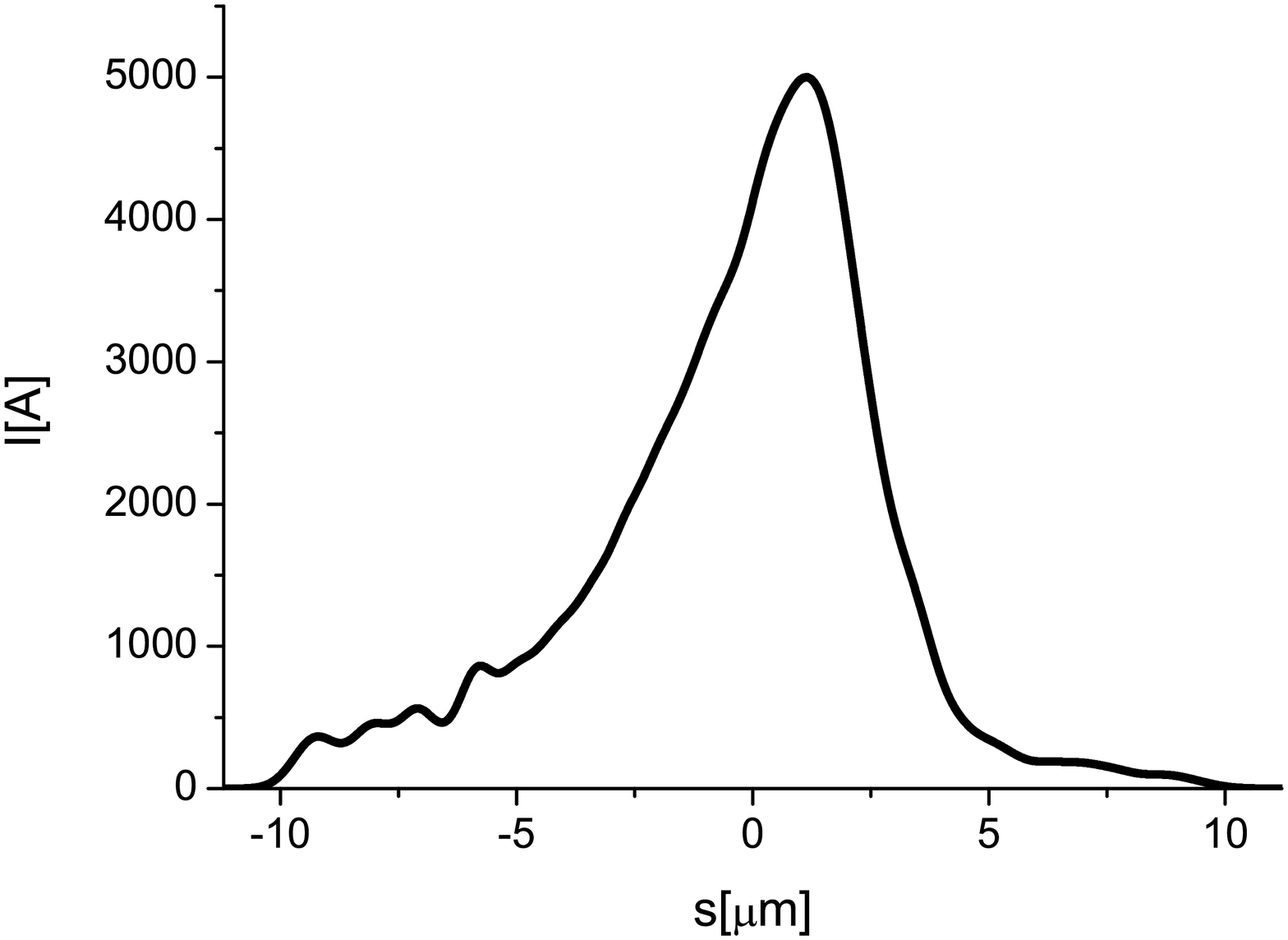}
\includegraphics[width=0.5\textwidth]{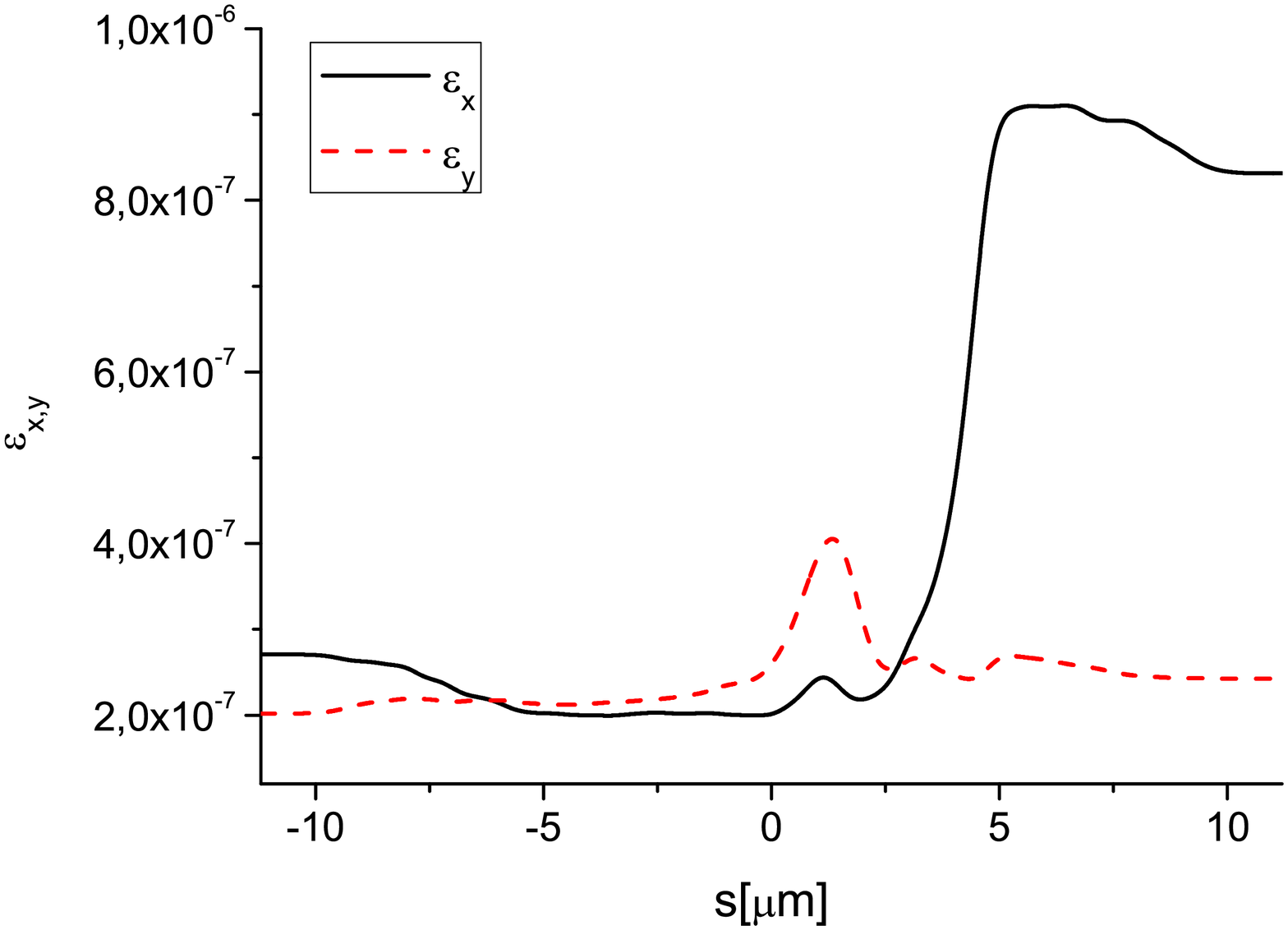}
\includegraphics[width=0.5\textwidth]{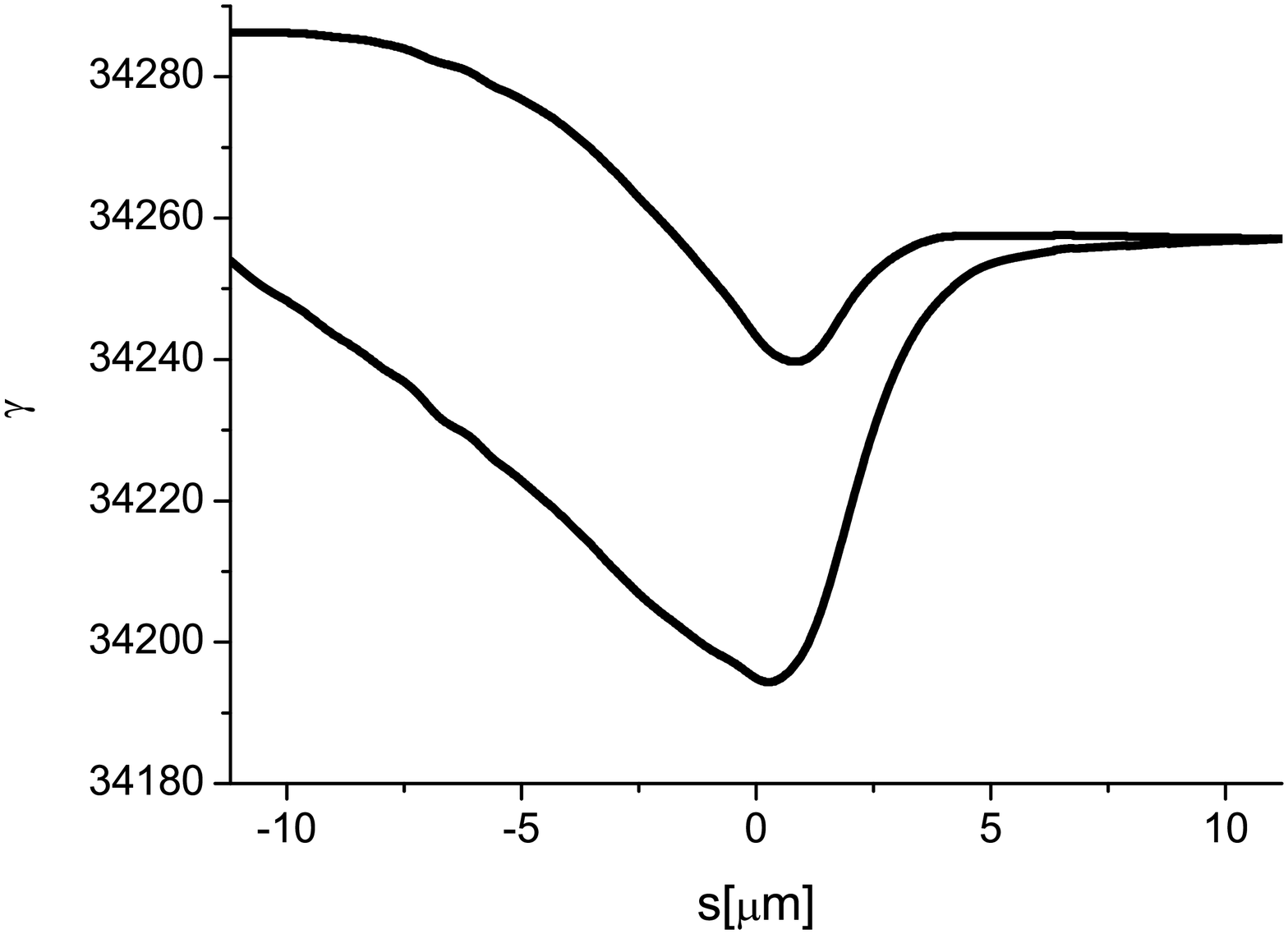}
\includegraphics[width=0.5\textwidth]{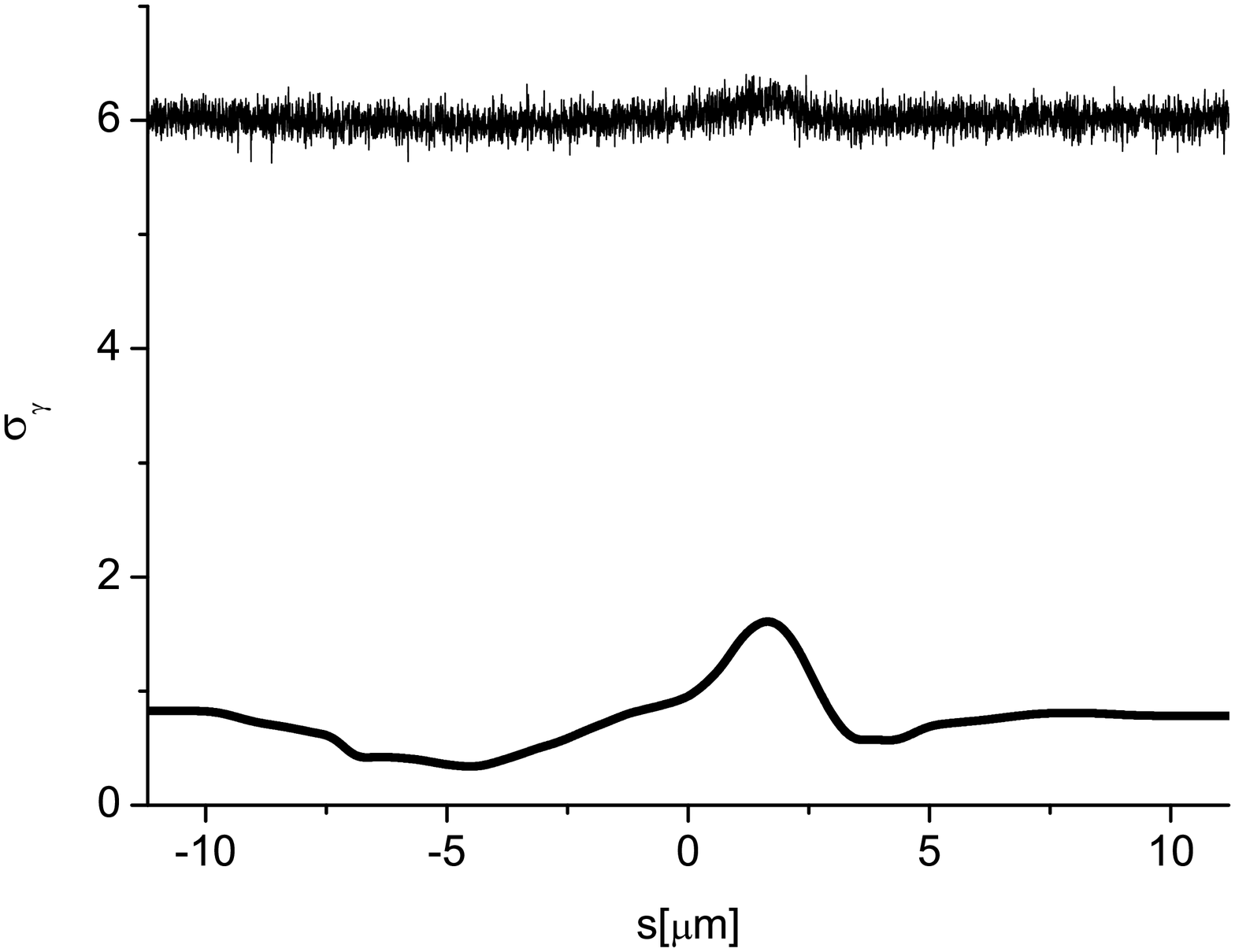}
\begin{center}
\includegraphics[width=0.5\textwidth]{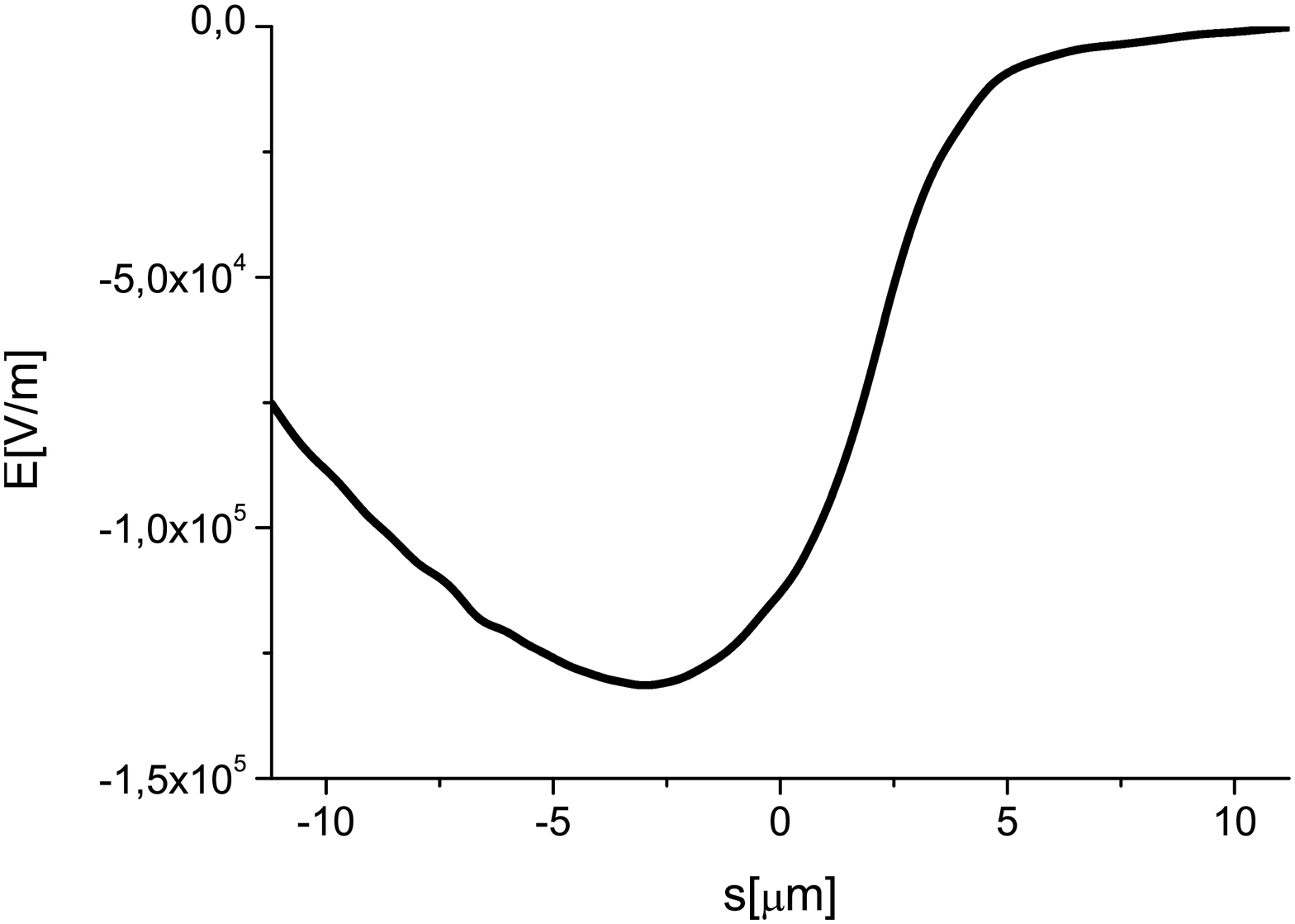}
\end{center}
\caption{Results from electron beam start-to-end simulations at the
entrance of the undulator system of the bio-imaging beamline
\cite{S2ER} for the hard X-ray case for the 17.5 GeV mode of
operation. (First Row, Left) Current profile. (First Row, Right)
Normalized emittance as a function of the position inside the
electron beam. (Second Row, Left) Energy profile along the beam,
lower curve. The effects of resistive wakefields along SASE1 are
illustrated by the comparison with the upper curve, referring to the
entrance of SASE1 (Second Row, Right) Electron beam energy spread
profile, upper curve. The effects of quantum diffusion along SASE1
are illustrated by the comparison with the lower curve, referring to
the entrance of SASE1. (Bottom row) Resistive wakefields in the
SASE3 undulator \cite{S2ER}.} \label{biof2f3}
\end{figure}

In order to quantitatively investigate the effects of the spatial
shift on the FEL amplification process, we performed numerical
simulations using the code Genesis \cite{GENE}. Simulations are
based on a statistical analysis consisting of $100$ runs. We
consider the setup in Fig. \ref{setup}, with reference to
\cite{OURCD} for details. In this article we are interested in the
energy range where hard X-ray self seeding can be implemented,
starting from $3.5$ keV up to $13$ keV. The C(111) reflections
(Bragg and Laue) will be used in the second chicane. C(113), C(004)
and C(333) reflections will be used in the third chicane.  Since we
are interested in energies starting from $3.5$ keV, we fix the
electron energy at $17.5$ GeV. The lower energy is used in the very
soft X-ray regime, see \cite{OURCD}, between $0.3$ keV and $0.5$
keV, which is outside the interests of this article, and where one
relies on a grating monochromator. In Fig. \ref{biof2f3} we plot the
results of start-to-end simulations \cite{S2ER}, while the main
undulator parameters are reported in Table \ref{tt1}. Our
simulations automatically include the influence of the spatial shift
of the photon beam induced by the crystal. We will present results
for the different reflections at given energy points.

\begin{table}
\caption{Undulator parameters}

\begin{small}\begin{tabular}{ l c c}
\hline & ~ Units &  ~ \\ \hline
Undulator period      & mm                  & 68     \\
Periods per cell      & -                   & 73   \\
Total number of cells & -                   & 40    \\
Intersection length   & m                   & 1.1   \\
Photon energy         & keV                 & 0.3-13 \\
\hline
\end{tabular}\end{small}
\label{tt1}
\end{table}

As is explained in detail in \cite{OURCD}, in the range between 3
keV and 5 keV the first chicane is not used and is switched off.
After the first 7 cells the electron and the photon beams are
separated with the help of the second magnetic chicane, and the
C(111) reflection is used to monochromatize the radiation. The seed
is amplified in the next 4 cells. After that, the electron and the
photon beam are separated again by the third chicane, and an X-ray
optical delay line allows for the introduction of a tunable delay of
the photon beam with respect to the electron beam. The following 6
cells use only a part of the electron beam as a lasing medium. A
magnetic chicane follows, which shifts the unspoiled part of the
electron bunch on top of the of the photon beam. In this way, a
fresh bunch technique can be implemented. Since the delays are
tunable, the photon pulse length can be tuned. Finally, radiation is
amplified into the last $23$ tapered cells to provide pulses with
about 2 TW power.

In partial difference with respect to what has been discussed in
\cite{OURCD}, for energies lower than $5$ keV we used the asymmetric
C(111) reflection (Bragg and Laue geometry), while in \cite{OURCD}
we exploited the C(111) symmetric Bragg reflection only. Moreover,
for energies larger than $5$ keV we rely on a crystal placed after
$11$ cells, and we take advantage of the C(113), the C(004) and the
C(333) reflections. In this case the fourth chicane is switched off.
If tunability of the pulse duration is requested in this energy
range, this is most easily achieved by providing additional delay
with the fourth magnetic chicane installed behind the hard X-ray
self-seeding setup.

\subsection{C(111) asymmetric Bragg reflection at $3.5$ keV}

\begin{figure}[tb]
\includegraphics[width=0.5\textwidth]{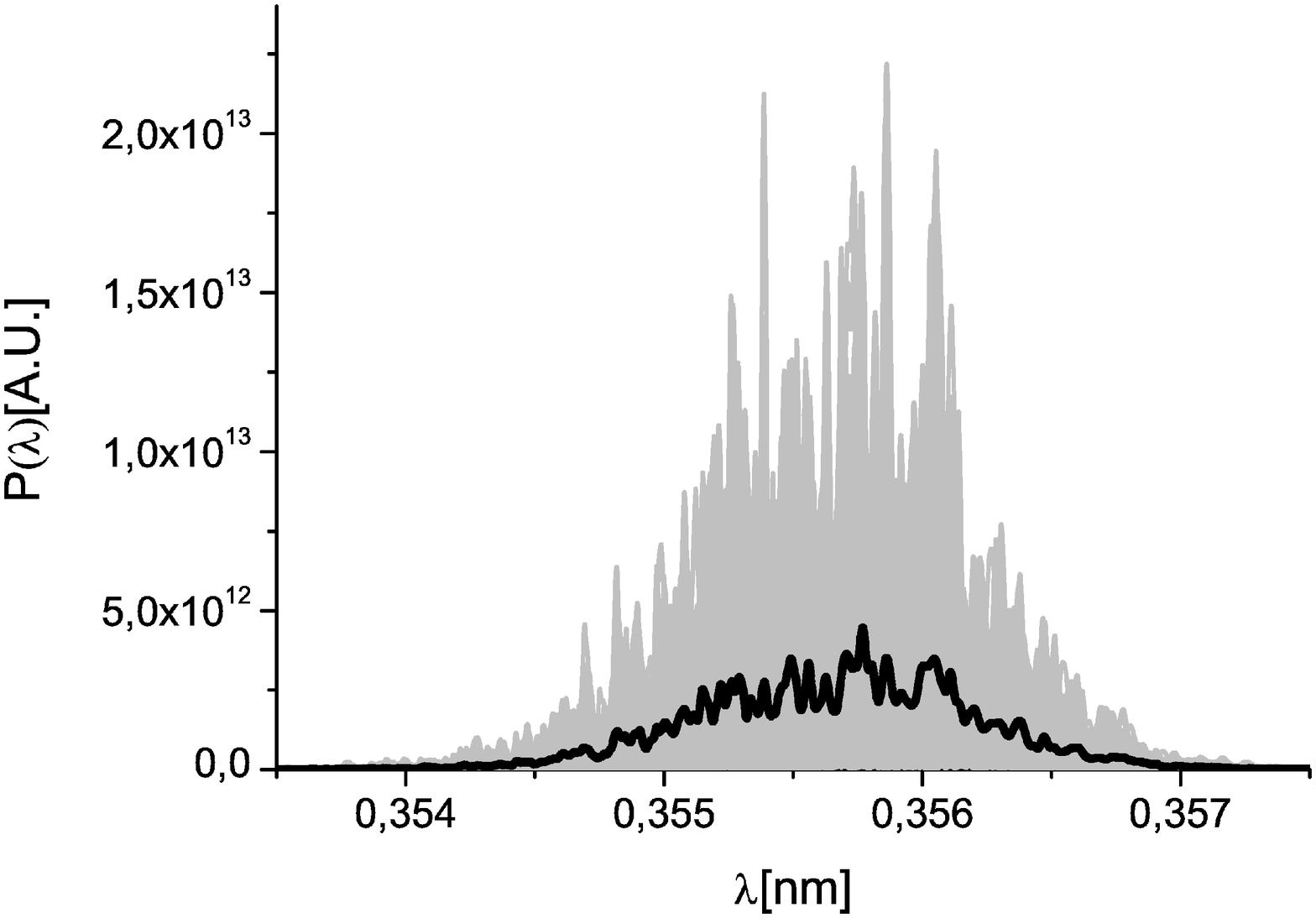}
\includegraphics[width=0.5\textwidth]{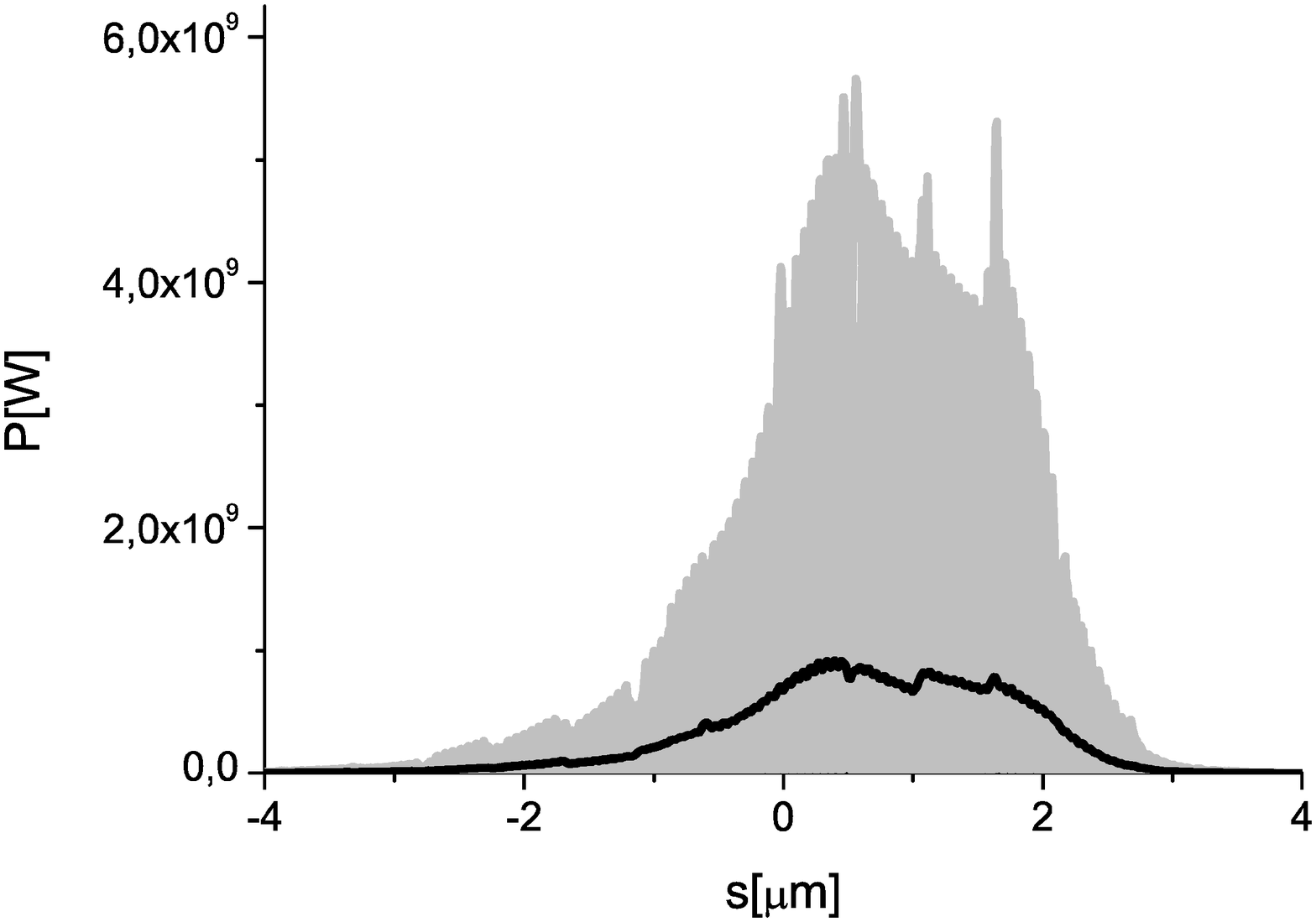}
\caption{Power and spectrum at $3.5$ keV before the second magnetic
chicane. Grey lines refer to single shot realizations, the black
line refers to the average over a hundred realizations.}
\label{biof175}
\end{figure}
As discussed above, the first chicane is switched off, so that the
first part of the undulator effectively consists of 7 uniform cells.
We begin our investigation by simulating the SASE power and spectrum
after the first part of the undulator, that is before the second
magnetic chicane in the setup. Results are shown in Fig.
\ref{biof175}.

\begin{figure}[tb]
\includegraphics[width=0.5\textwidth]{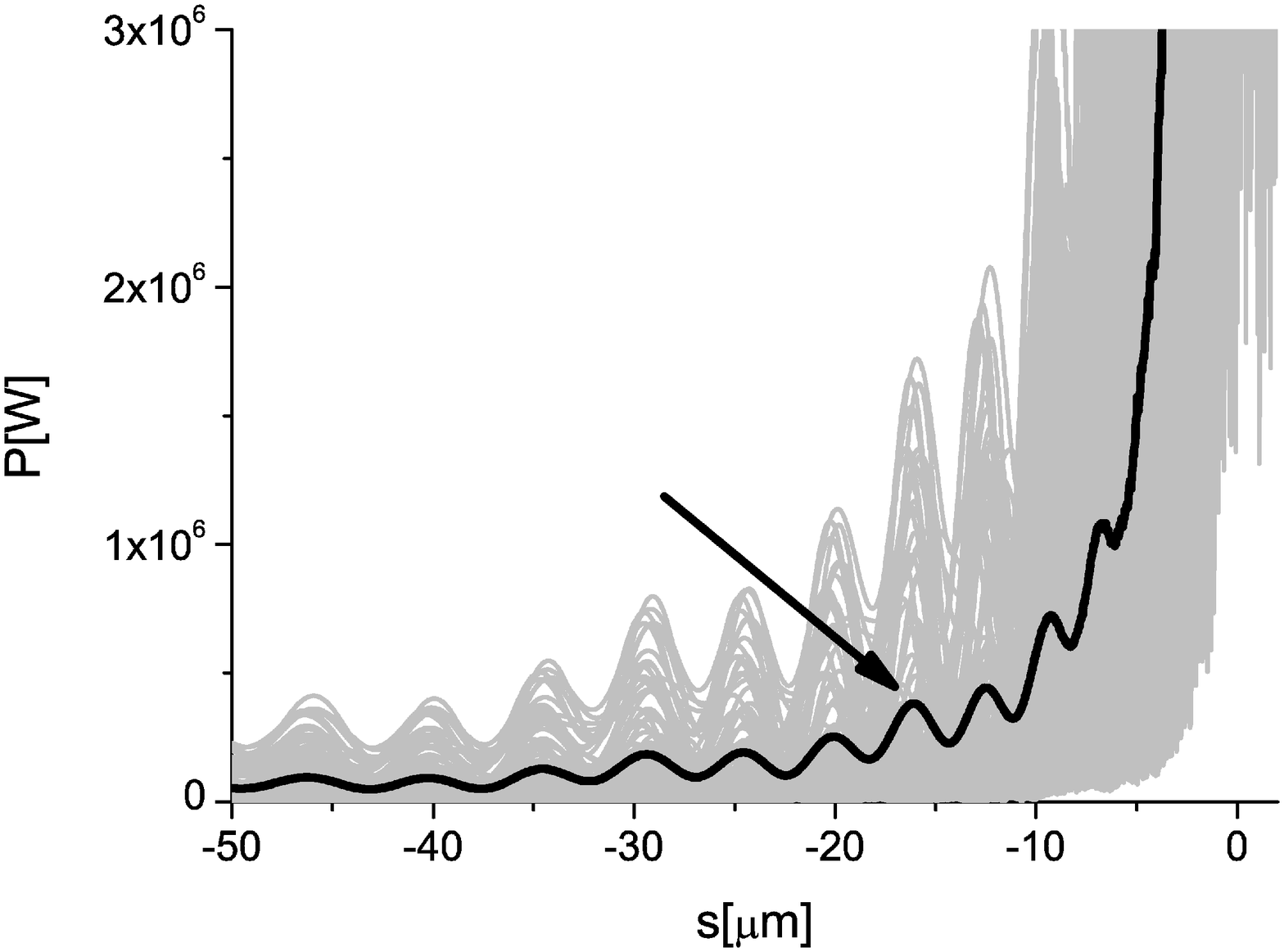}
\includegraphics[width=0.5\textwidth]{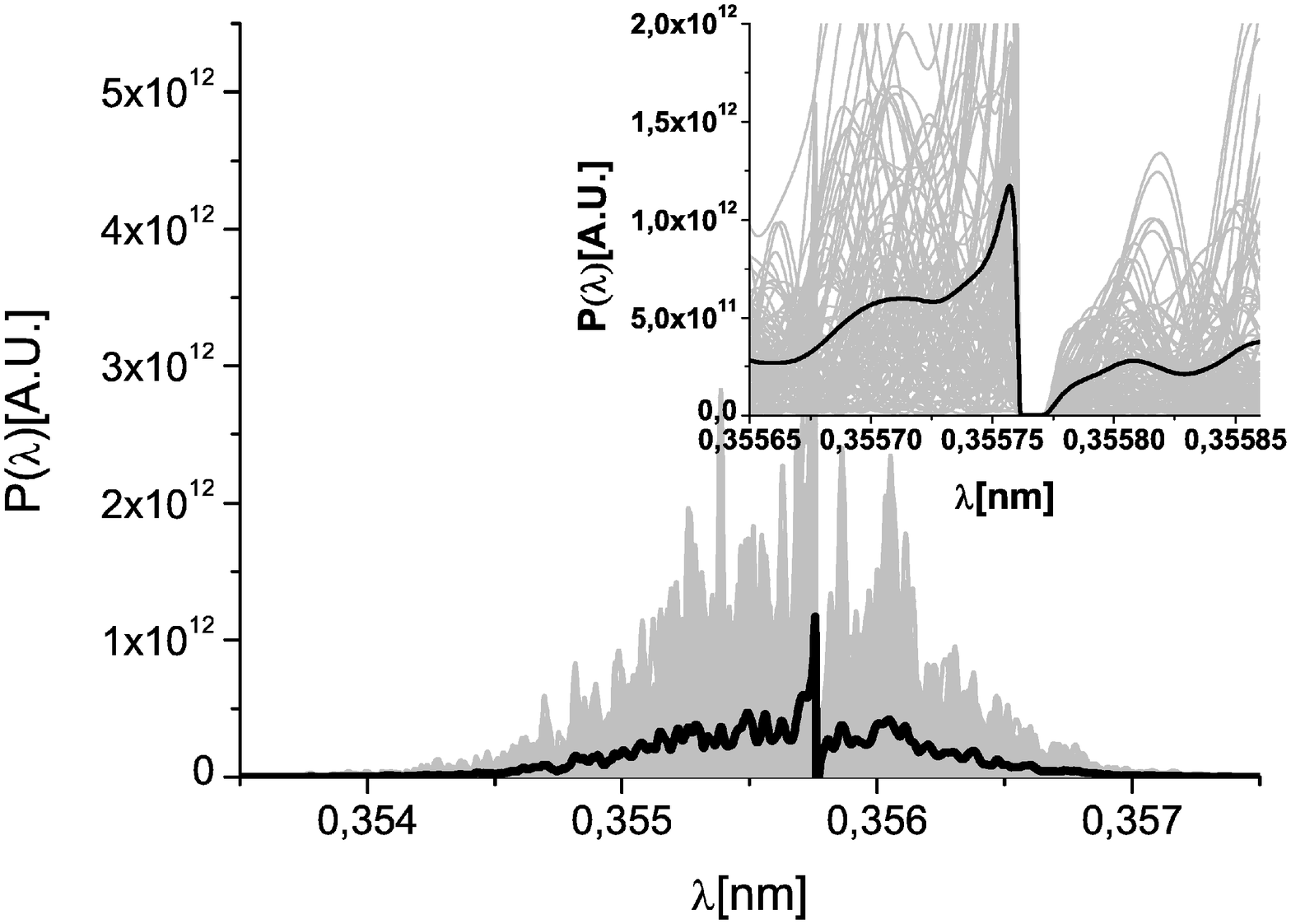}
\caption{Power and spectrum at $3.5$ keV after the single crystal
self-seeding X-ray monochromator. A $100~\mu$m thick diamond crystal
( C(111) Bragg reflection, $\sigma$-polarization ) is used. Grey
lines refer to single shot realizations, the black line refers to
the average over a hundred realizations. The black arrow indicates
the seeding region.} \label{biof183p5}
\end{figure}

\begin{figure}[tb]
\includegraphics[width=0.75\textwidth]{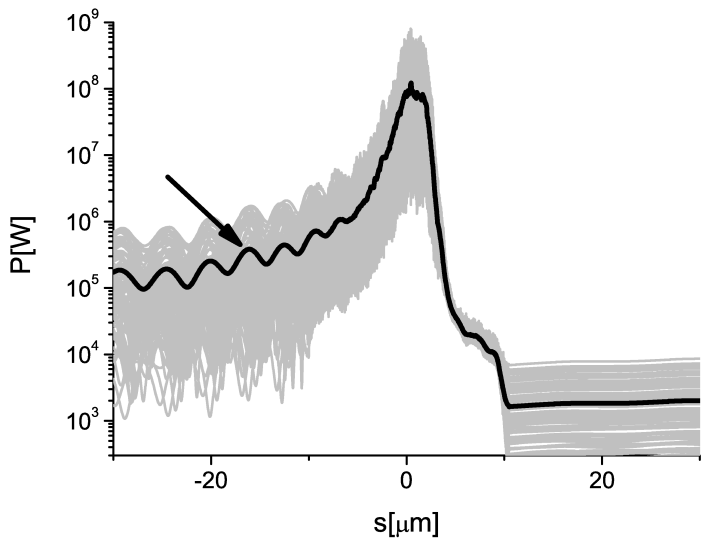}
\caption{Logarithmic plot of the power at $3.5$ keV after the single
crystal self-seeding X-ray monochromator.  The black arrow indicates
the seeding region. The region on the right hand side of the plot
(before the FEL pulse) is nominally zero because of causality
reasons. Differences with respect to zero give back the accuracy of
our calculations.} \label{3caus}
\end{figure}
The second magnetic chicane is switched on, and the single-crystal
X-ray monochromator is set into the photon beam. We take advantage
of the C(111) Bragg reflection, $\sigma$-polarization, Fig.
\ref{biof183p5}. As discussed before, the modulus and the phase of
the crystal transmissivity are related by Kramers-Kroening
relations.  The numerical accuracy with which causality is satisfied
in our simulations can be shown by a logarithmic plot of the FEL
pulse power after the crystal, Fig. \ref{3caus}. The peak in the
center is the main FEL pulse. On the left side one can identify the
seed pulse.  The large jump between the power after transmission (on
the right side of the peak in the center) is due to causality, and
is strictly connected with the fact that the phase in the
transmission function is properly accounted for. For example, if a
transmission without phase were used, one would have obtained a
symmetric behavior, which is not causal at all. The small ($\sim
10^{-3}$) but visible departure from the exact zero appearing in the
right-hand side of Fig. \ref{3caus} is related with the accuracy of
our calculations. This accuracy is acceptable for most purposes.

\begin{figure}[tb]
\includegraphics[width=0.5\textwidth]{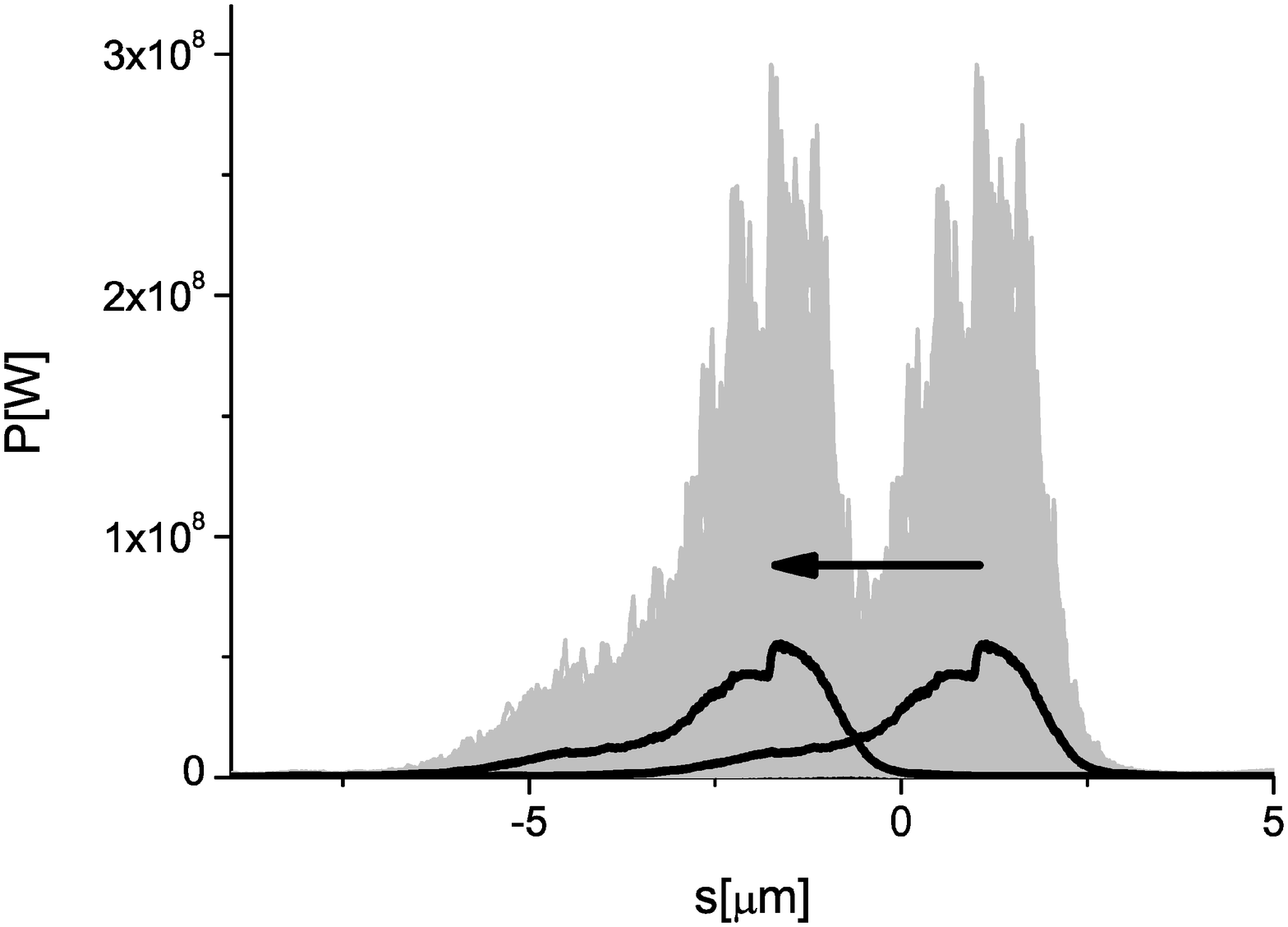}
\includegraphics[width=0.5\textwidth]{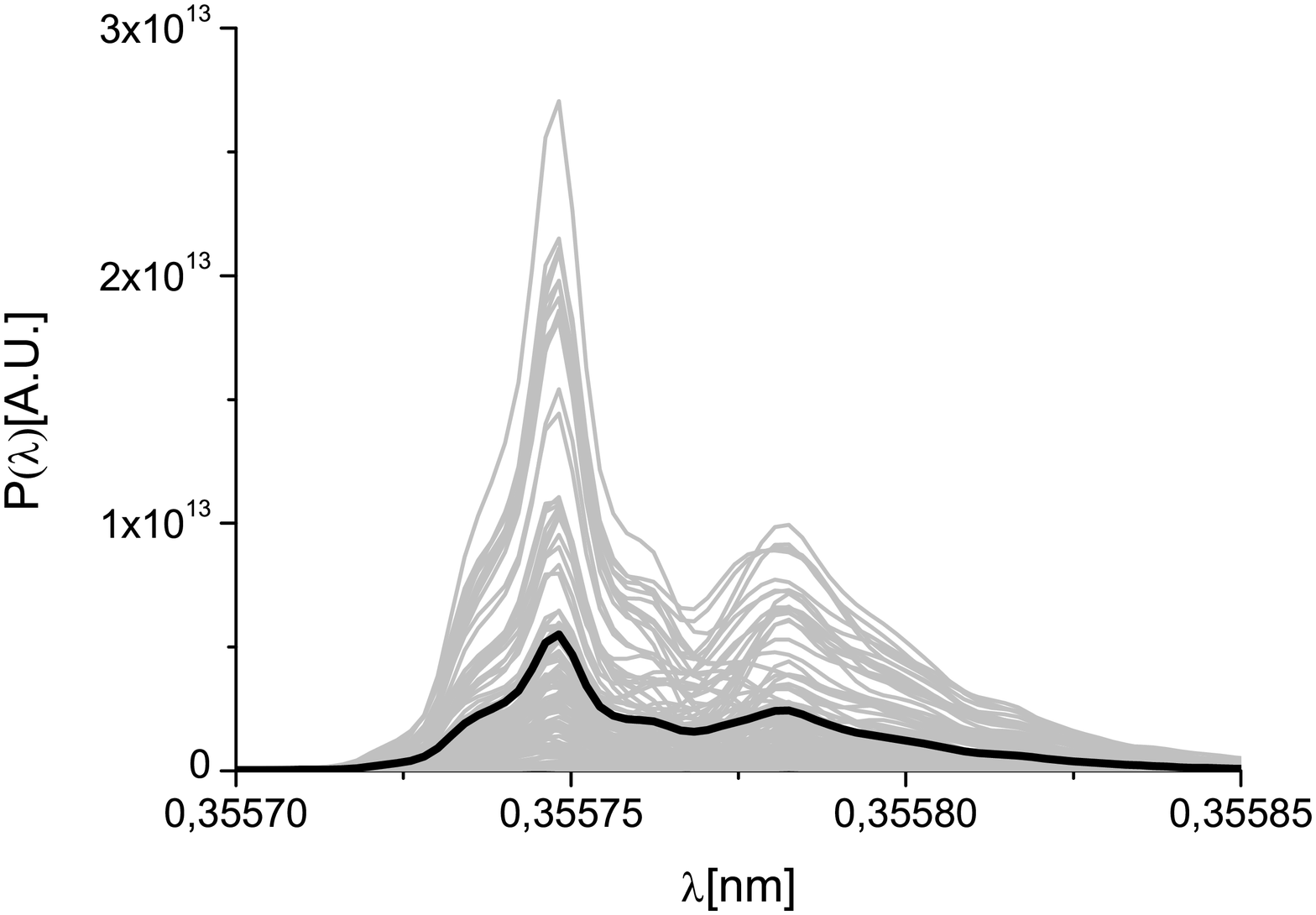}
\caption{Power and spectrum at $3.5$ keV after the third chicane
equipped with the X-ray optical delay line, delaying the radiation
pulse with respect to the electron bunch. Grey lines refer to single
shot realizations, the black line refers to the average over a
hundred realizations.} \label{biof203p5}
\end{figure}
Following the seeding setup, the electron bunch amplifies the seed
in the following 4 undulator cells. After that, a third chicane is
used to allow for the installation of an x-ray optical delay line,
which delays the radiation pulse with respect to the electron bunch.
The power and spectrum of the radiation pulse after the optical
delay line are shown in Fig. \ref{biof203p5}, where the effect of
the optical delay is illustrated.

\begin{figure}[tb]
\includegraphics[width=0.5\textwidth]{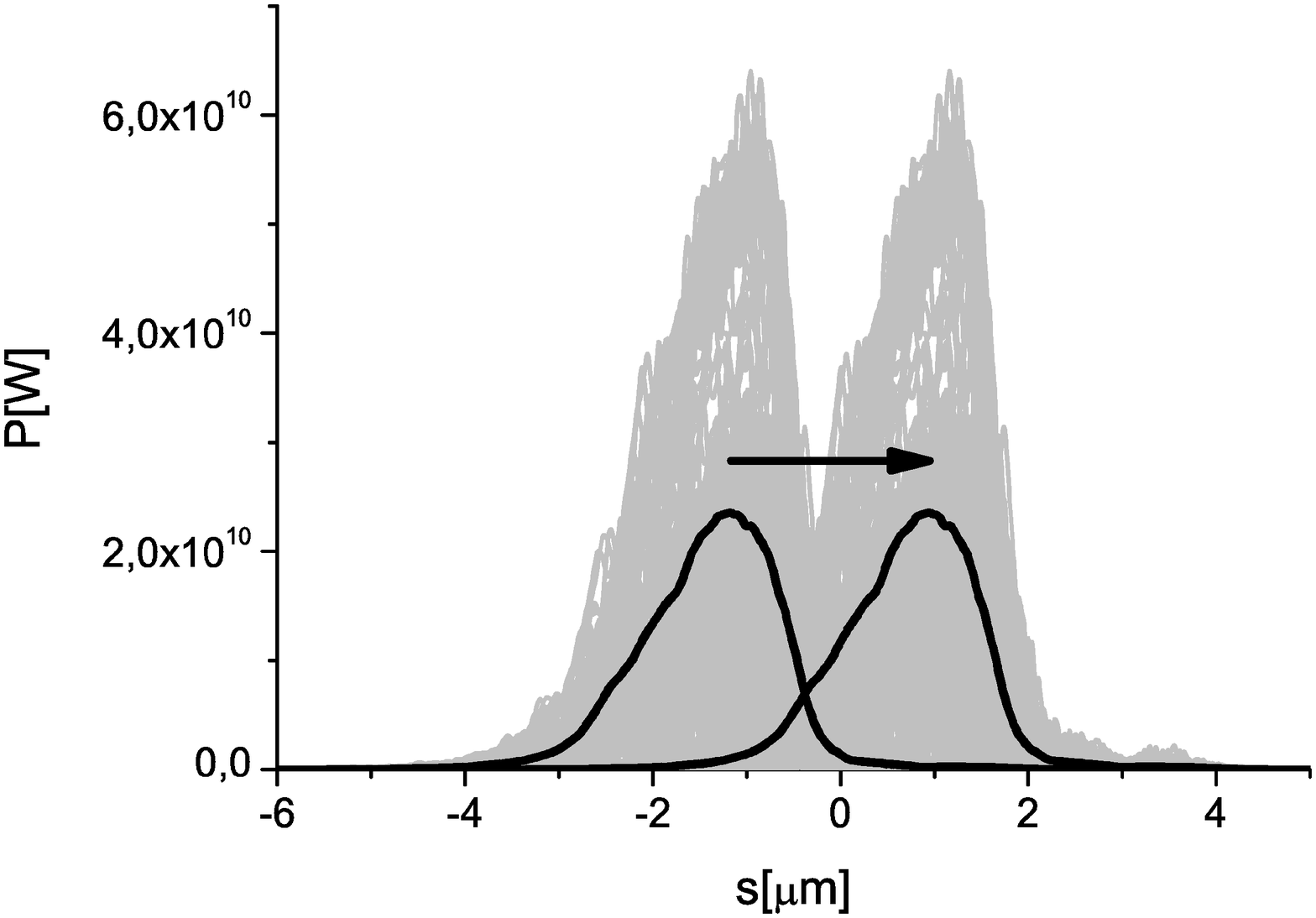}
\includegraphics[width=0.5\textwidth]{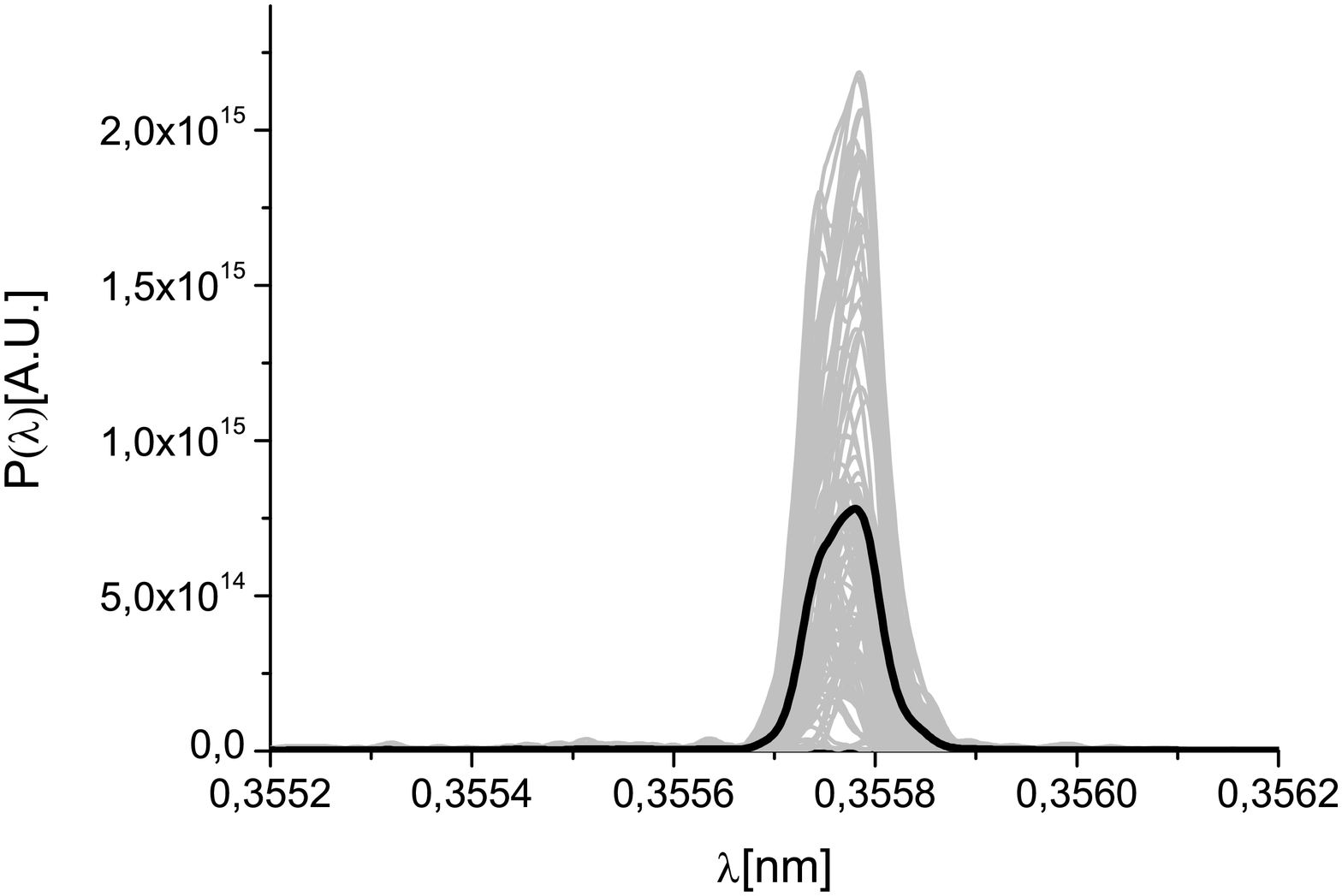}
\caption{Power and spectrum at $3.5$ keV after the last magnetic
chicane. Grey lines refer to single shot realizations, the black
line refers to the average over a hundred realizations.}
\label{biof215}
\end{figure}
Due to the presence of the optical delay, only part of the electron
beam is used to further amplify the radiation pulse in the following
6 undulator cells. The electron beam part which is not exploited is
fresh, and can further lase. In order to allow for that, after
amplification, the electron beam passes through the final magnetic
chicane, which delays the electron beam. The power and spectrum of
the radiation pulse after the last magnetic chicane are shown in
Fig. \ref{biof215}. By delaying the electron bunch, the magnetic
chicane effectively shifts forward the photon beam with respect to
the electron beam. Tunability of such shift allows the selection of
different photon pulse length.

\begin{figure}[tb]
\begin{center}
\includegraphics[width=0.5\textwidth]{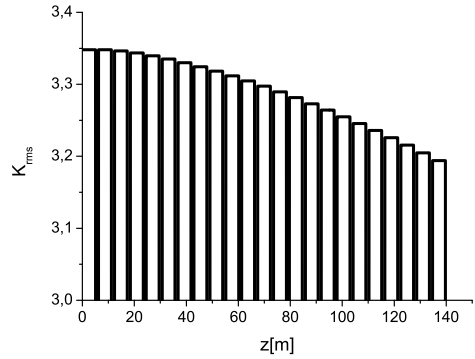}
\end{center}
\caption{Tapering law at $3.5$ keV.} \label{biof225}
\end{figure}
The last part of the undulator is composed by $23$ cells. It is
partly tapered post-saturation, to increase the region where
electrons and radiation interact properly to the advantage of the
radiation pulse. Tapering is implemented by changing the $K$
parameter of the undulator segment by segment according to Fig.
\ref{biof225}. The tapering law used in this work has been
implemented on an empirical basis.

\begin{figure}[tb]
\includegraphics[width=0.5\textwidth]{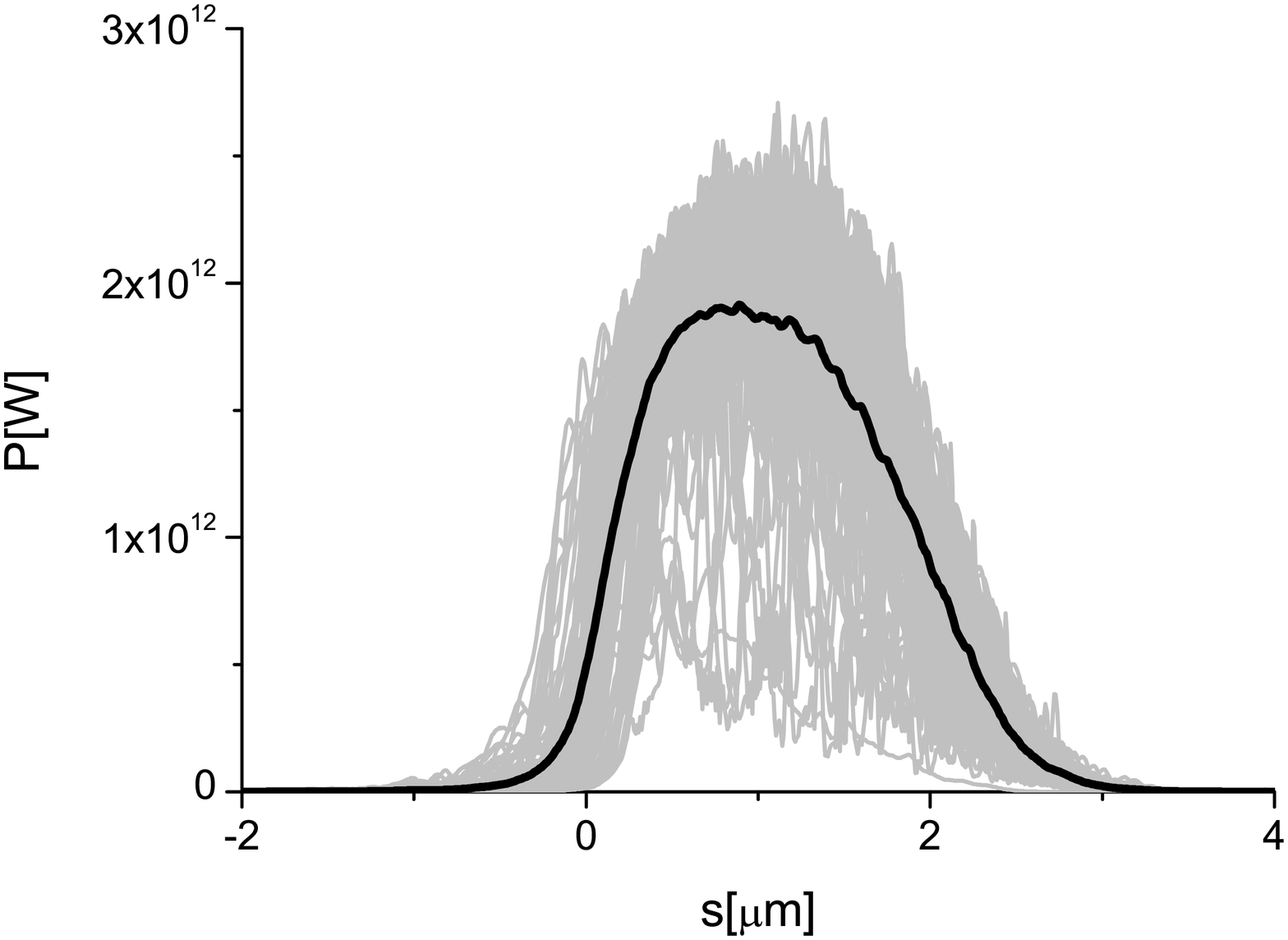}
\includegraphics[width=0.5\textwidth]{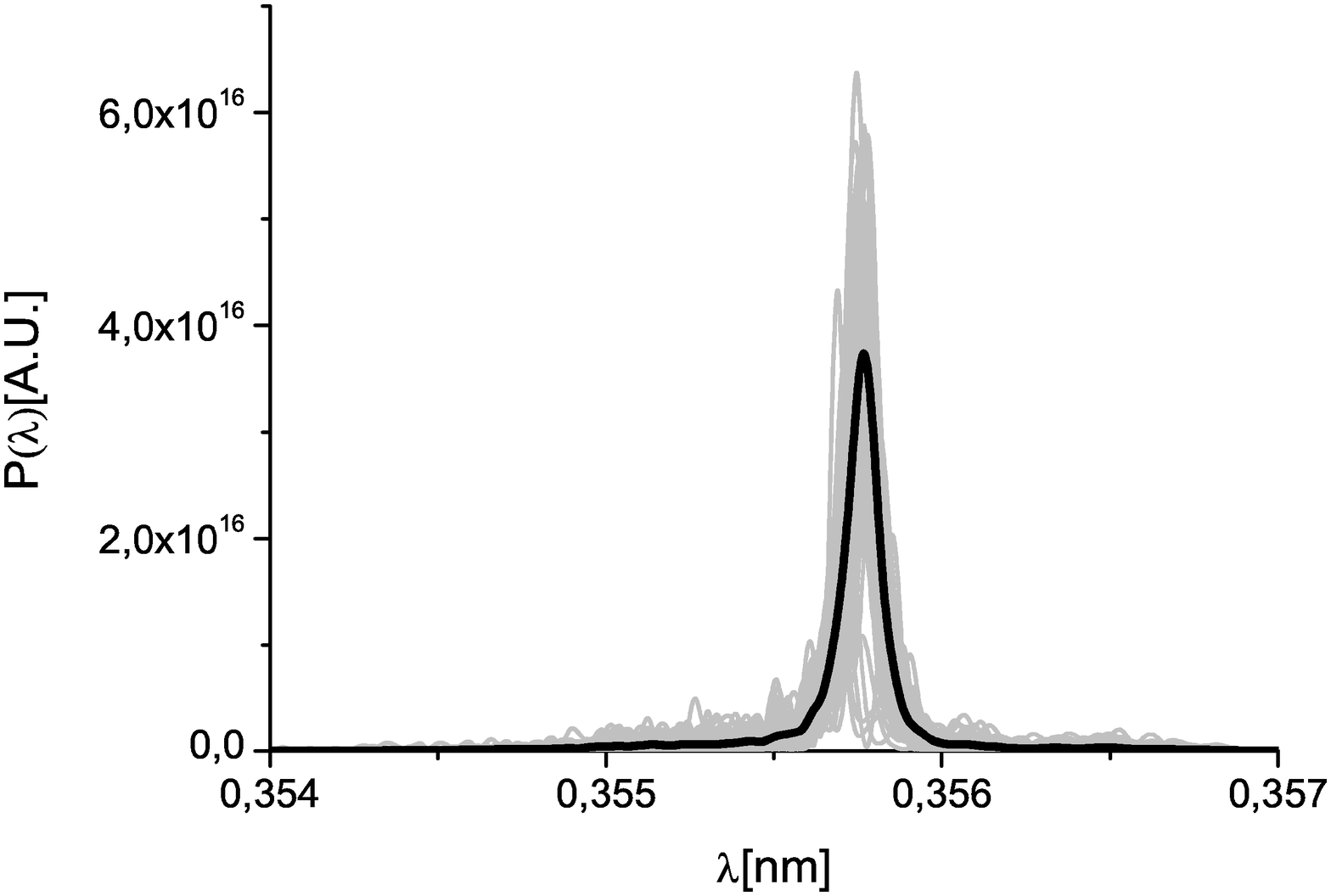}
\caption{Final output. Power and spectrum after tapering at $3.5$
keV. Grey lines refer to single shot realizations, the black line
refers to the average over a hundred realizations.} \label{biof235}
\end{figure}
The use of tapering together with monochromatic radiation is
particularly effective, since the electron beam does not experience
brisk changes of the ponderomotive potential during the slippage
process. The final output is presented in Fig. \ref{biof235} in
terms of power and spectrum. As one can see, simulations indicate an
output power of about $2$ TW.

\begin{figure}[tb]
\includegraphics[width=0.5\textwidth]{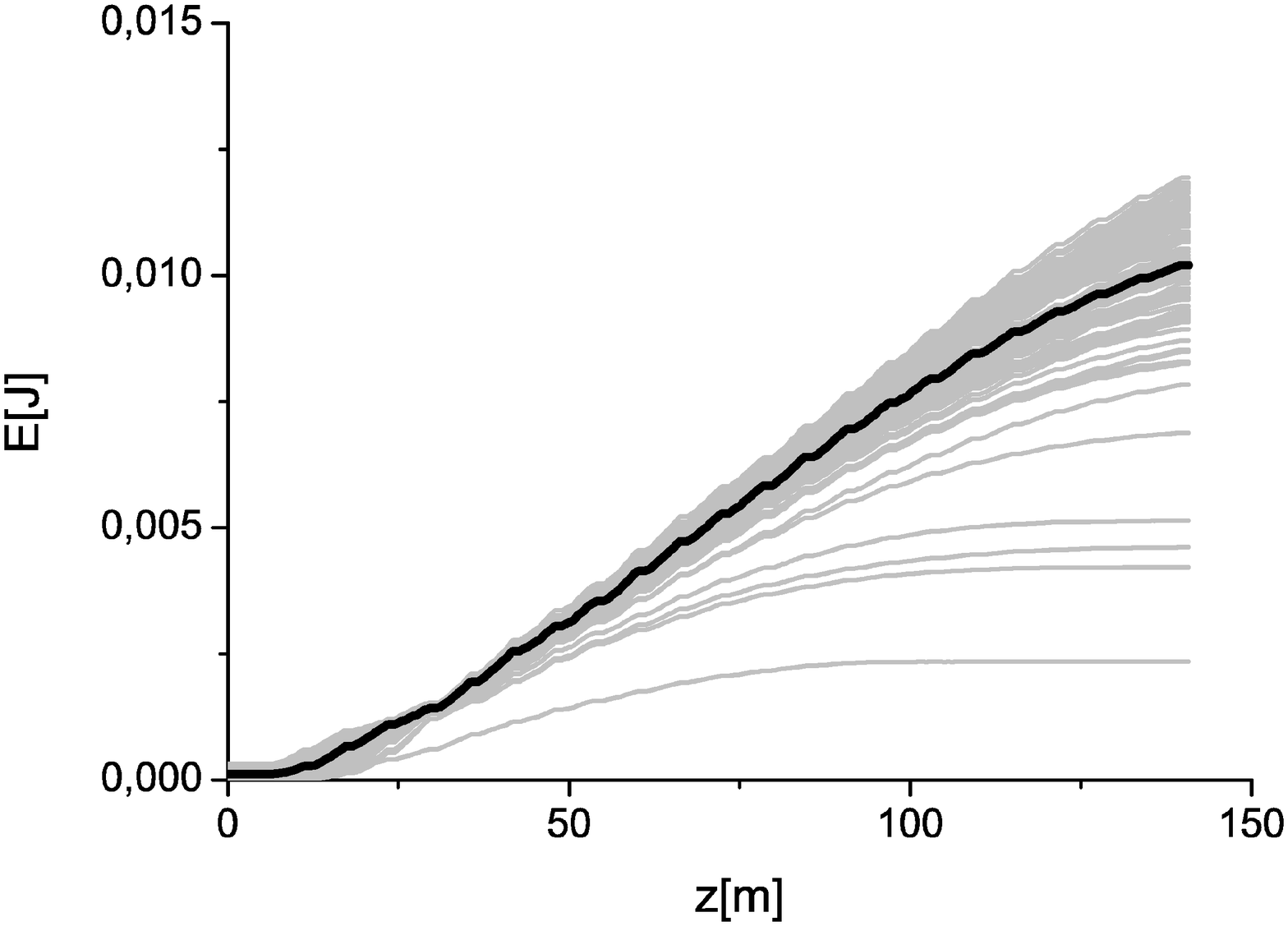}
\includegraphics[width=0.5\textwidth]{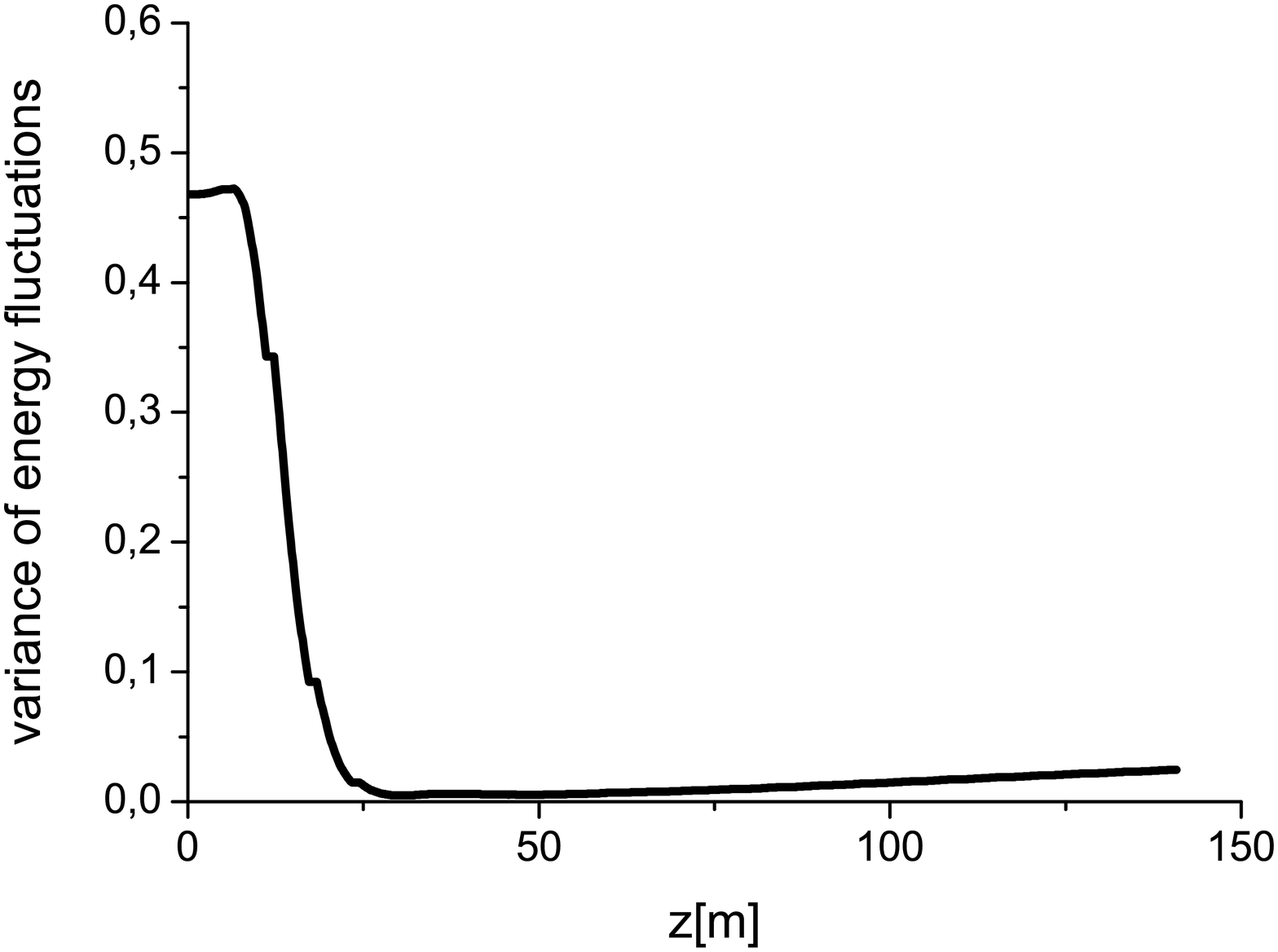}
\caption{Final output. Energy and energy variance of output pulses
at $3.5$ keV. In the left plot, grey lines refer to single shot
realizations, the black line refers to the average over a hundred
realizations.} \label{biof245}
\end{figure}

\begin{figure}[tb]
\includegraphics[width=0.5\textwidth]{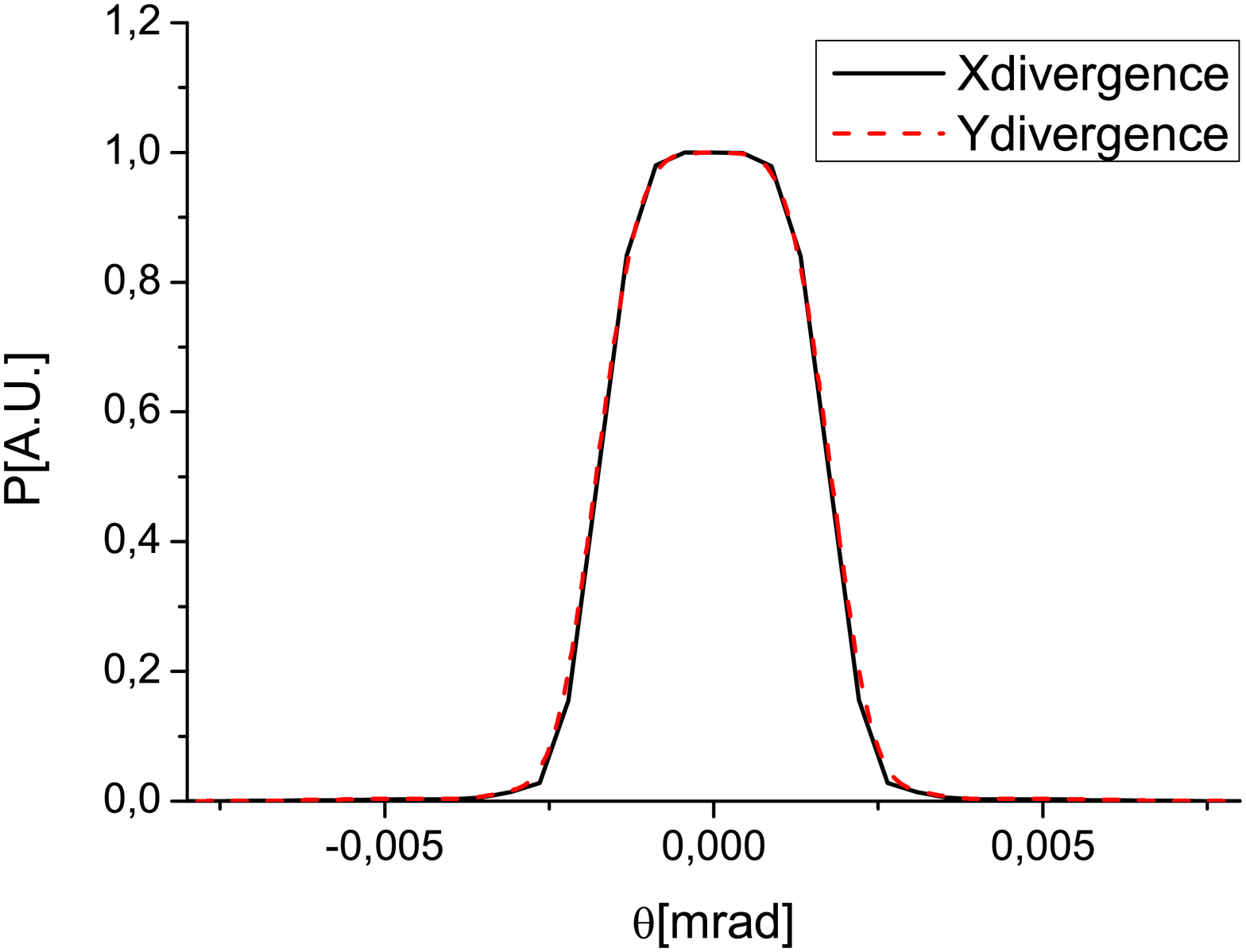}
\includegraphics[width=0.5\textwidth]{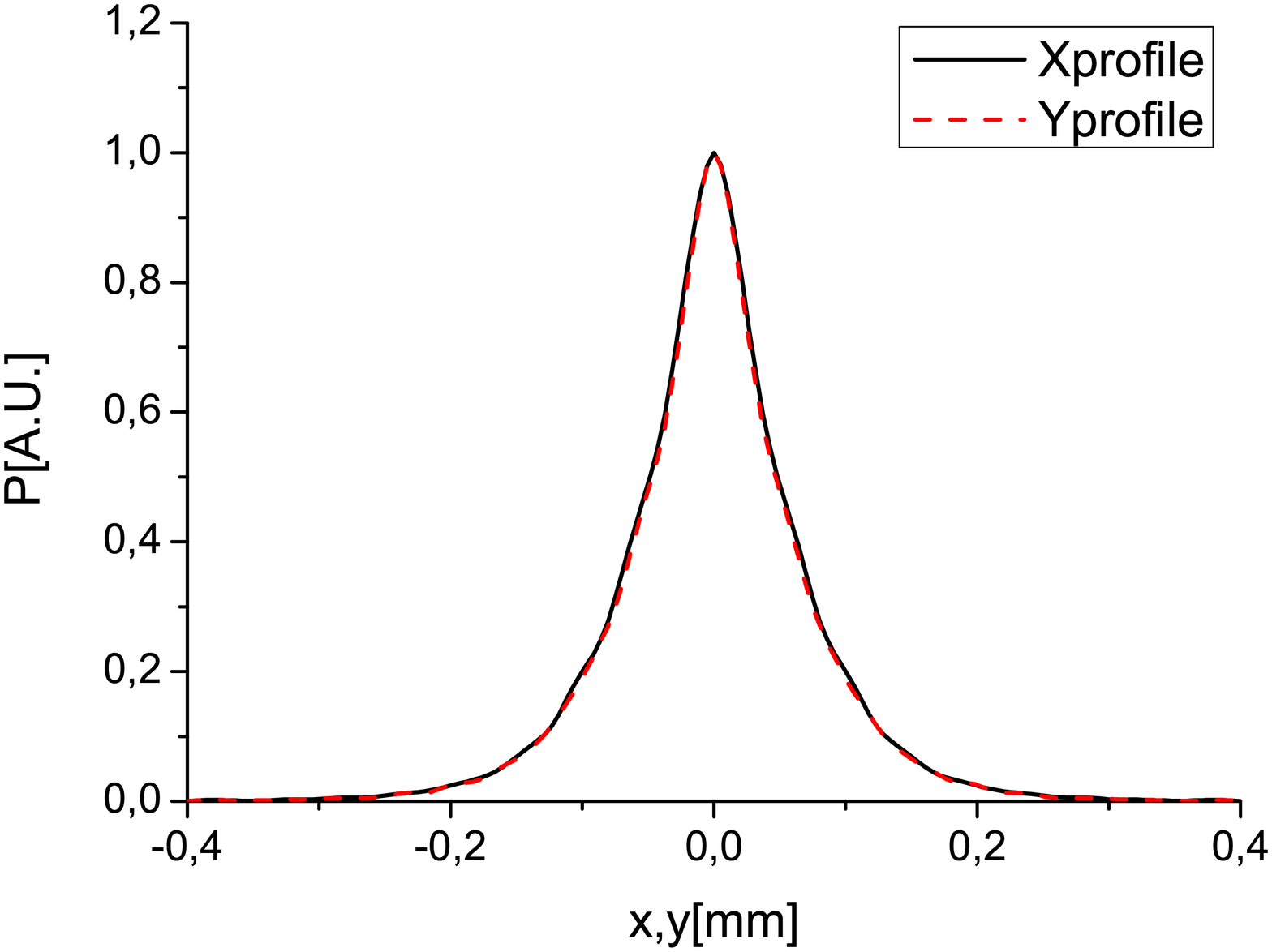}
\caption{Final output. X-ray radiation pulse energy distribution per
unit surface and angular distribution of the X-ray pulse energy at
$3.5$ keV at the exit of output undulator.} \label{biof253p5}
\end{figure}
The energy of the radiation pulse and the energy variance are shown
in Fig. \ref{biof245} as a function of the position along the
undulator. The divergence and the size of the radiation pulse at the
exit of the final undulator are shown, instead, in Fig.
\ref{biof253p5}. In order to calculate the size, an average of the
transverse intensity profiles is taken. In order to calculate the
divergence, the spatial Fourier transform of the field is
calculated.

\subsection{C(111) asymmetric Laue reflection at $5.0$ keV}

Operation at $5$ keV is identical to the case for $3.5$ keV. The
only difference is that now the Laue C(111) reflection is used.

\begin{figure}[tb]
\includegraphics[width=0.5\textwidth]{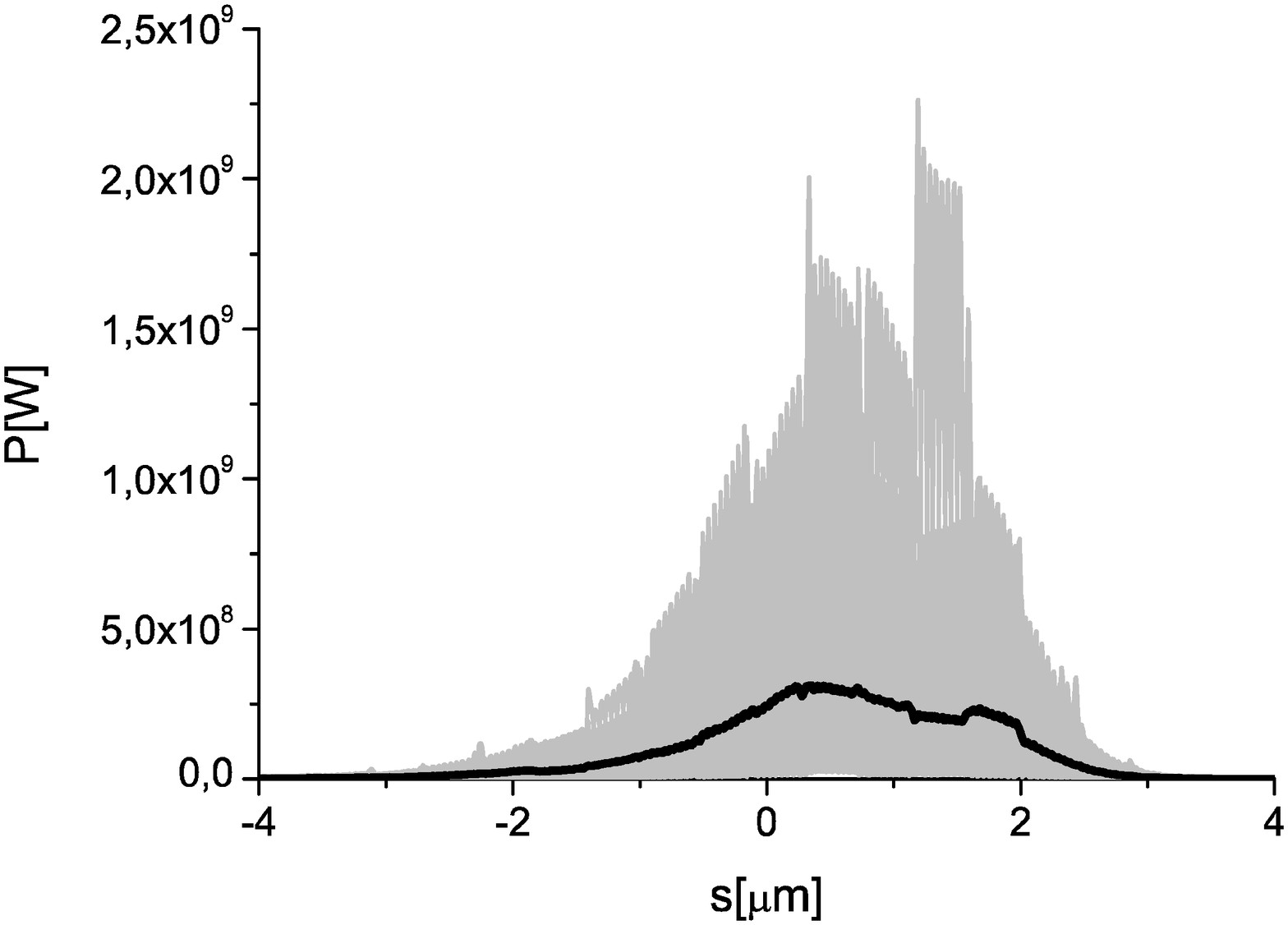}
\includegraphics[width=0.5\textwidth]{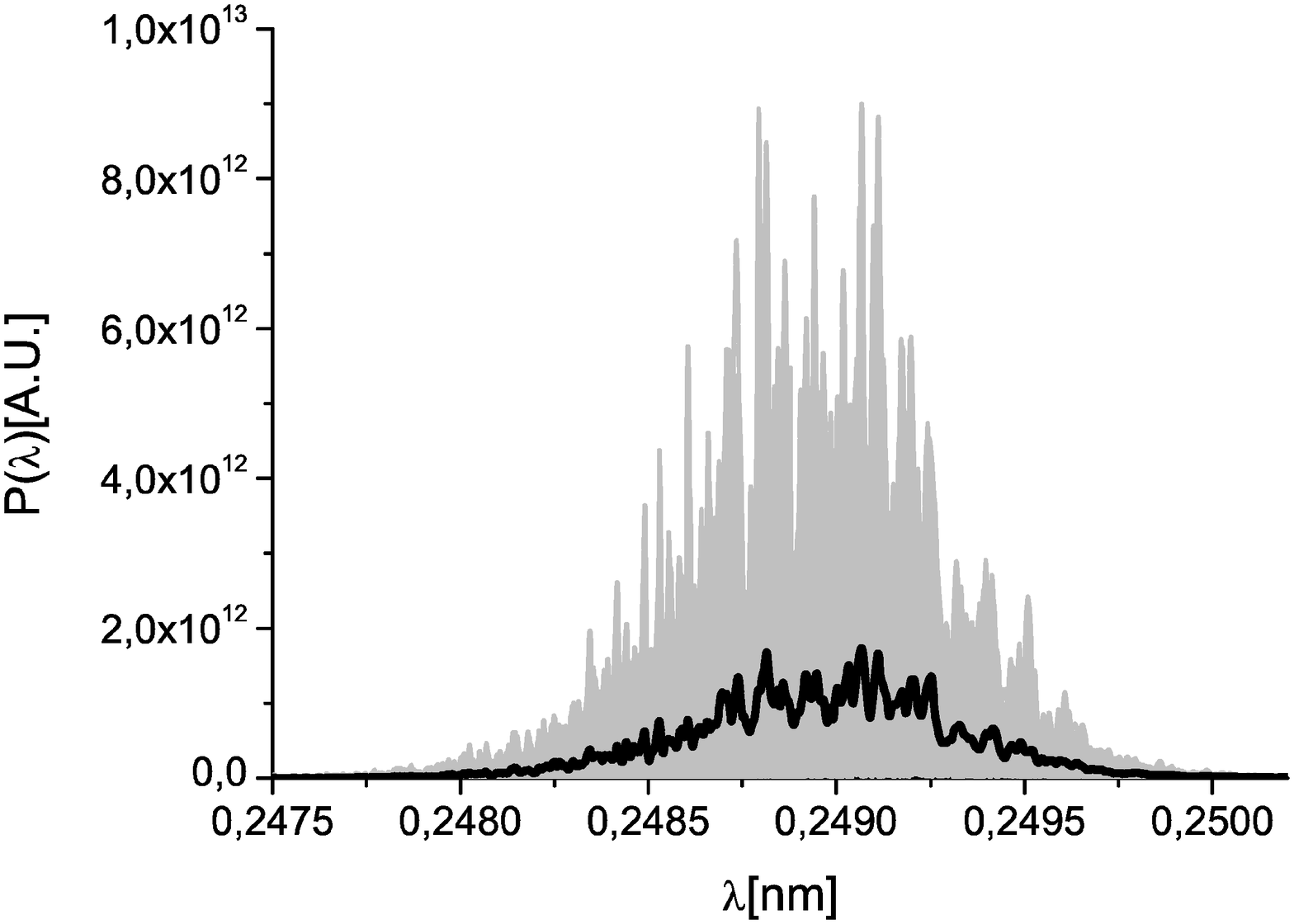}
\caption{Power and spectrum at $5$ keV before the second magnetic
chicane. Grey lines refer to single shot realizations, the black
line refers to the average over a hundred realizations.}
\label{biof175B}
\end{figure}
\begin{figure}[tb]
\includegraphics[width=0.5\textwidth]{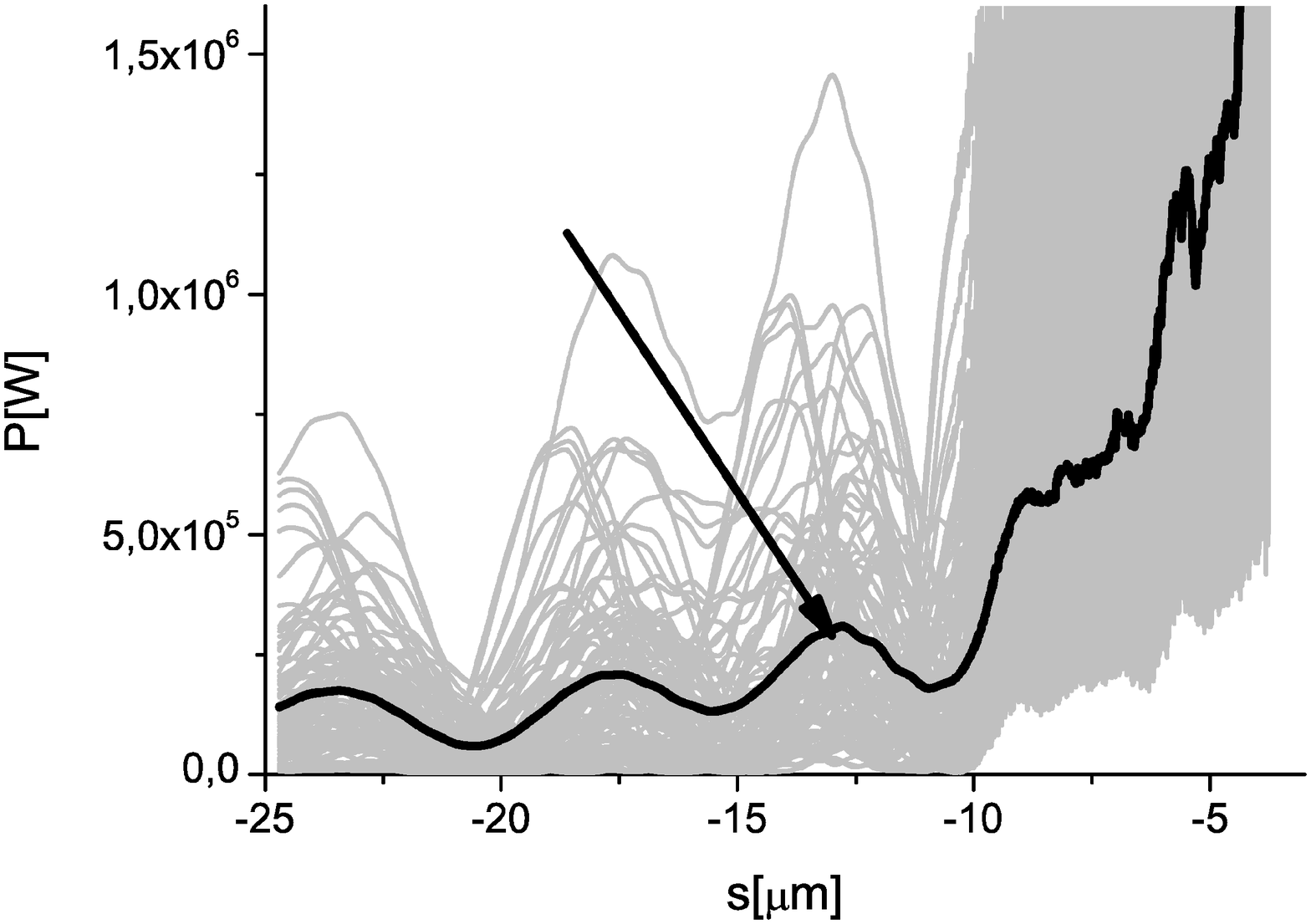}
\includegraphics[width=0.5\textwidth]{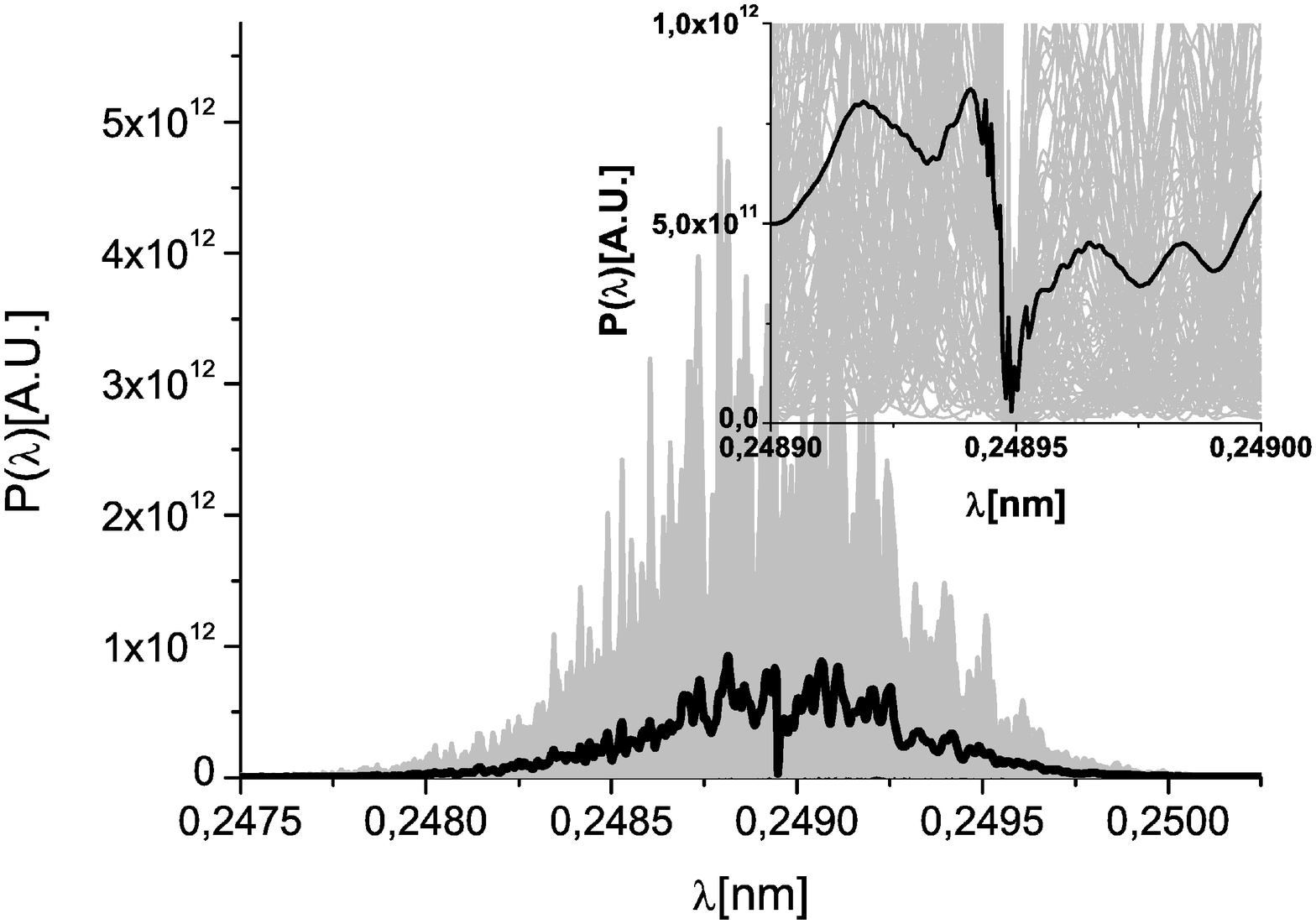}
\caption{Power and spectrum at $5$ keV after the single crystal
self-seeding X-ray monochromator. A $100~\mu$m thick diamond crystal
( C(111) Laue reflection, $\sigma$-polarization ) is used. Grey
lines refer to single shot realizations, the black line refers to
the average over a hundred realizations. The black arrow indicates
the seeding region.} \label{biof183p5B}
\end{figure}

As before,  the first chicane is switched off, so that the first
part of the undulator effectively consists of 7 uniform cells. After
the first part of the undulator, one has the power and spectrum
shown in Fig. \ref{biof175B}. The second magnetic chicane is
switched on, and the single-crystal X-ray monochromator is set into
the photon beam. We take advantage of the C(111) Laue reflection,
$\sigma$-polarization, Fig. \ref{biof183p5B}.

\begin{figure}[tb]
\includegraphics[width=0.75\textwidth]{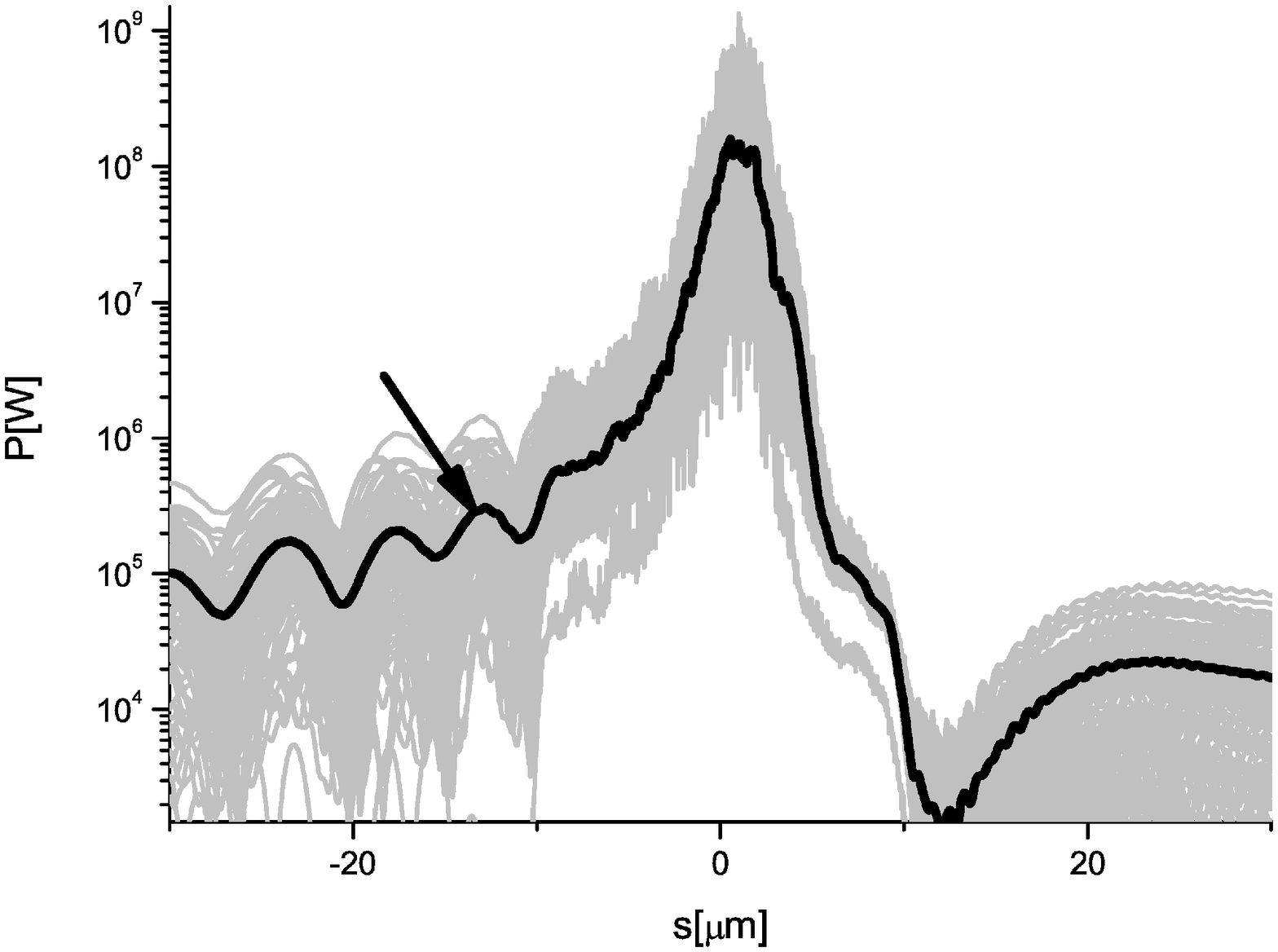}
\caption{Logarithmic plot of the power at $5$ keV after the single
crystal self-seeding X-ray monochromator.  The black arrow indicates
the seeding region. The region on the right hand side of the plot
(before the FEL pulse) is nominally zero because of causality
reasons. Differences with respect to zero give back the accuracy of
our calculations.} \label{5caus}
\end{figure}

\begin{figure}[tb]
\includegraphics[width=0.5\textwidth]{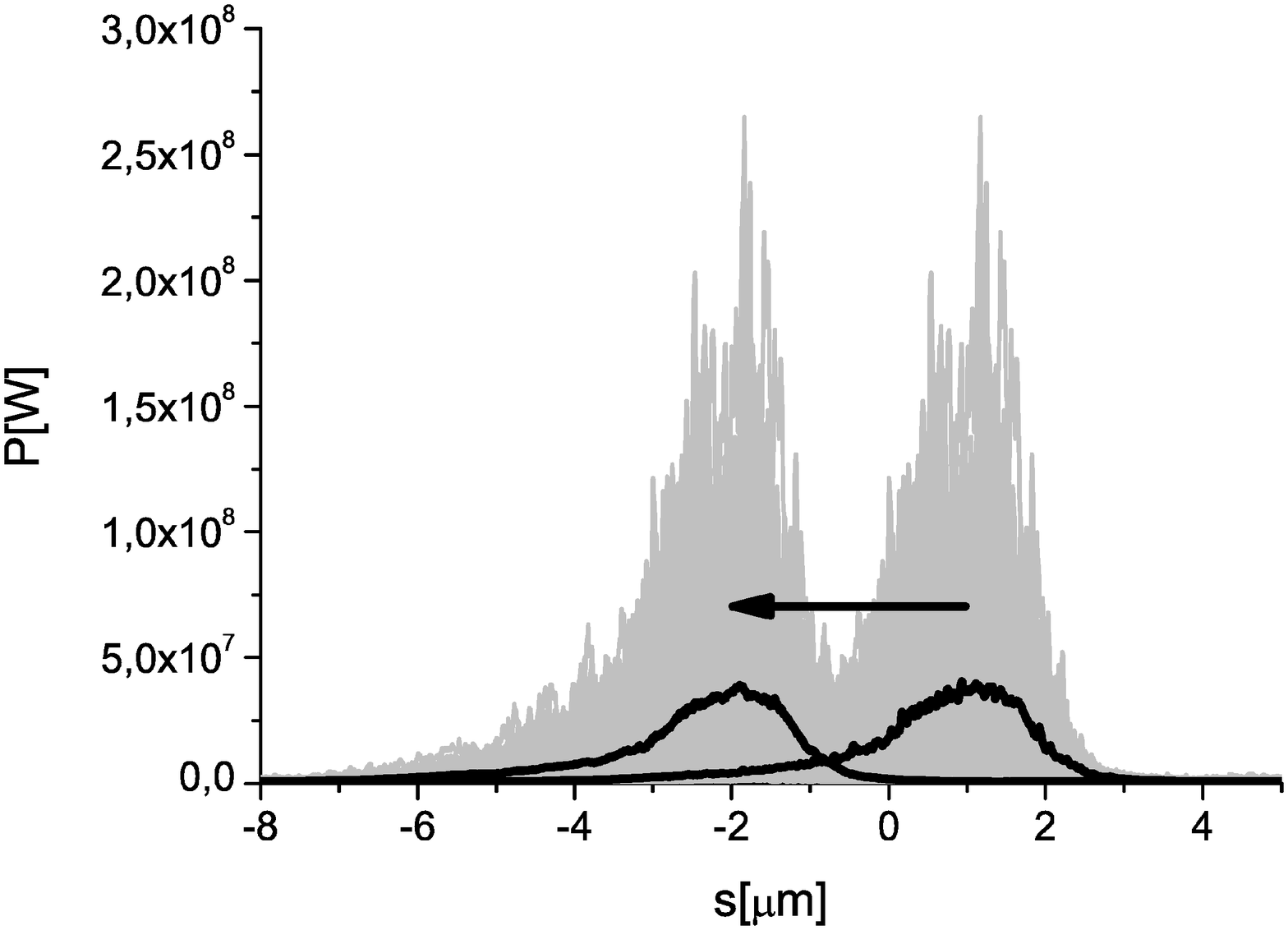}
\includegraphics[width=0.5\textwidth]{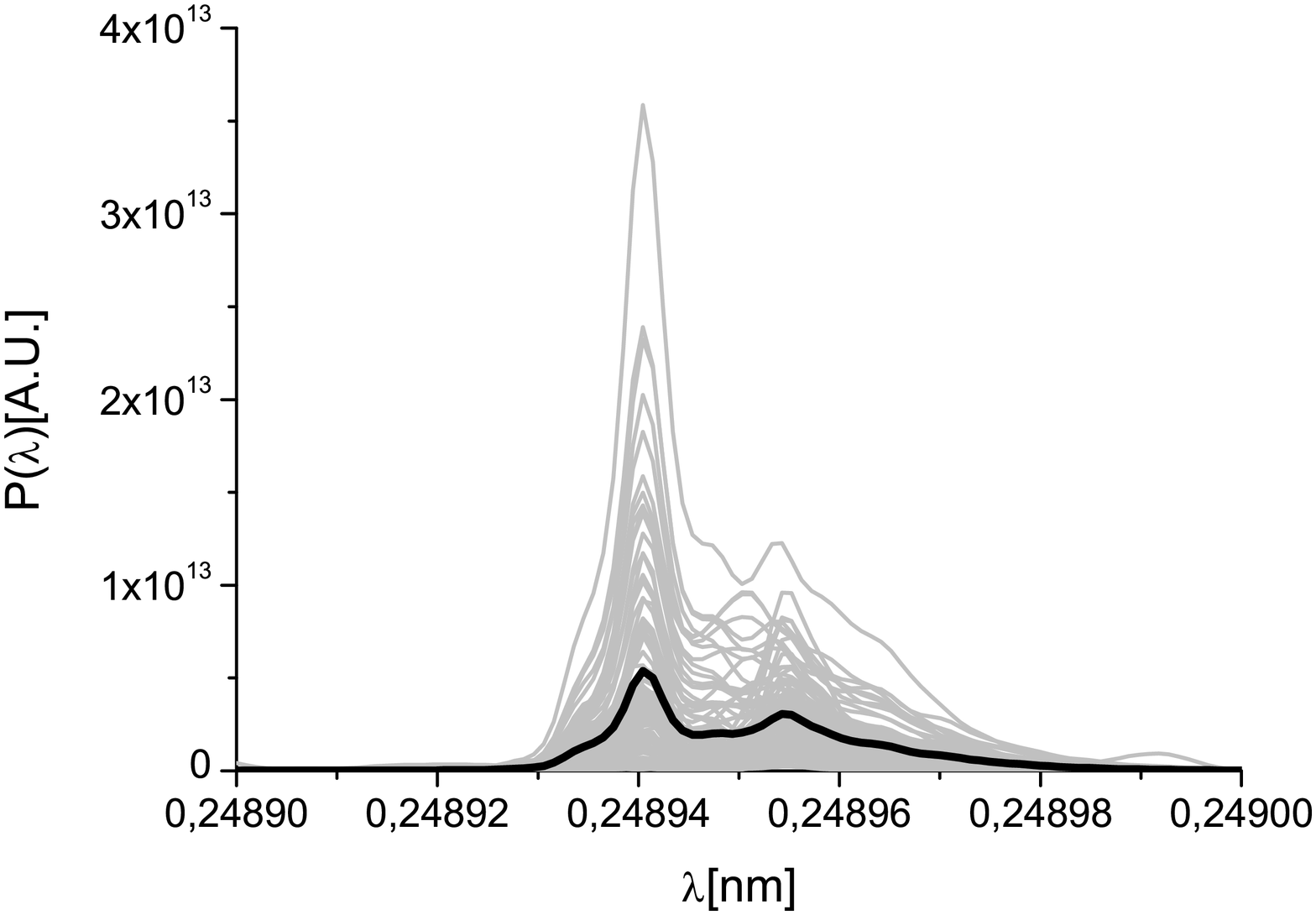}
\caption{Power and spectrum at $5$ keV after the third chicane
equipped with the X-ray optical delay line, delaying the radiation
pulse with respect to the electron bunch. Grey lines refer to single
shot realizations, the black line refers to the average over a
hundred realizations.} \label{biof203p5B}
\end{figure}

Exactly as before, following the seeding setup, the electron bunch
amplifies the seed in the following 4 undulator cells. After that, a
third chicane is used to allow for the installation of an x-ray
optical delay line, which delays the radiation pulse with respect to
the electron bunch. The power and spectrum of the radiation pulse
after the optical delay line are shown in Fig. \ref{biof203p5B},
where the effect of the optical delay is illustrated. The numerical
accuracy with which causality is satisfied in our simulations can be
shown by a logarithmic plot of the FEL pulse power after the
crystal, Fig. \ref{5caus}. The peak in the center is the main FEL
pulse. On the left side one can identify the seed pulse. One the
right side, before the FEL pulse, one has nominally zero power.

\begin{figure}[tb]
\includegraphics[width=0.5\textwidth]{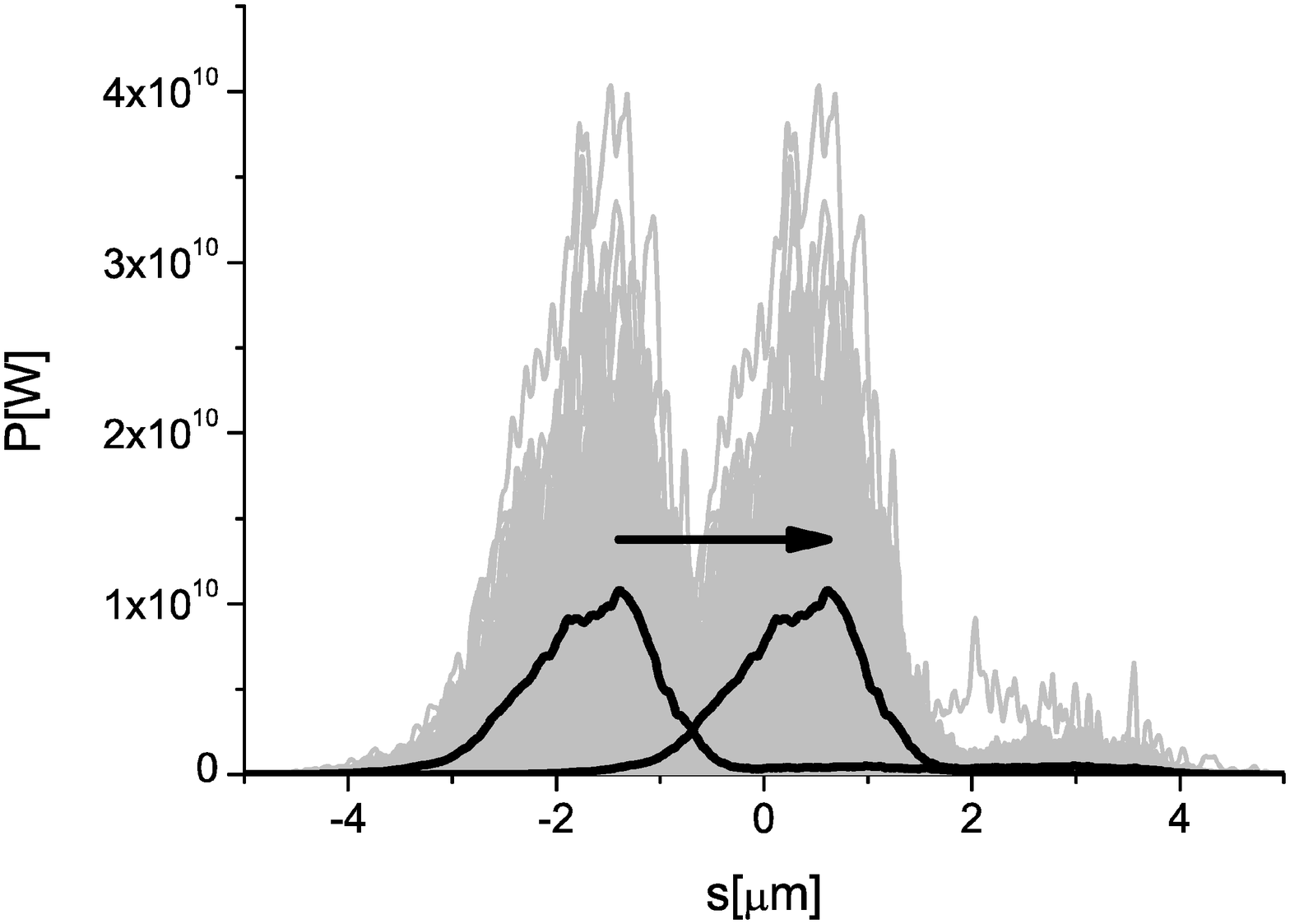}
\includegraphics[width=0.5\textwidth]{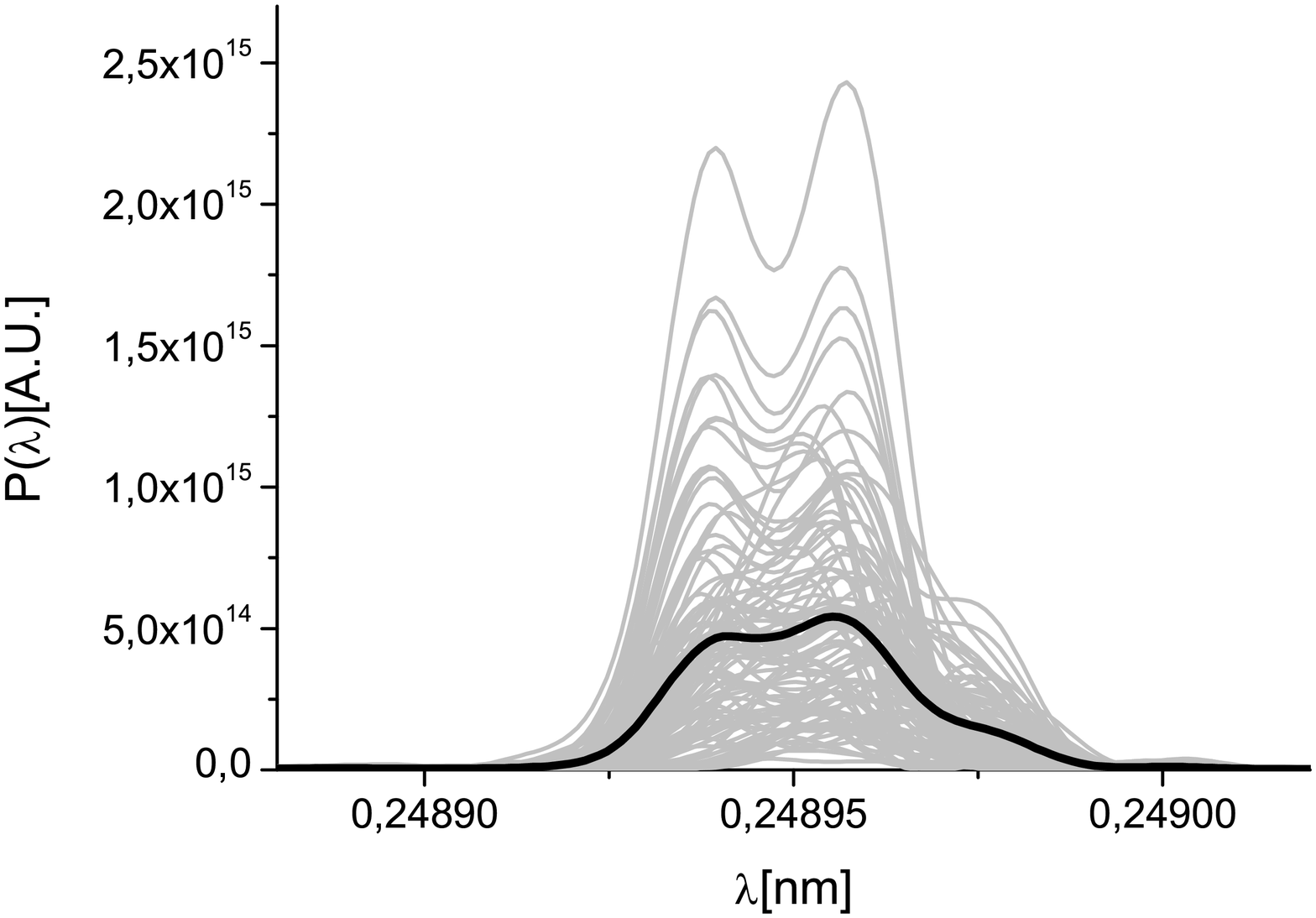}
\caption{Power and spectrum  at $5$ keV after the last magnetic
chicane. Grey lines refer to single shot realizations, the black
line refers to the average over a hundred realizations.}
\label{biof215B}
\end{figure}
Due to the presence of the optical delay, only part of the electron
beam is used to further amplify the radiation pulse in the following
6 undulator cells. The electron beam part which is not used is
fresh, and can be used for further lasing. In order to do so, after
amplification, the electron beam passes through the final magnetic
chicane, which delays the electron beam. The power and spectrum of
the radiation pulse after the last magnetic chicane are shown in
Fig. \ref{biof215B}. By delaying the electron bunch, the magnetic
chicane effectively shifts forward the photon beam with respect to
the electron beam. Tunability of such shift allows the selection of
different photon pulse length.

\begin{figure}[tb]
\begin{center}
\includegraphics[width=0.5\textwidth]{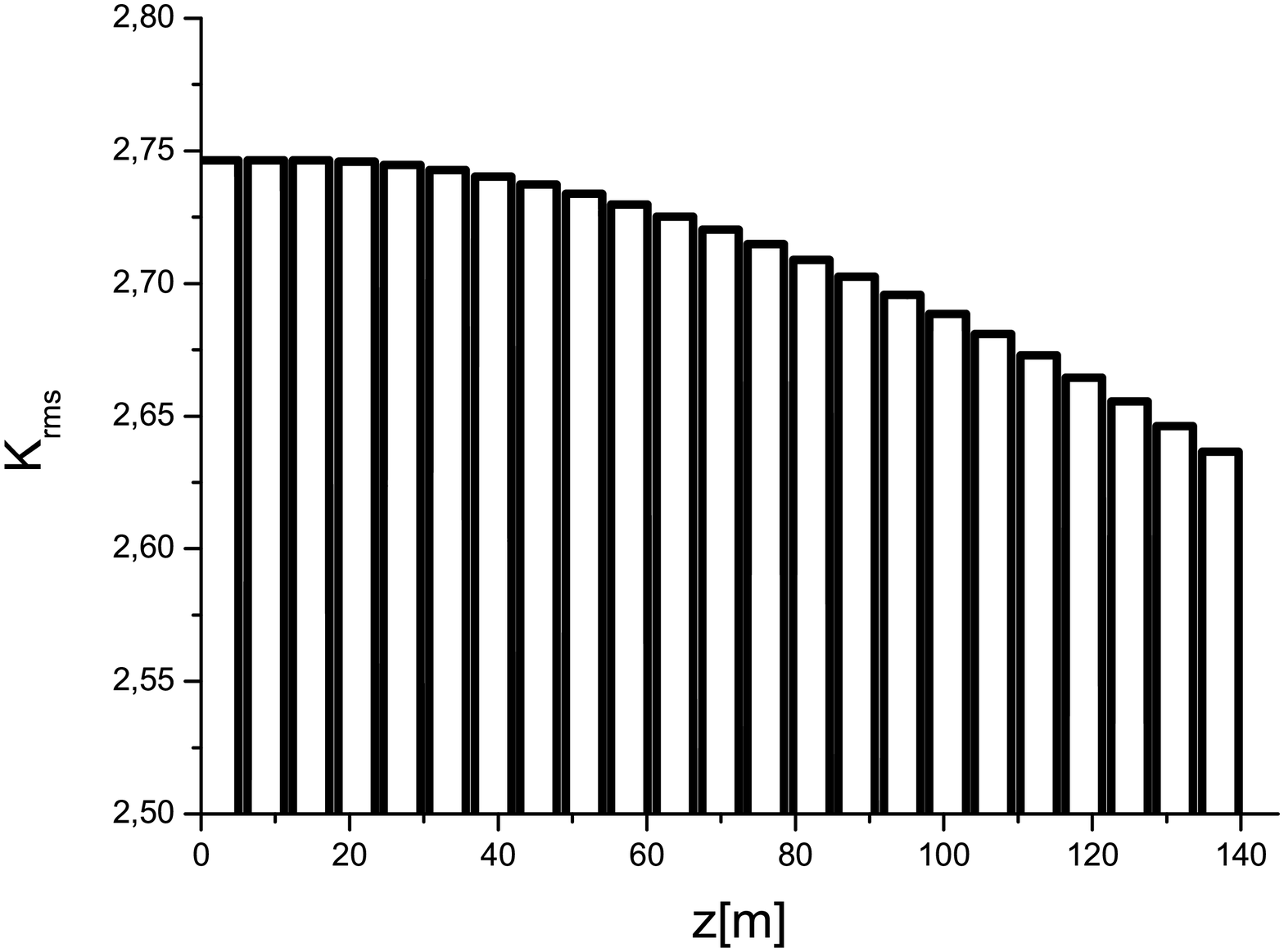}
\end{center}
\caption{Tapering law at $5$ keV.} \label{biof225B}
\end{figure}
The last part of the undulator is composed by $23$ cells. It is
partly tapered post-saturation, to increase the region where
electrons and radiation interact properly to the advantage of the
radiation pulse. Tapering is implemented by changing the $K$
parameter of the undulator segment by segment according to Fig.
\ref{biof225B}. The tapering law used in this work has been
implemented on an empirical basis.

\begin{figure}[tb]
\includegraphics[width=0.5\textwidth]{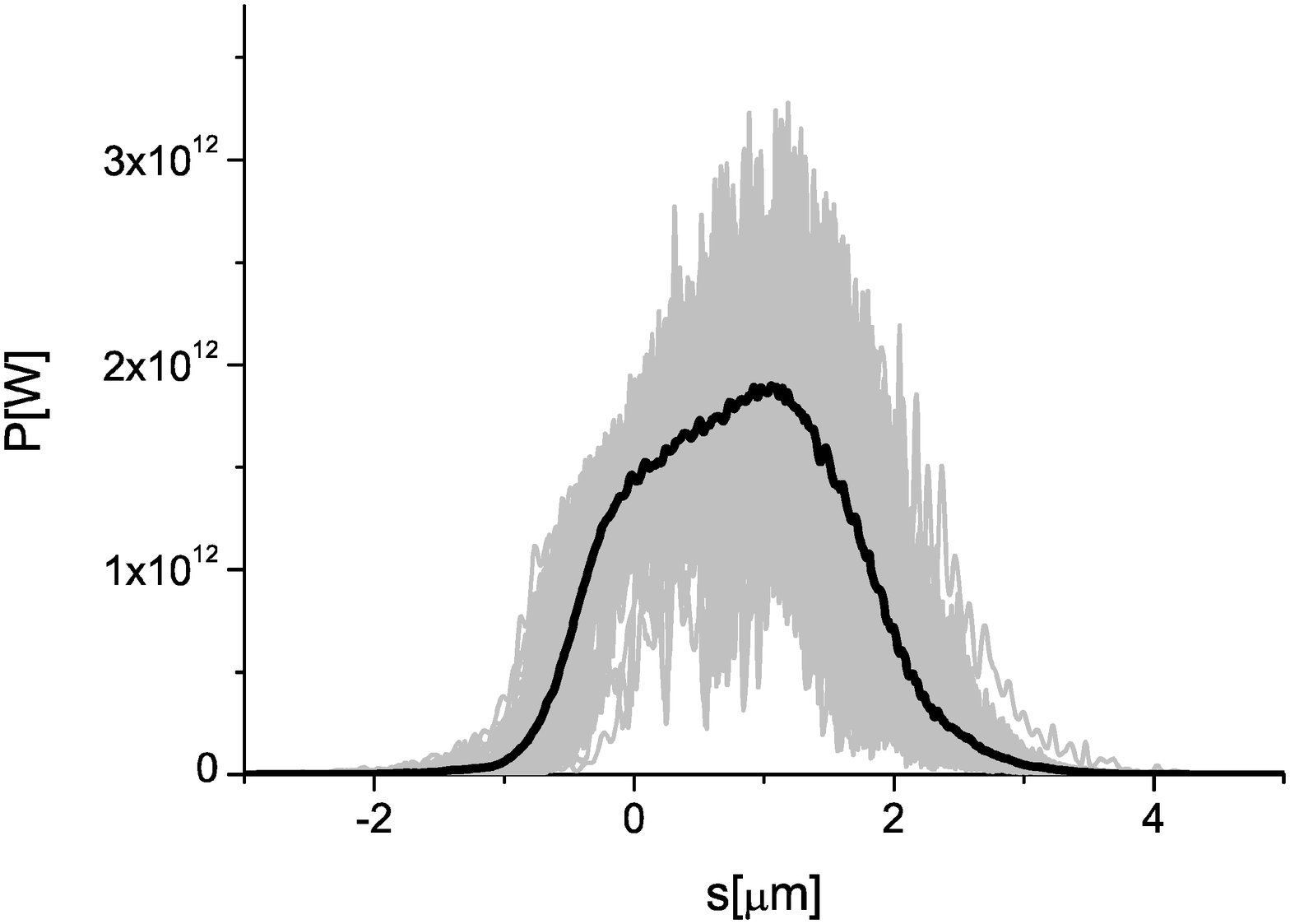}
\includegraphics[width=0.5\textwidth]{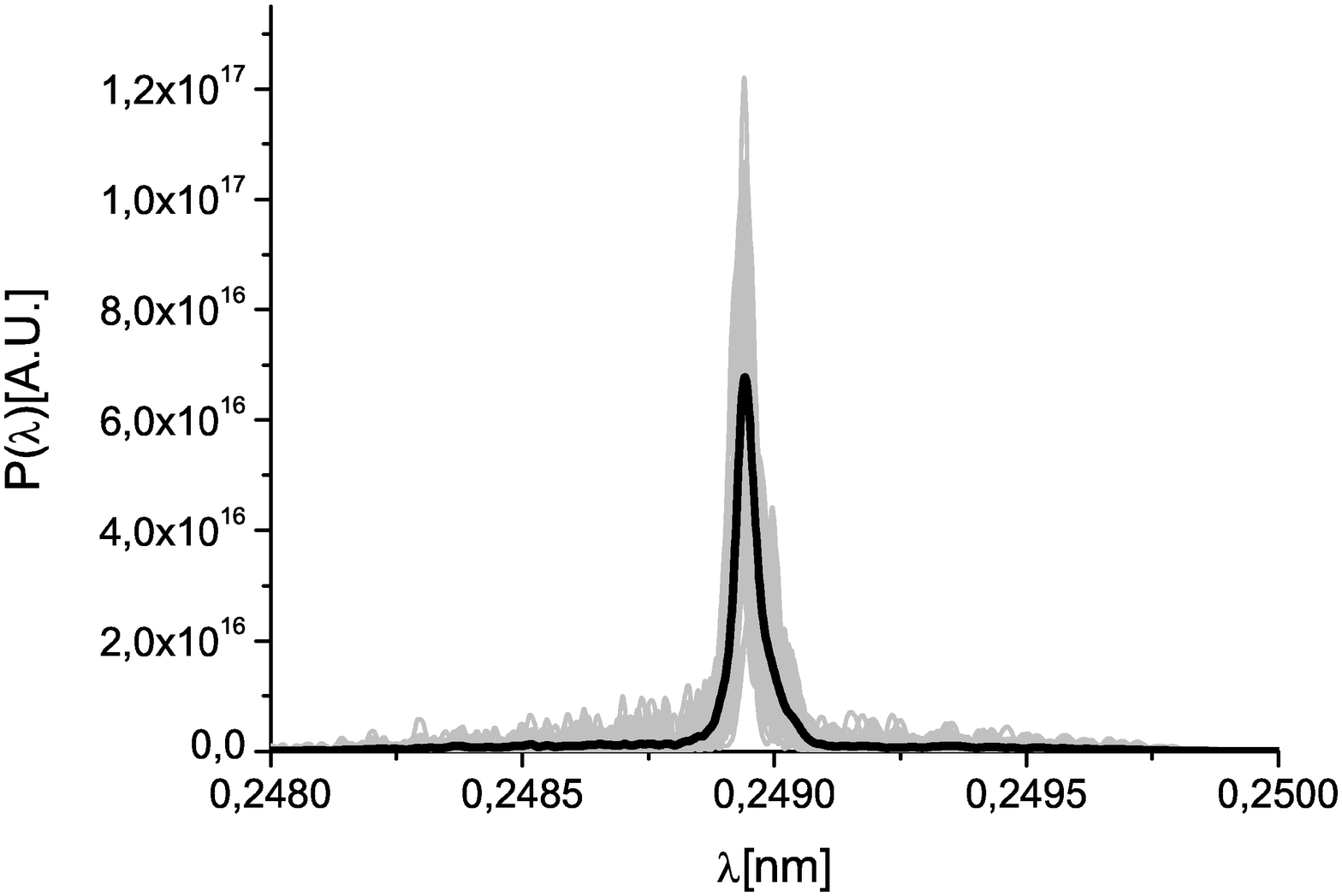}
\caption{Final output. Power and spectrum after tapering at $5$ keV.
Grey lines refer to single shot realizations, the black line refers
to the average over a hundred realizations.} \label{biof235B}
\end{figure}
As explained before, the use of tapering together with monochromatic
radiation is particularly effective. The final output is presented
in Fig. \ref{biof235B} in terms of power and spectrum. As one can
see, simulations indicate an output power of about $2$ TW.

\begin{figure}[tb]
\includegraphics[width=0.5\textwidth]{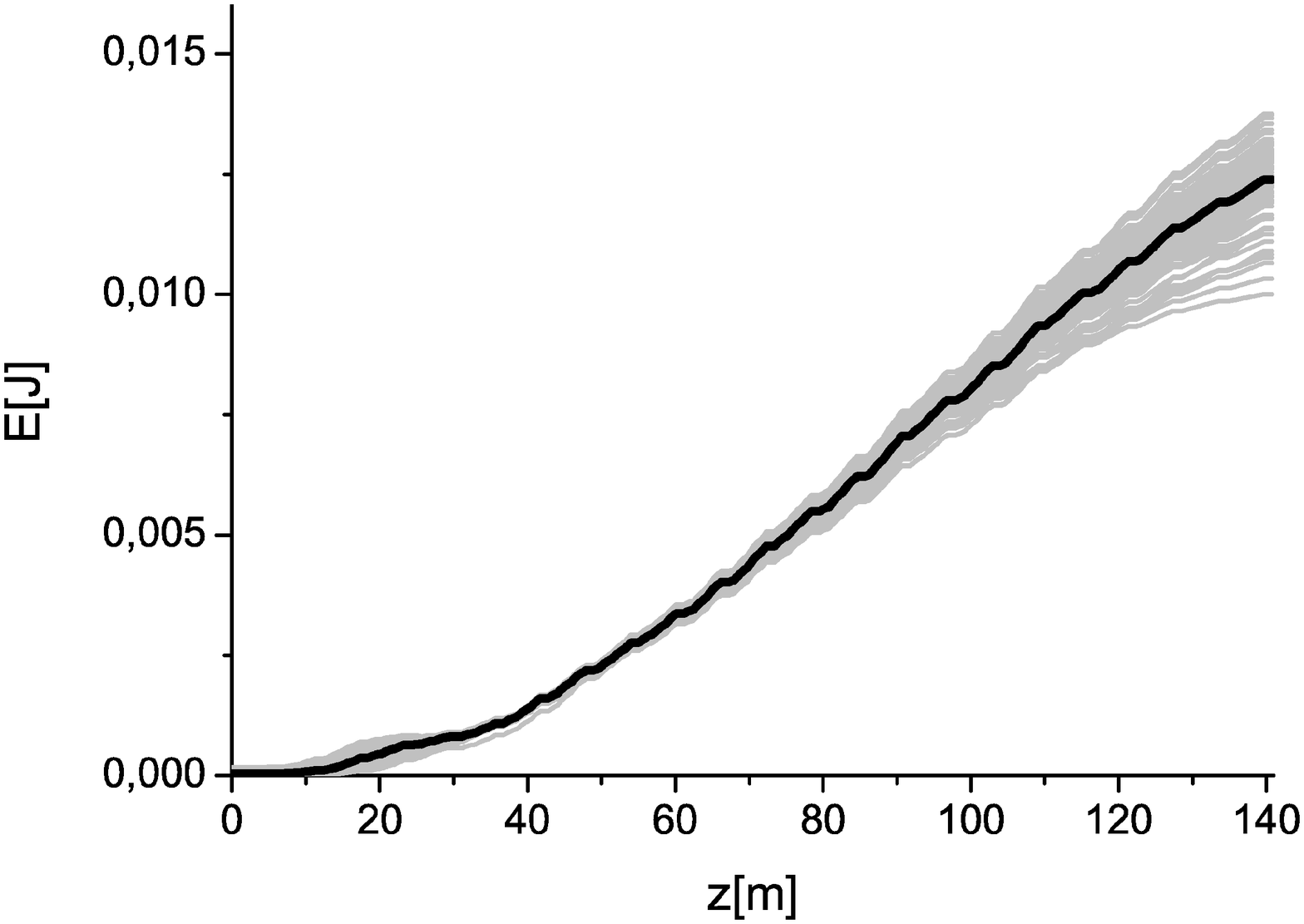}
\includegraphics[width=0.5\textwidth]{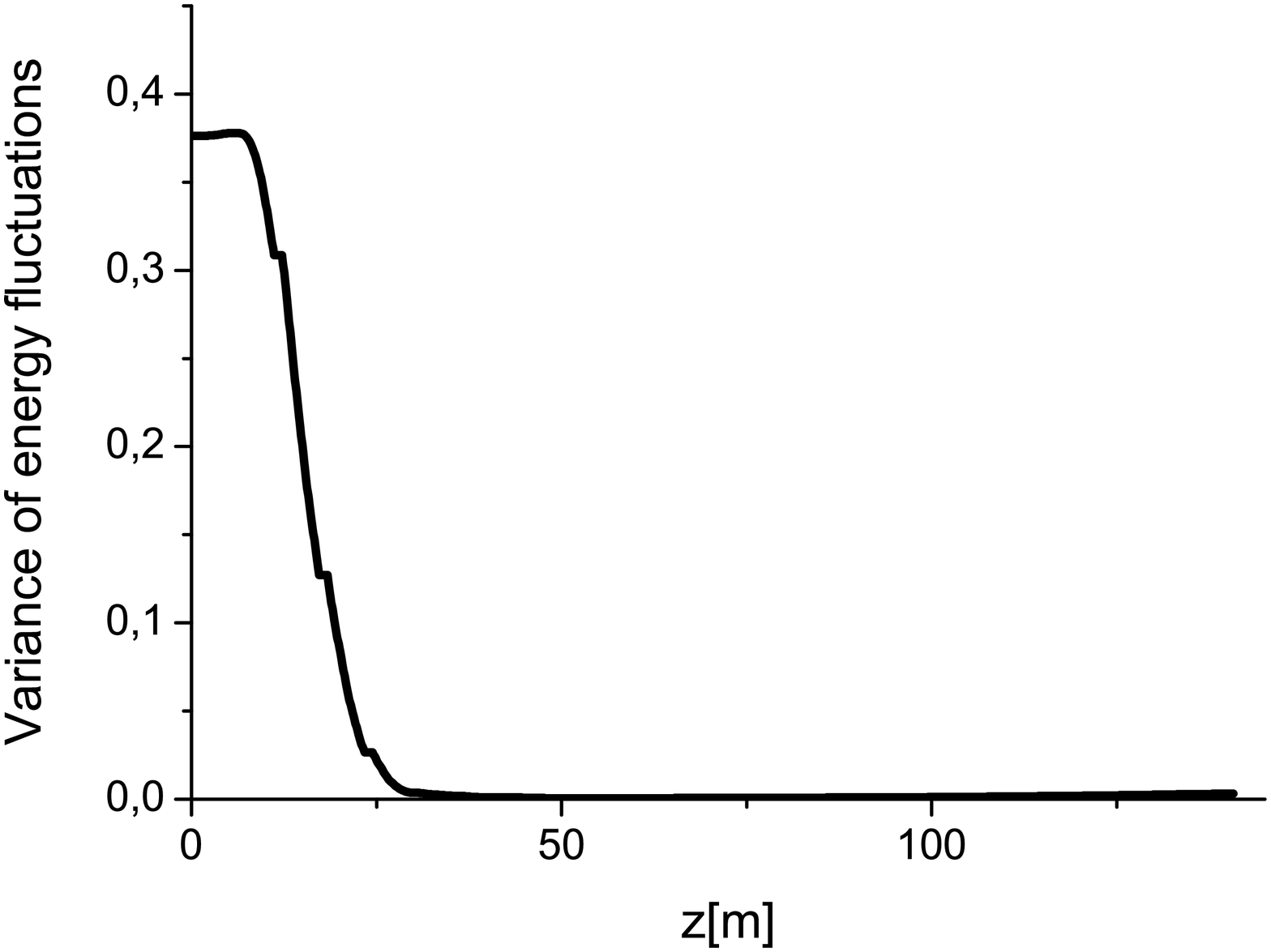}
\caption{Final output. Energy and energy variance of output pulses
at $5$ keV. In the left plot, grey lines refer to single shot
realizations, the black line refers to the average over a hundred
realizations.} \label{biof245B}
\end{figure}

\begin{figure}[tb]
\includegraphics[width=0.5\textwidth]{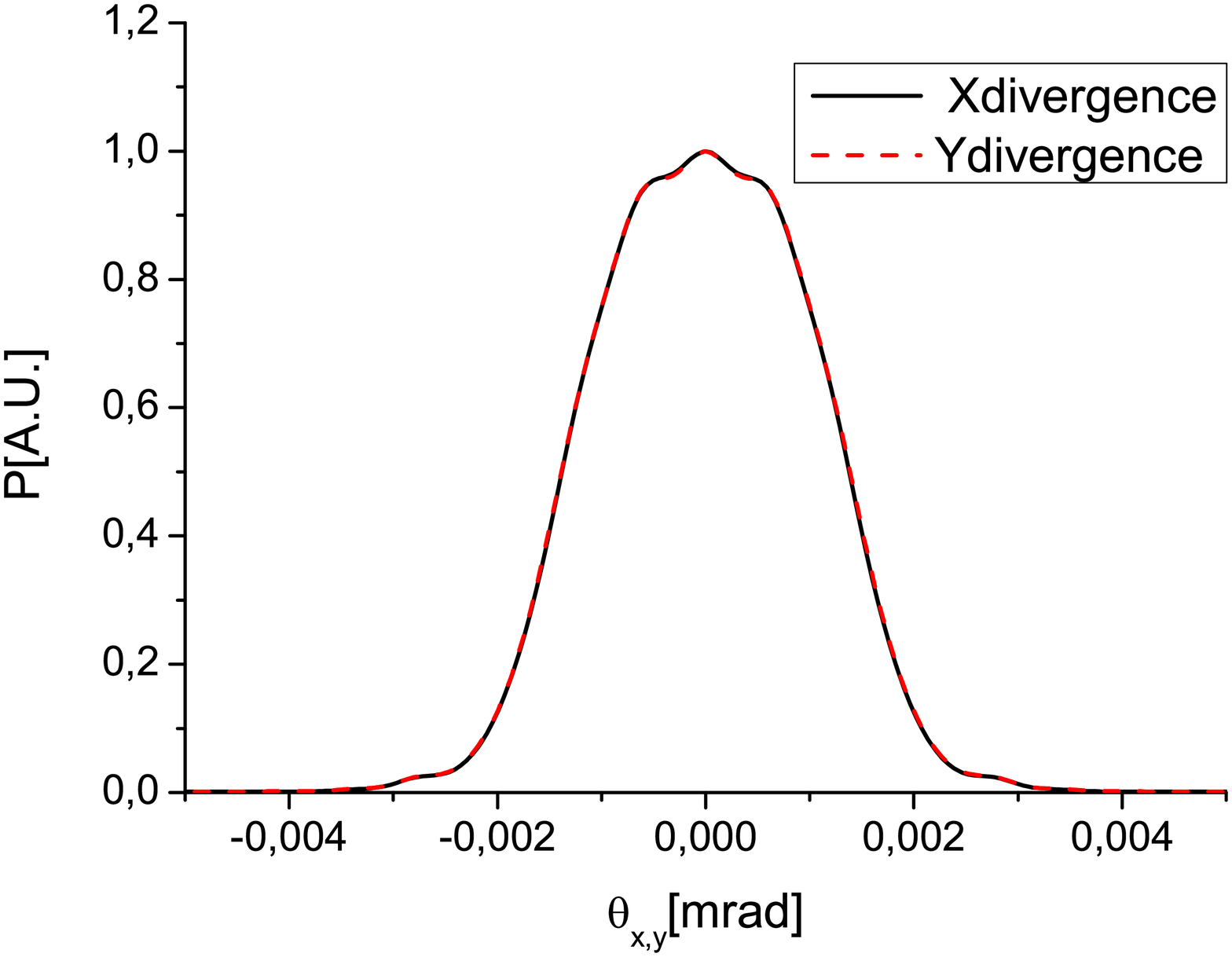}
\includegraphics[width=0.5\textwidth]{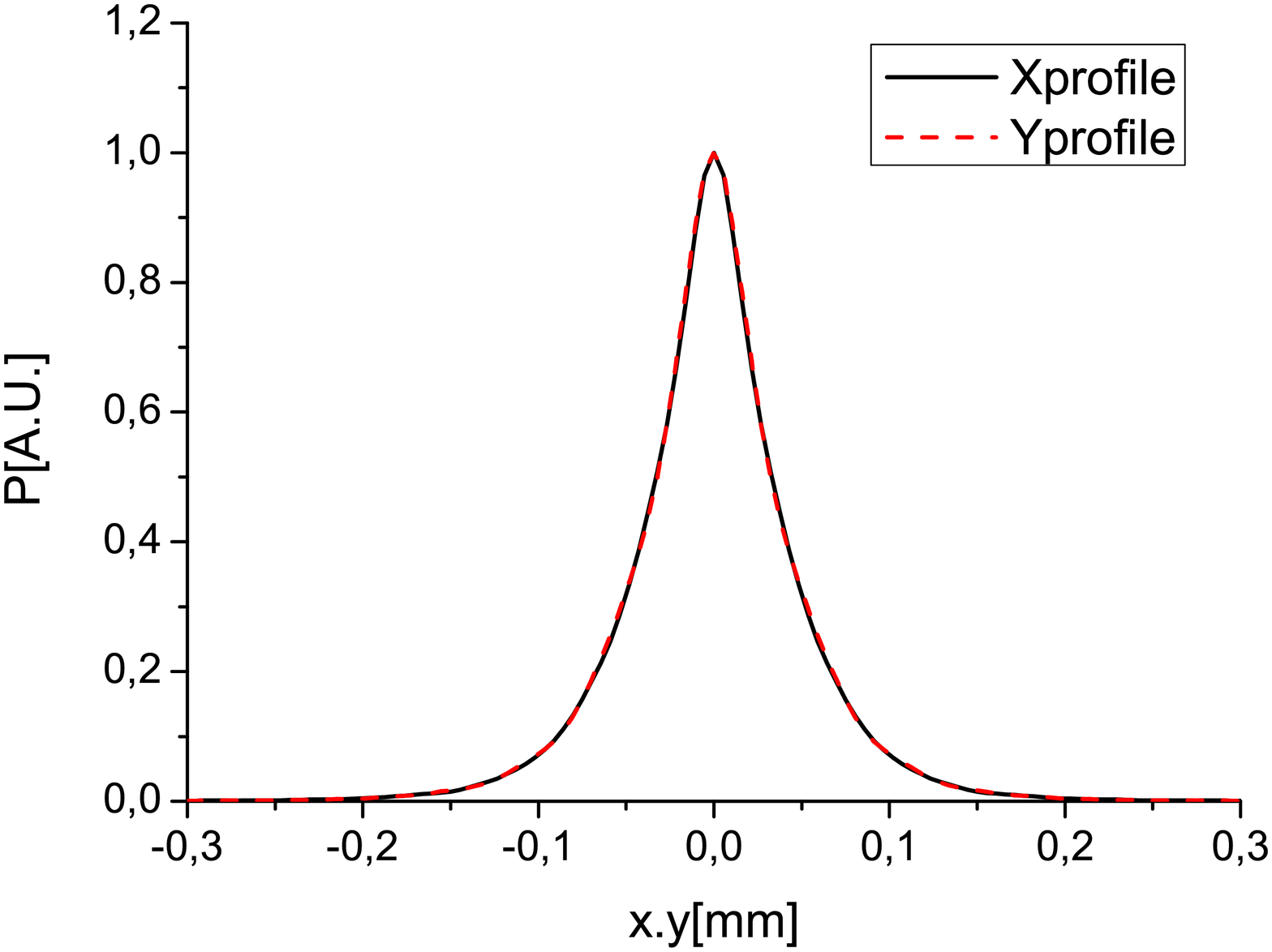}
\caption{Final output. X-ray radiation pulse energy distribution per
unit surface and angular distribution of the X-ray pulse energy at
$5$ keV at the exit of output undulator.} \label{biof253p5B}
\end{figure}
The energy of the radiation pulse and the energy variance are shown
in Fig. \ref{biof245B} as a function of the position along the
undulator. The divergence and the size of the radiation pulse at the
exit of the final undulator are shown, instead, in Fig.
\ref{biof253p5B}.

\subsection{C(113) asymmetric Bragg reflection at $7$ keV}

\begin{figure}[tb]
\includegraphics[width=0.5\textwidth]{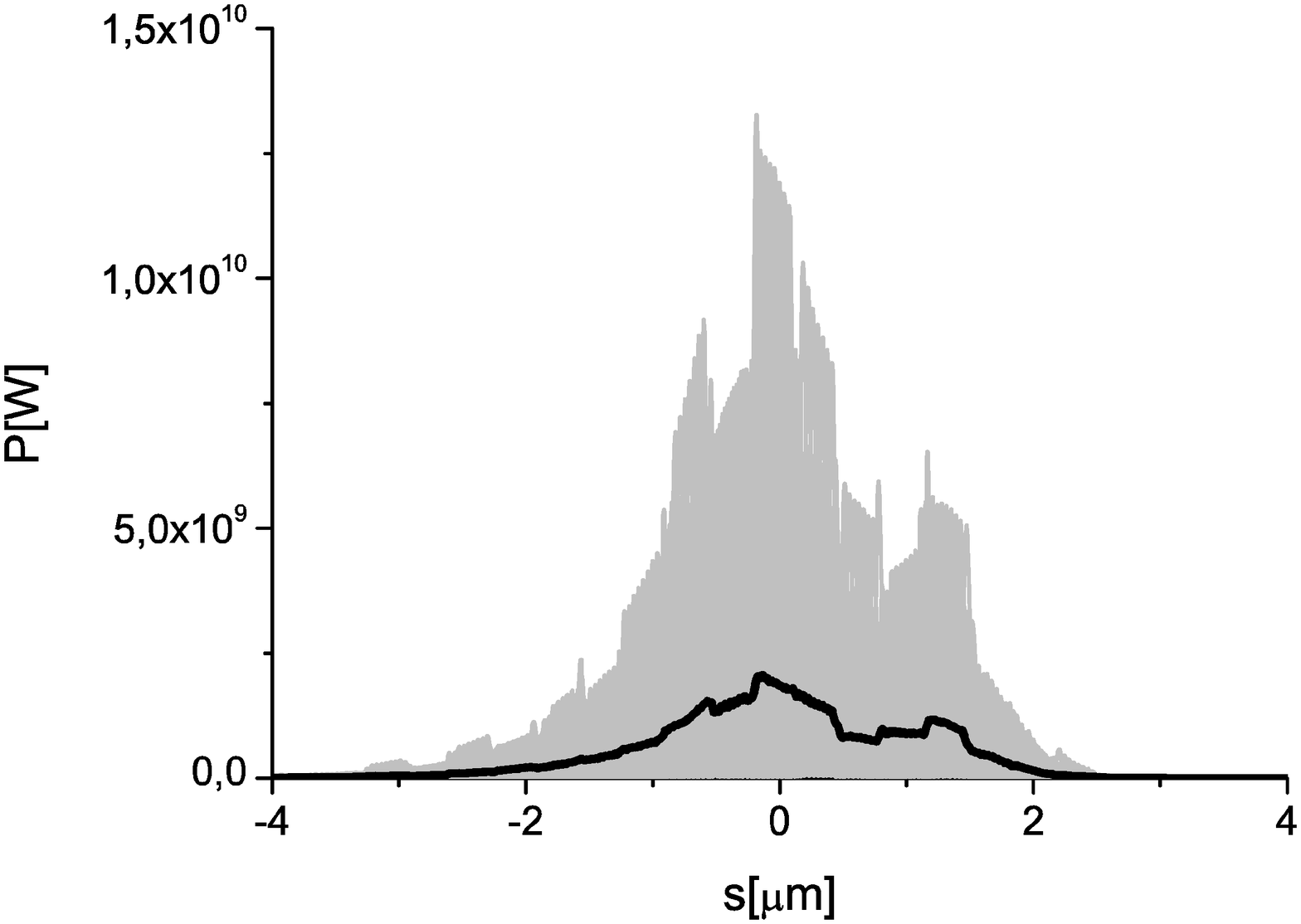}
\includegraphics[width=0.5\textwidth]{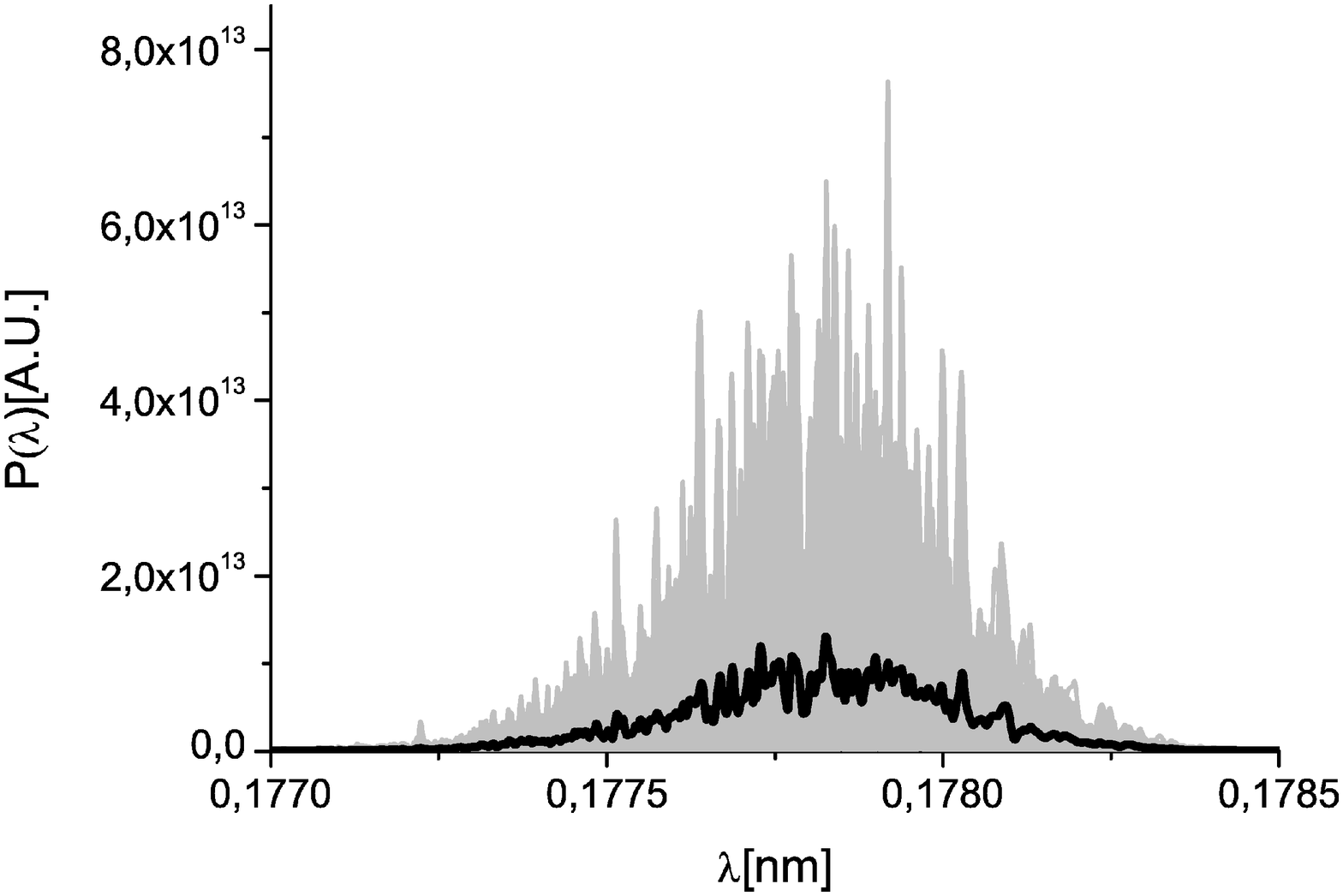}
\caption{Power and spectrum at $7$ keV before the third magnetic
chicane. Grey lines refer to single shot realizations, the black
line refers to the average over a hundred realizations.}
\label{biofC113}
\end{figure}

We now consider the energy point at $7$ keV. In this case a simpler
scheme with a single hard x-ray self-seeding setup is considered.
The first two magnetic chicanes are switched off. Therefore, the
electron beam lases in SASE mode along the first $11$ undulator
cells before passing through the single-crystal monochromator
filter. The input power and spectrum are shown in Fig.
\ref{biofC113}.

\begin{figure}[tb]
\includegraphics[width=0.5\textwidth]{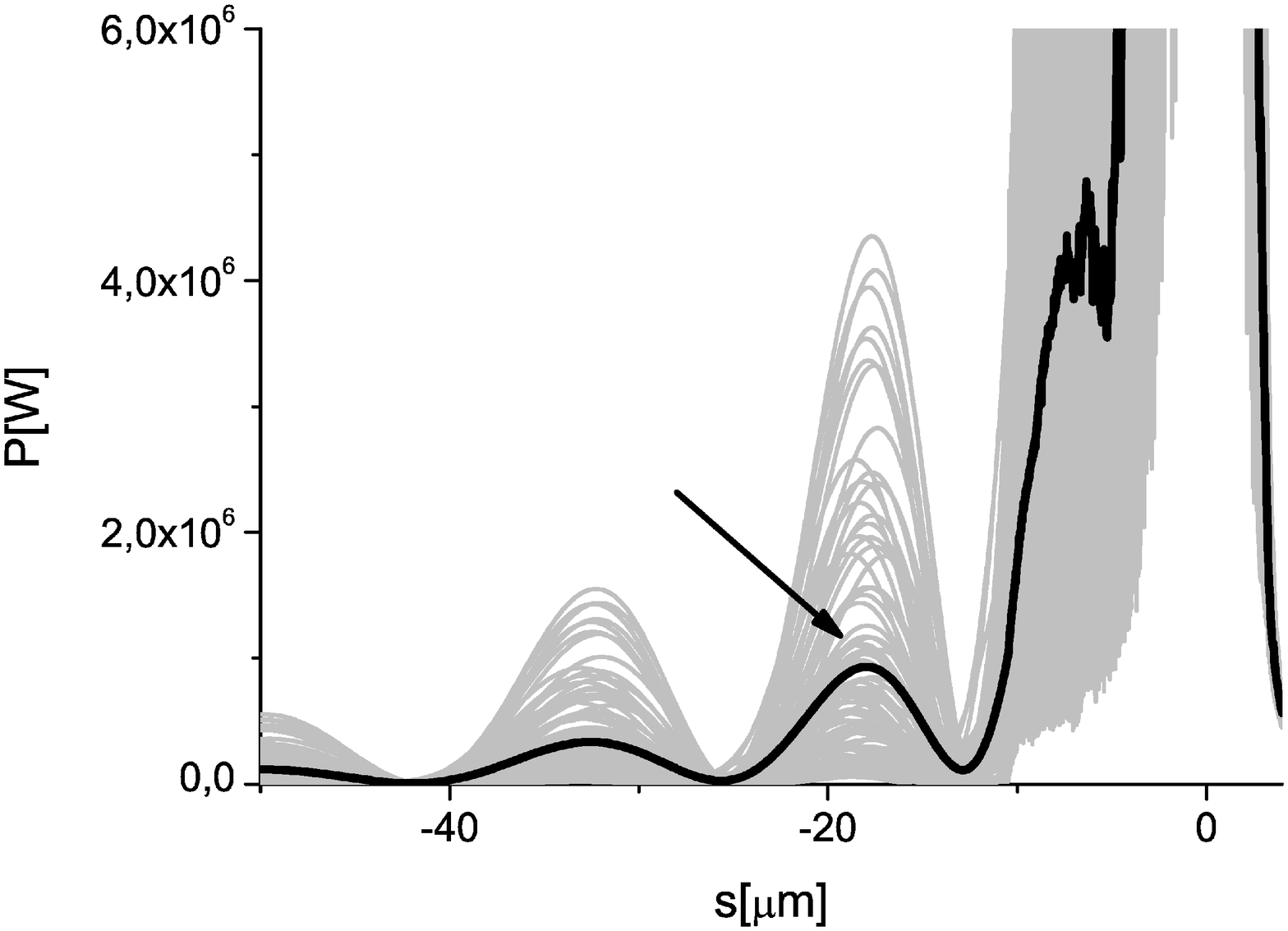}
\includegraphics[width=0.5\textwidth]{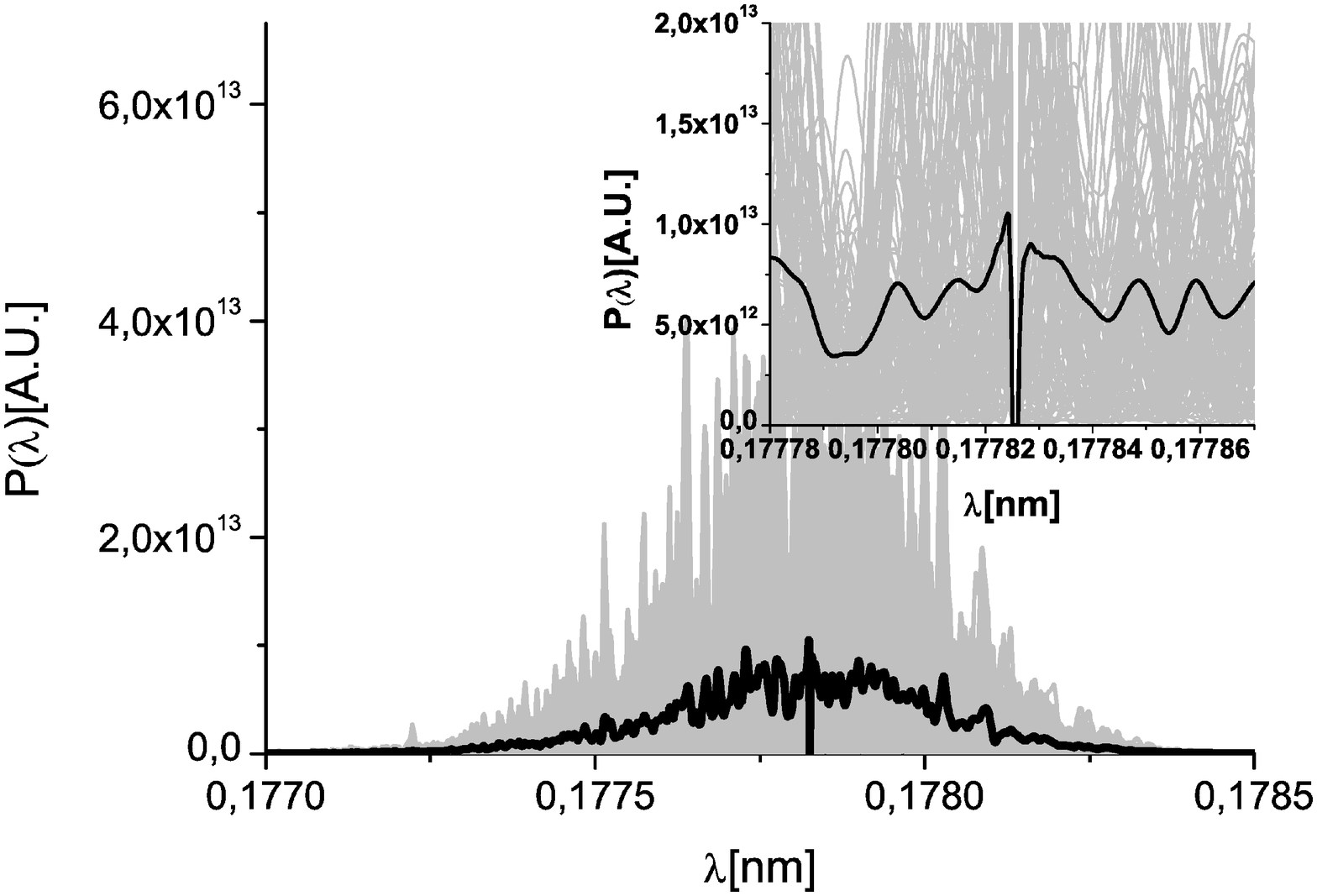}
\caption{Power and spectrum at $7$ keV after the single crystal
self-seeding X-ray monochromator. A $100~\mu$m thick diamond crystal
( C(113) Bragg reflection, $\sigma$-polarization ) is used. Grey
lines refer to single shot realizations, the black line refers to
the average over a hundred realizations. The black arrow indicates
the seeding region. } \label{biofh8}
\end{figure}
The effect of the filtering process is illustrated, both in the time
and in the frequency domain, in Fig. \ref{biofh8}.

\begin{figure}[tb]
\includegraphics[width=0.75\textwidth]{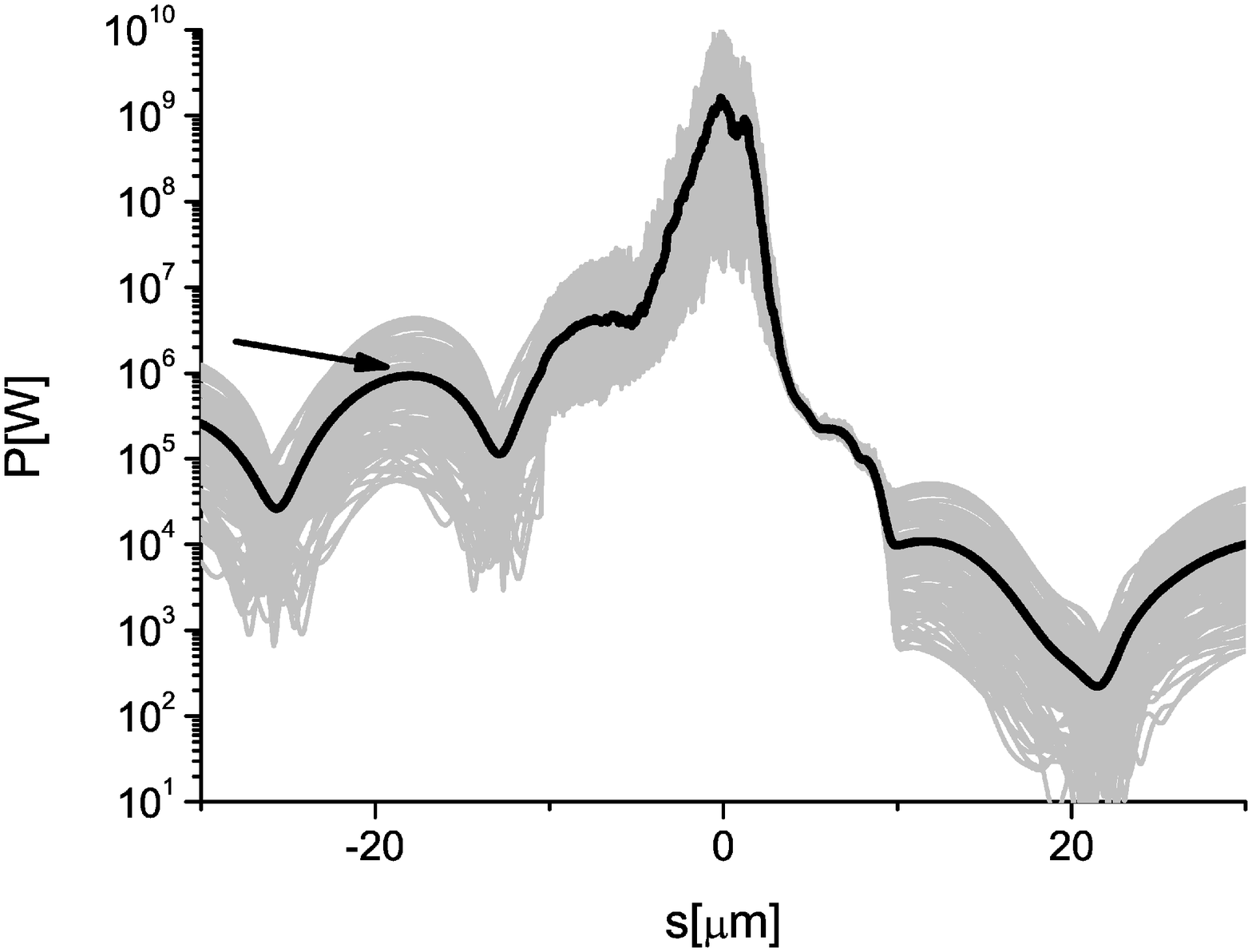}
\caption{Logarithmic plot of the power at $7$ keV after the single
crystal self-seeding X-ray monochromator.  The black arrow indicates
the seeding region. The region on the right hand side of the plot
(before the FEL pulse) is nominally zero because of causality
reasons. Differences with respect to zero give back the accuracy of
our calculations.} \label{7caus}
\end{figure}
The numerical accuracy with which causality is satisfied in our
simulations can be shown by a logarithmic plot of the FEL pulse
power after the crystal, Fig. \ref{7caus}. The peak in the center is
the main FEL pulse. On the left side one can identify the seed
pulse. One the right side, before the FEL pulse, one has nominally
zero power.

\begin{figure}[tb]
\begin{center}
\includegraphics[width=0.5\textwidth]{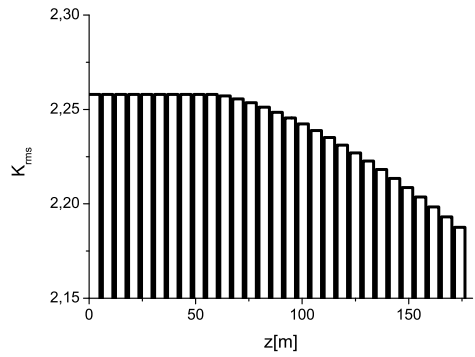}
\end{center}
\caption{Tapering law at $7$ keV.} \label{biofh3}
\end{figure}
The seed is amplified up to saturation in the output undulator. As
already discussed, we can use post-saturation tapering to increase
the output power level. The tapering configuration in Fig.
\ref{biofh3} is optimized for maximum output power level.

\begin{figure}[tb]
\includegraphics[width=0.5\textwidth]{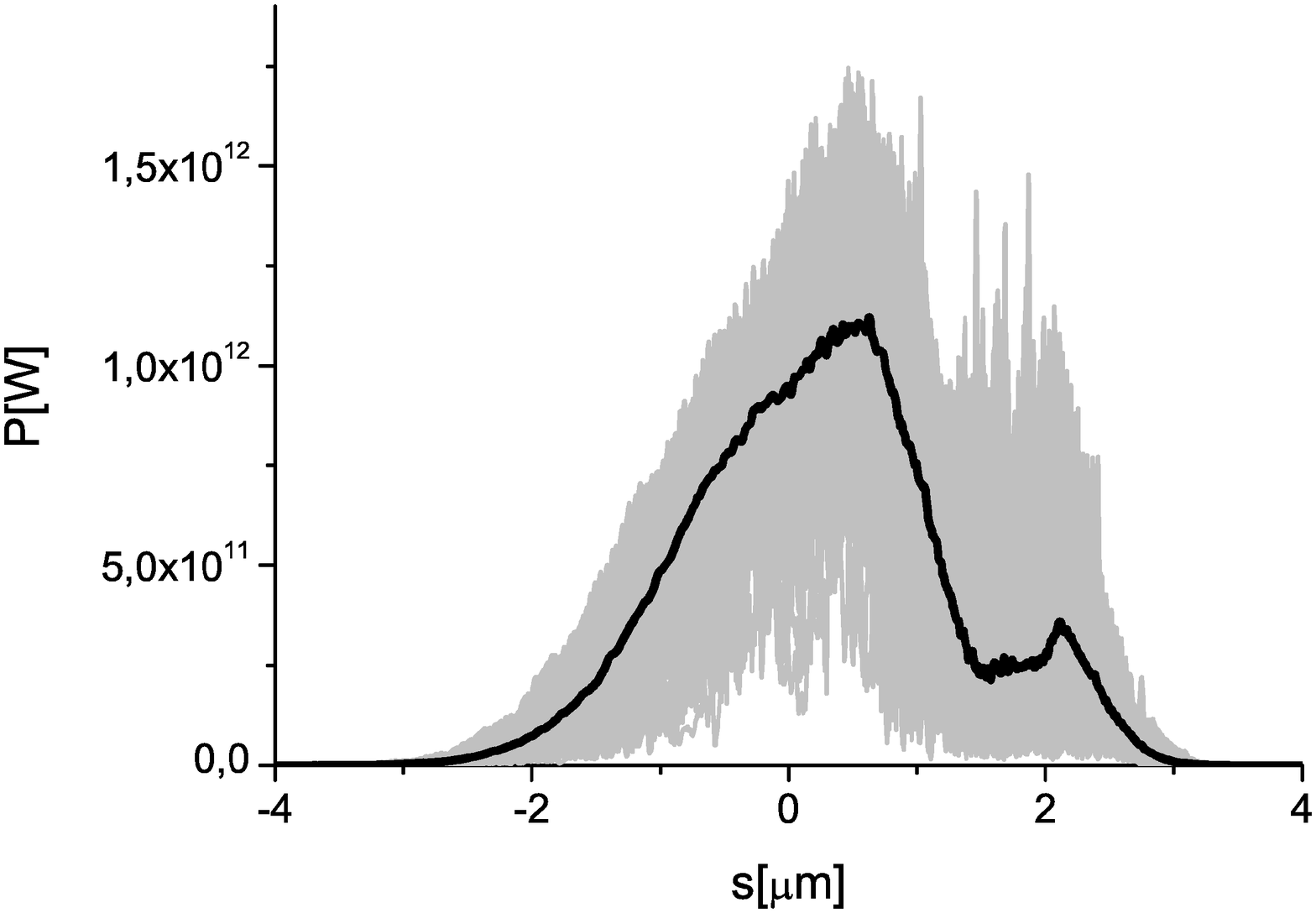}
\includegraphics[width=0.5\textwidth]{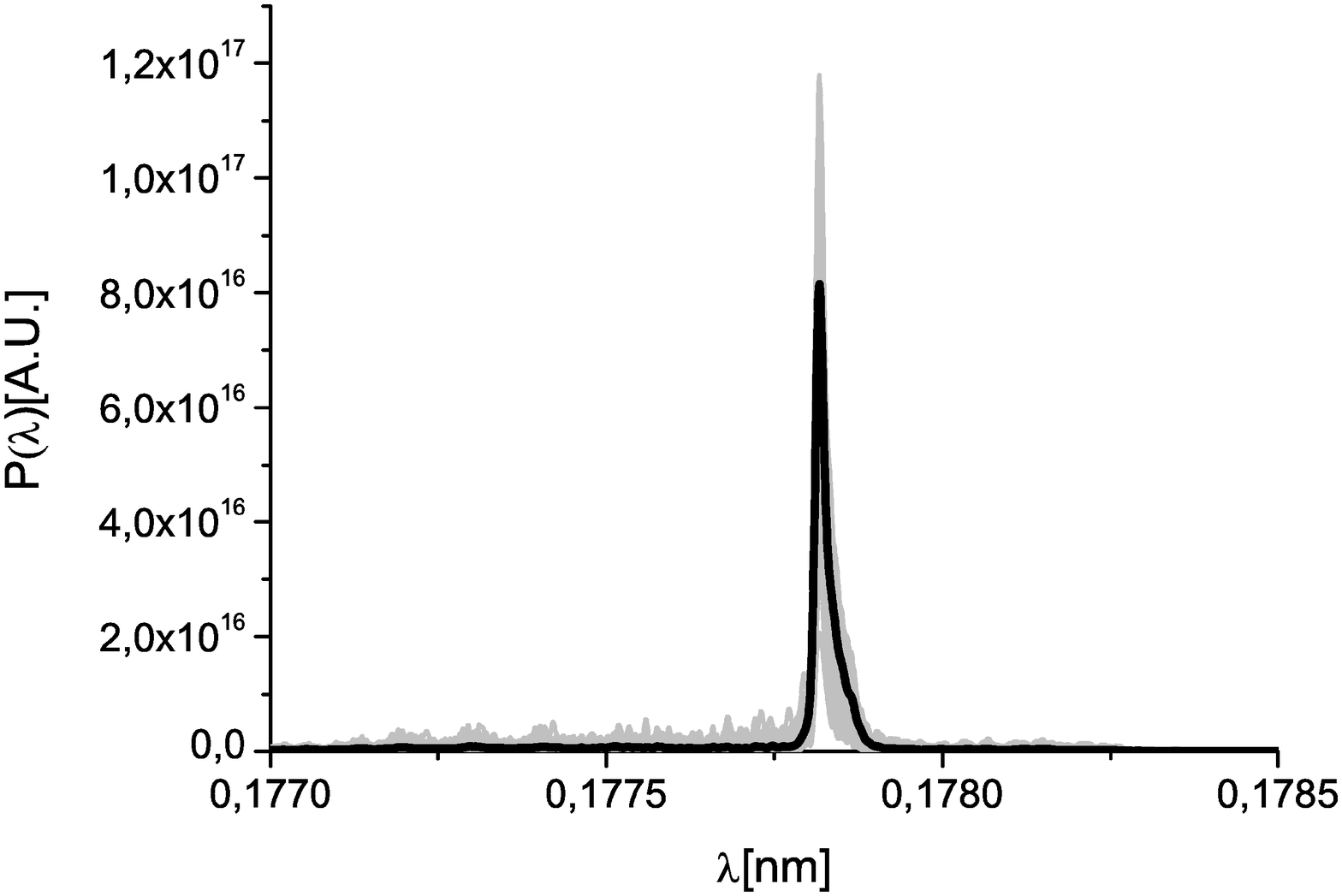} \caption{Final
output in the case of tapered output undulator at $7$ keV. Power and
spectrum are shown. Grey lines refer to single shot realizations,
the black line refers to the average over a hundred realizations.}
\label{biofh6}
\end{figure}

\begin{figure}[tb]
\includegraphics[width=0.5\textwidth]{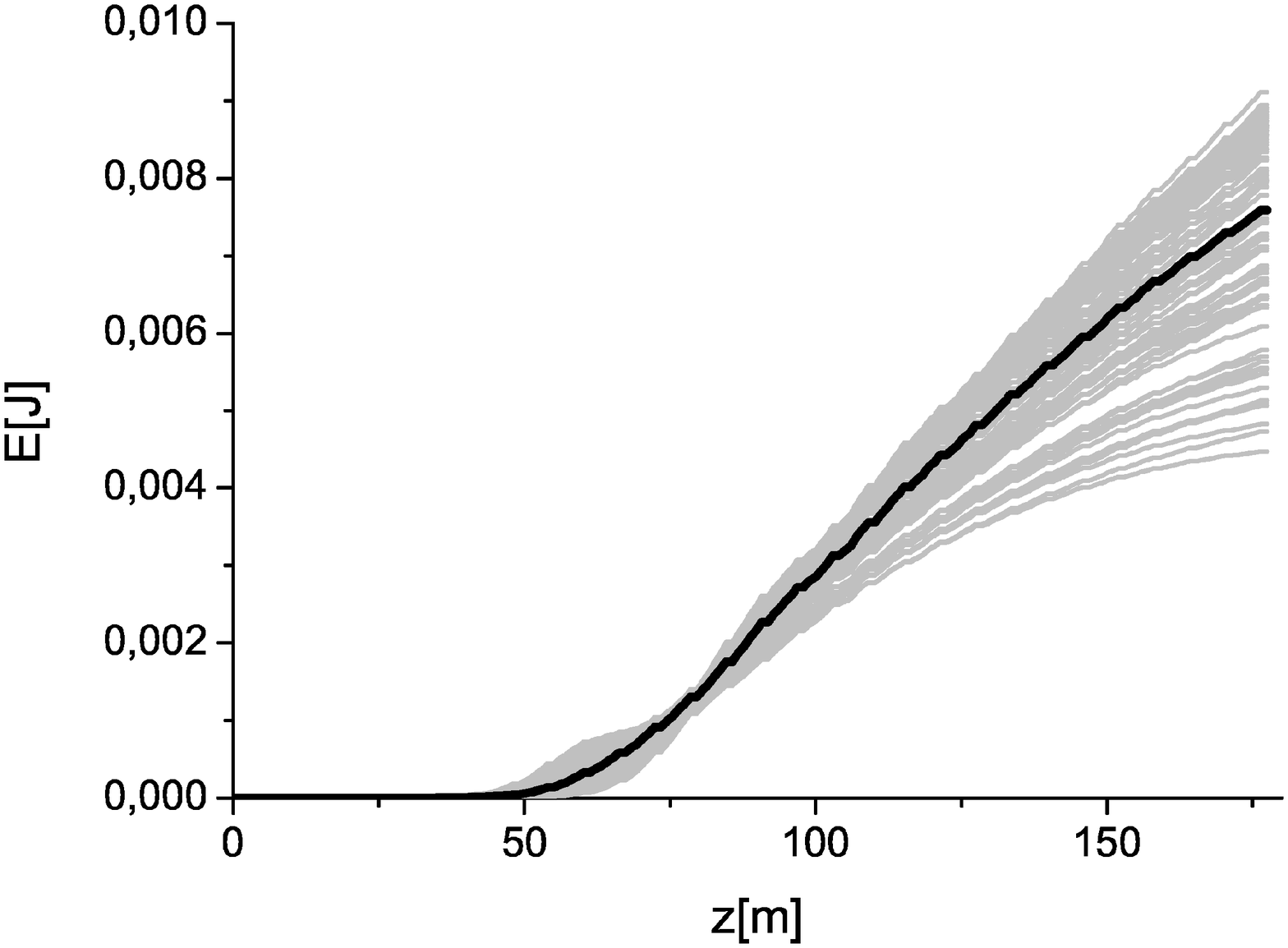}
\includegraphics[width=0.5\textwidth]{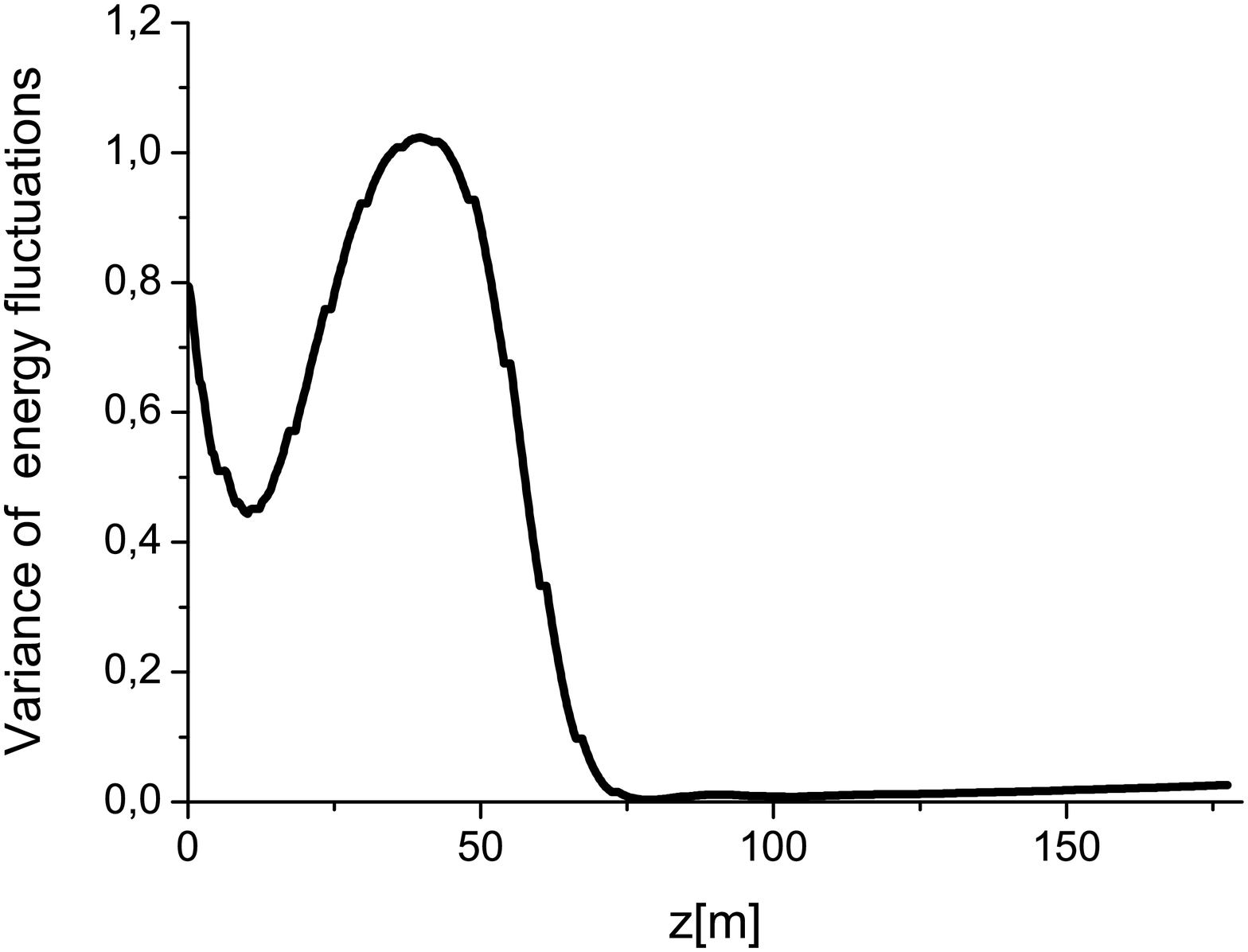}
\caption{Energy and energy variance of output pulses in the case of
tapered output undulator at $7$ keV. In the left plot, grey lines
refer to single shot realizations, the black line refers to the
average over a hundred realizations.} \label{biofh4}
\end{figure}

\begin{figure}[tb]
\includegraphics[width=0.5\textwidth]{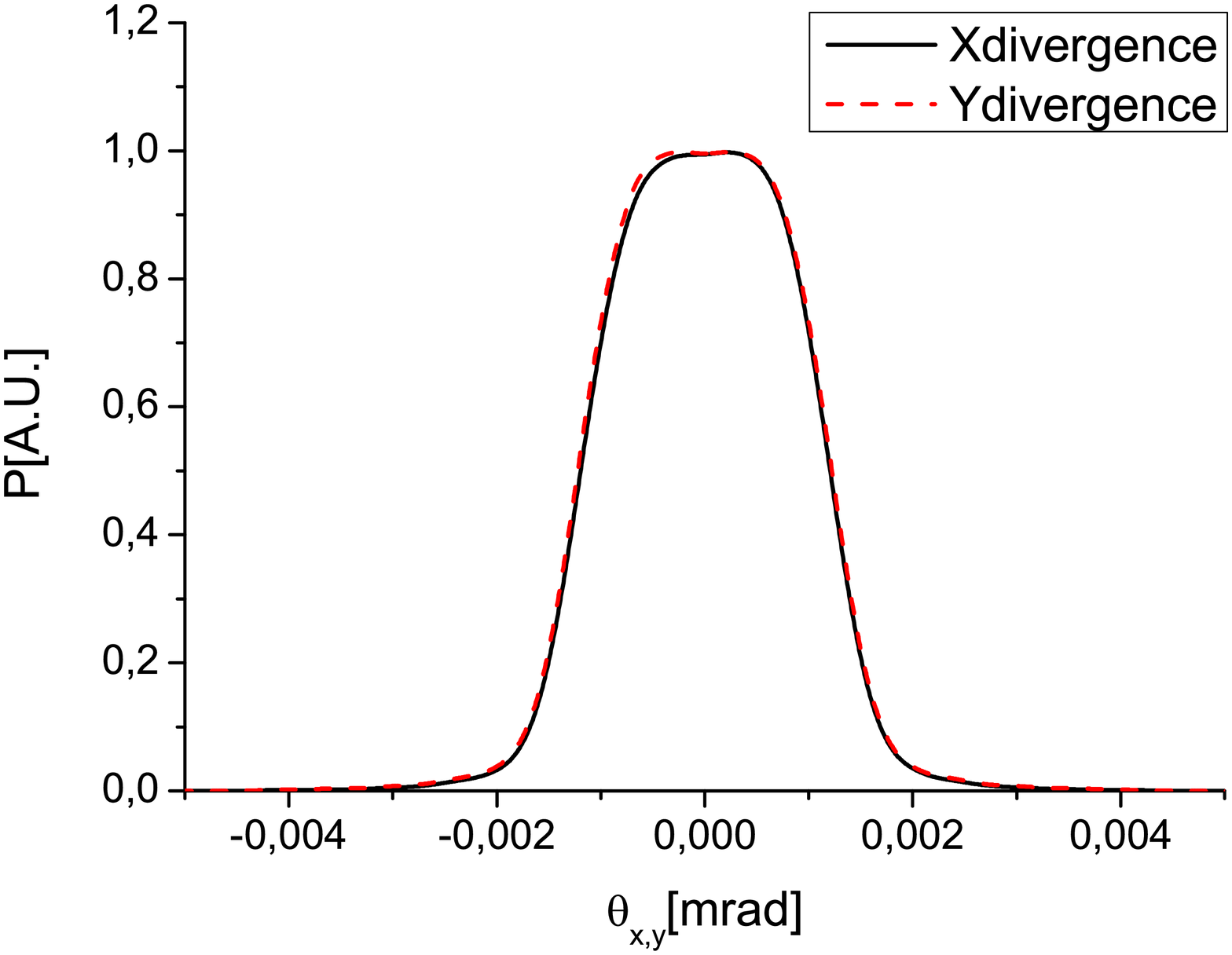}
\includegraphics[width=0.5\textwidth]{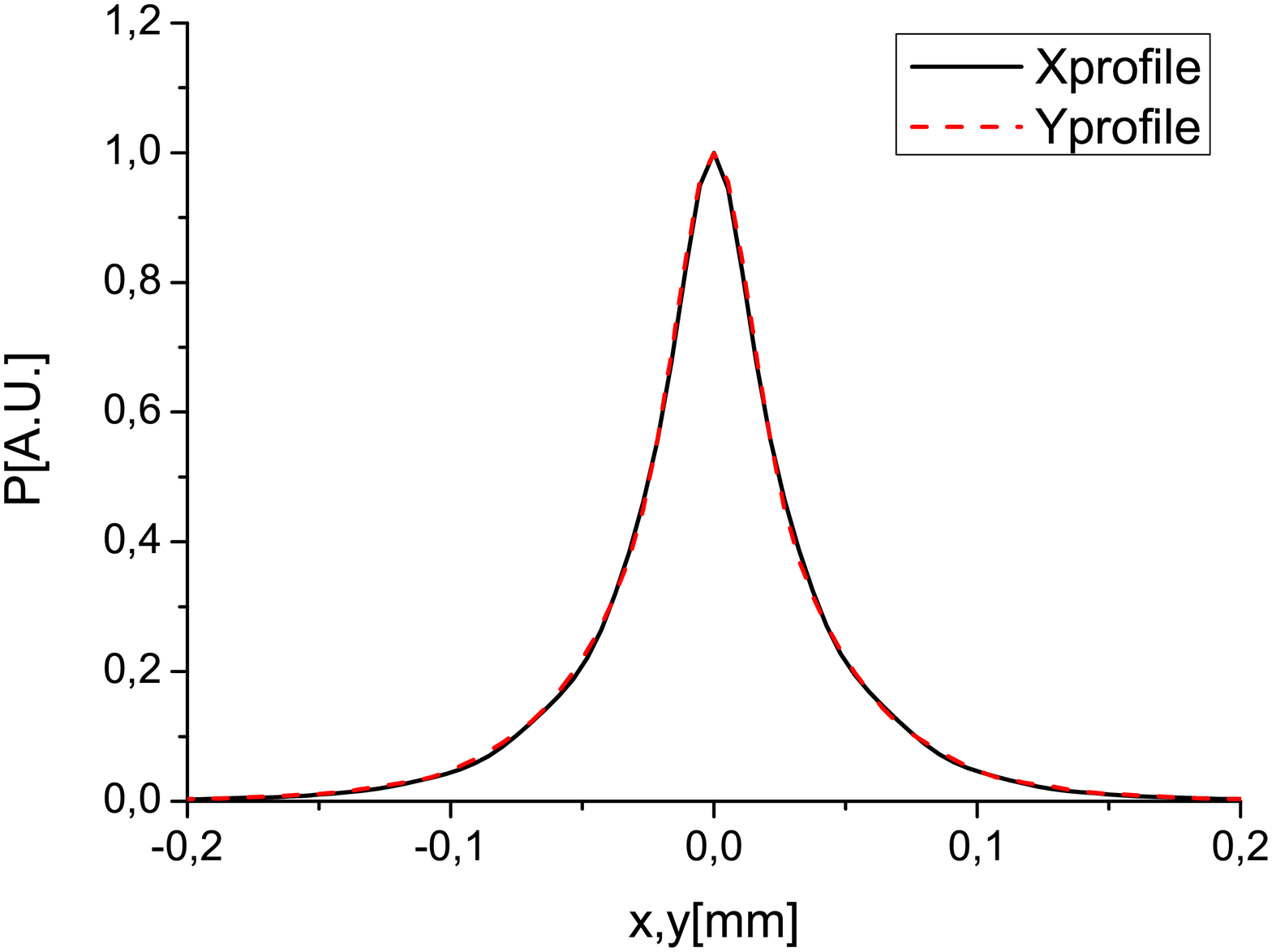}
\caption{Final output. X-ray radiation pulse energy distribution per
unit surface and angular distribution of the X-ray pulse energy at
$7$ keV at the exit of output undulator.} \label{biofh5}
\end{figure}
The output characteristics, in terms of power and spectrum, are
shown in Fig. \ref{biofh6}. The output power is increased of about a
factor $20$, allowing one to reach about one TW. The spectral width
remains almost unvaried. The output level has been optimized by
changing the tapering law, resulting in Fig. \ref{biofh3}, and by
changing the electron beam transverse size along the undulator, as
suggested in \cite{LAST}. Optimization was performed empirically.
The evolution of the energy per pulse and of the energy fluctuations
as a function of the undulator length are shown in Fig.
\ref{biofh4}. Finally, the transverse radiation distribution and
divergence at the exit of the output undulator are shown in Fig.
\ref{biofh5}.

\subsection{C(004) symmetric Bragg reflection at $8$ keV}

A feasibility study dealing with this energy range can be found in
\cite{OURCC}. No changes are foreseen in this case.

\subsection{C(333) asymmetric Bragg reflection at $9$ keV}

\begin{figure}[tb]
\includegraphics[width=0.5\textwidth]{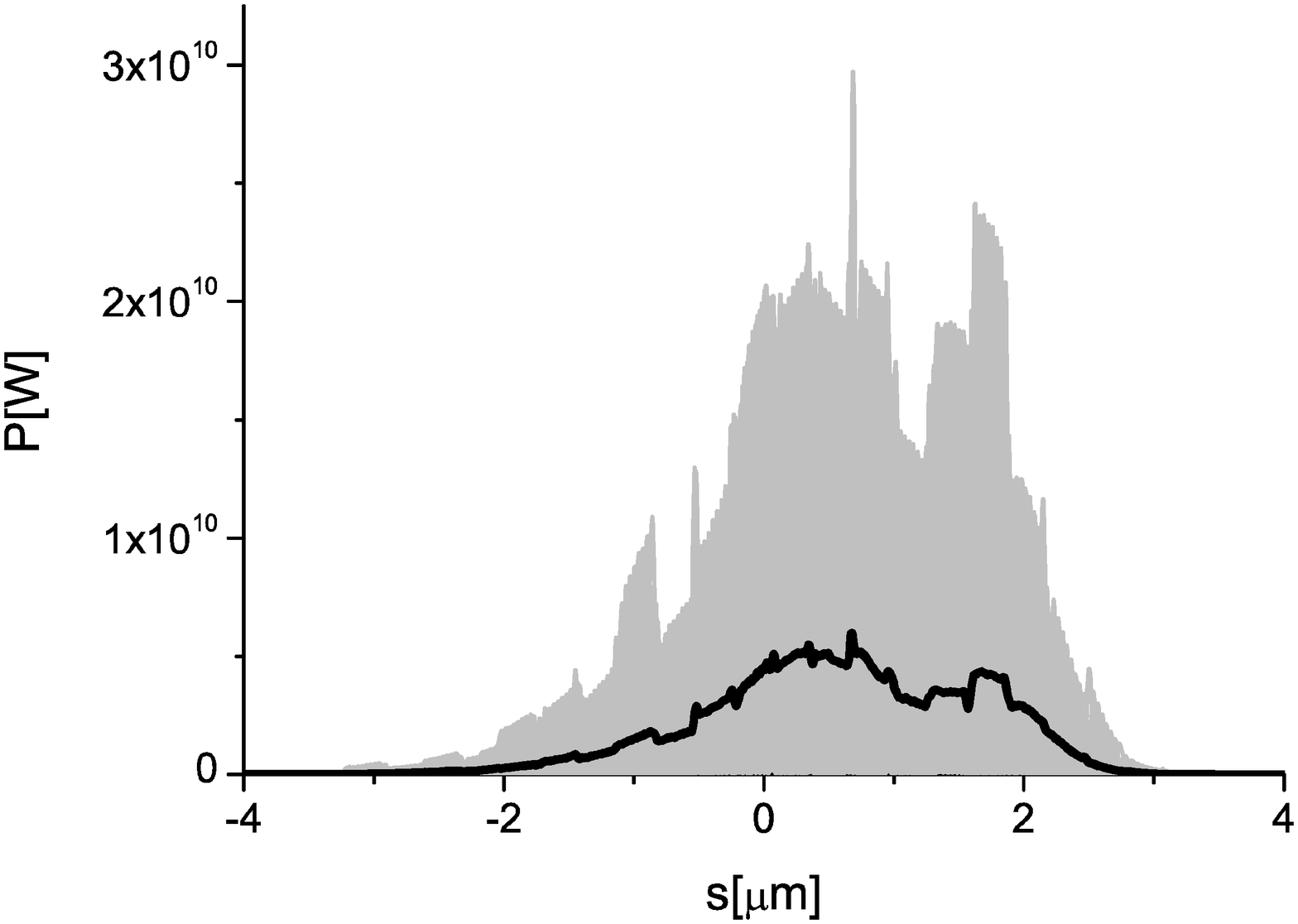}
\includegraphics[width=0.5\textwidth]{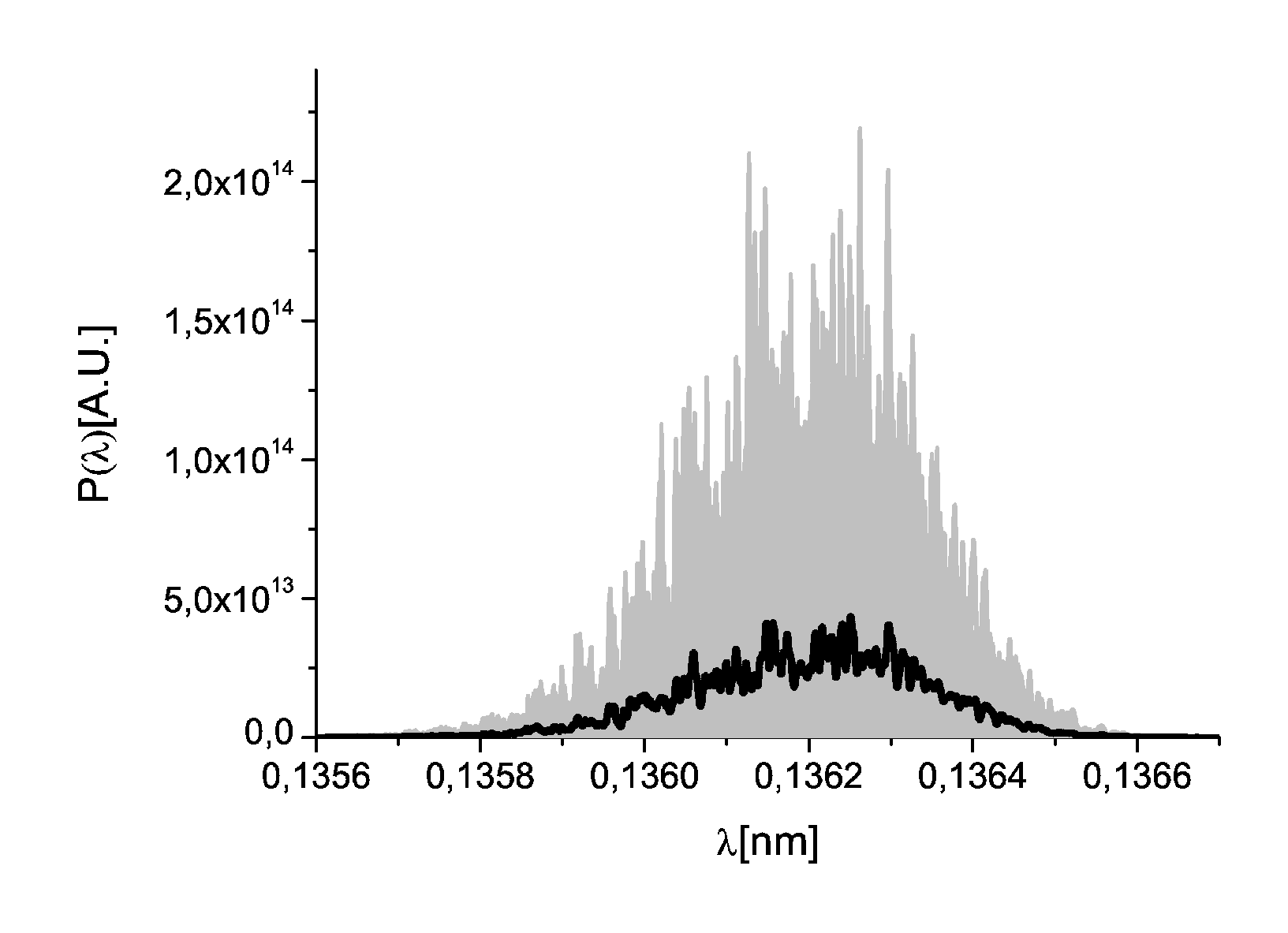}
\caption{Power and spectrum at $9$ keV before the third magnetic
chicane. Grey lines refer to single shot realizations, the black
line refers to the average over a hundred realizations.}
\label{biofC333}
\end{figure}

Finally, we consider the energy point at $9$ keV, which relies on
the C(333) asymmetric Bragg reflection. Again, the first two
magnetic chicanes are switched off. The electron beam lases in SASE
mode along the first $11$ undulator cells before passing through the
single-crystal monochromator filter. Input power and spectrum are
shown in Fig. \ref{biofC333}

\begin{figure}[tb]
\includegraphics[width=0.5\textwidth]{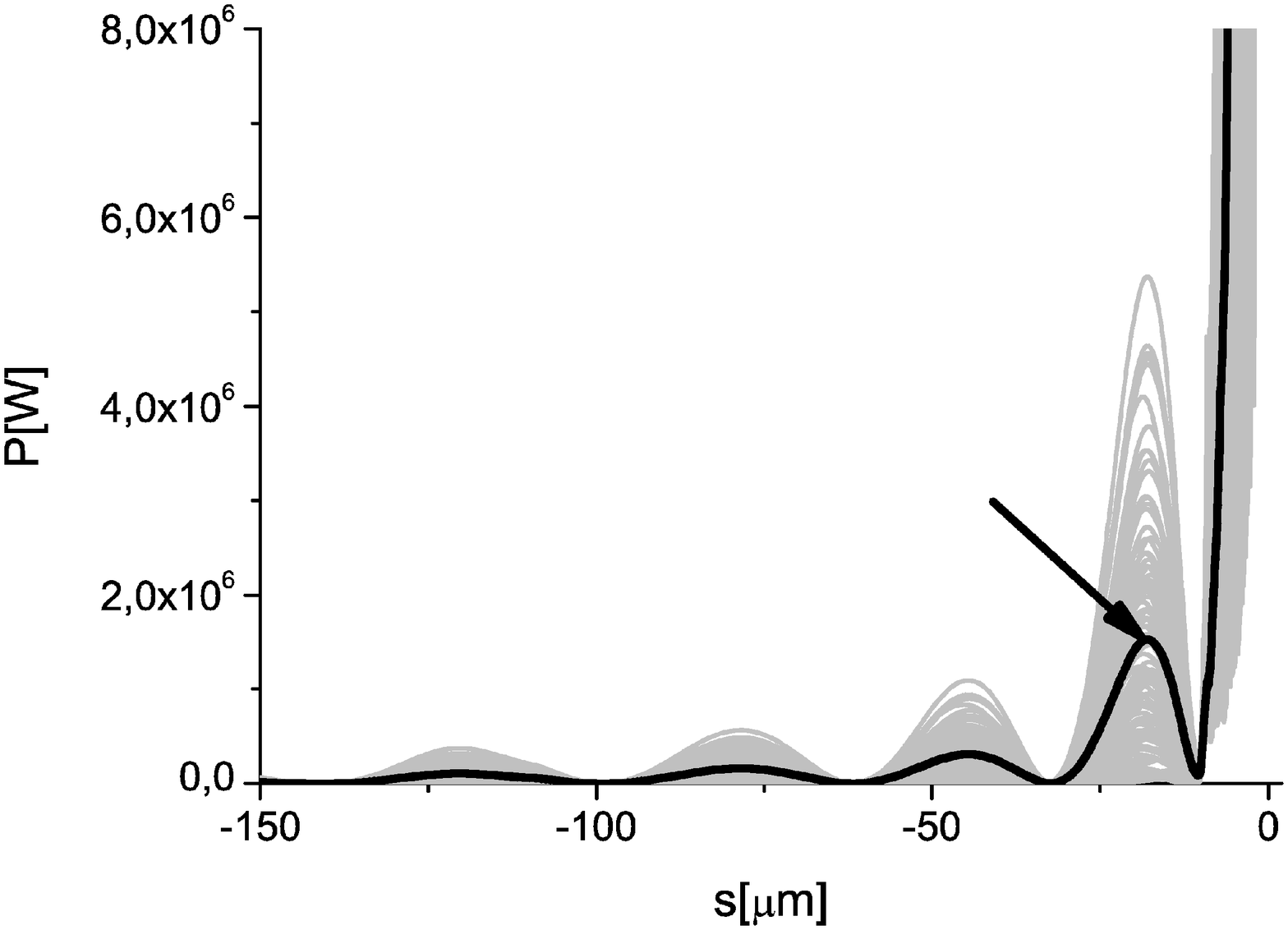}
\includegraphics[width=0.5\textwidth]{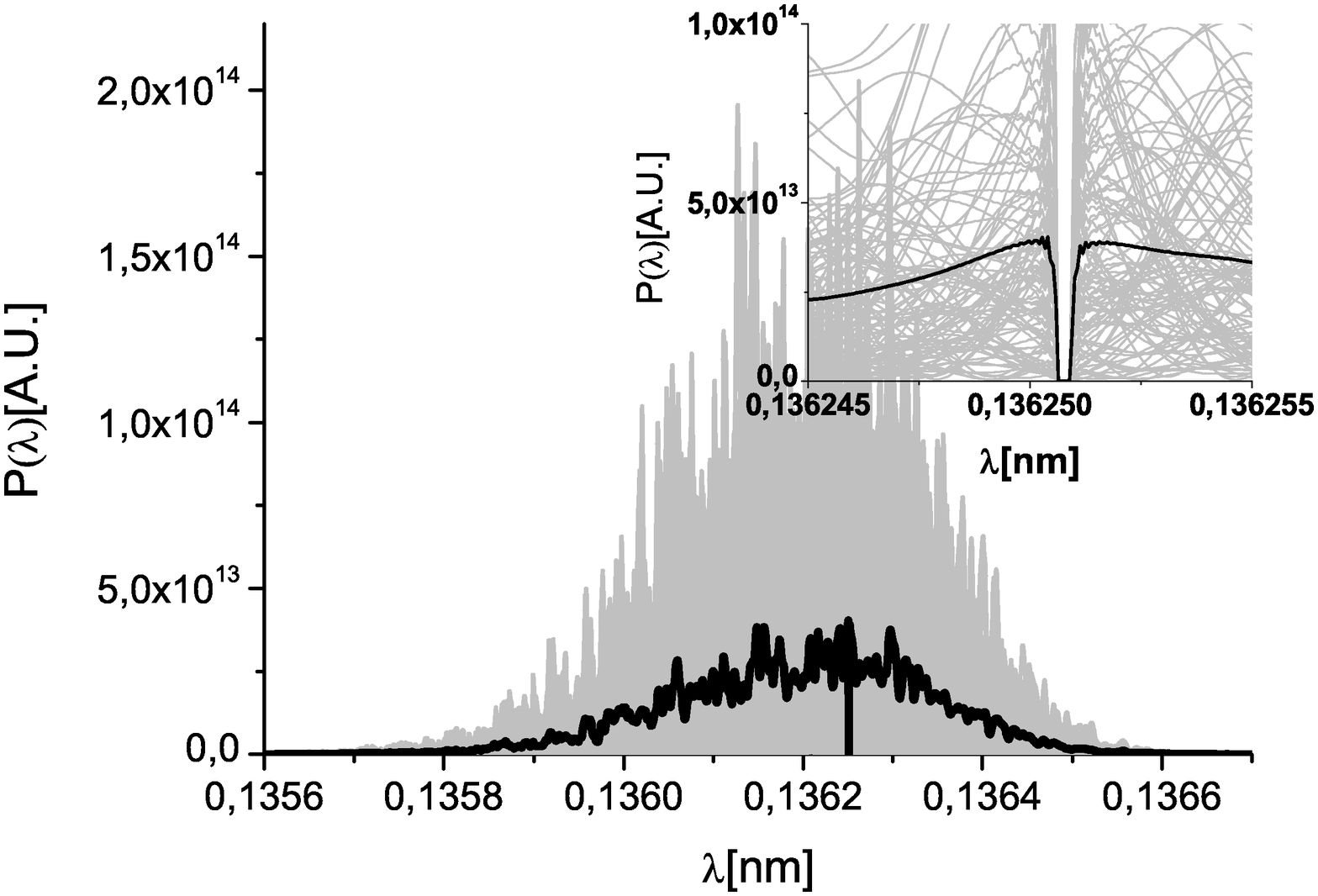}
\caption{Power and spectrum at $9$ keV after the single crystal
self-seeding X-ray monochromator. A $100~\mu$m thick diamond crystal
( C(333) Bragg reflection, $\sigma$-polarization ) is used. Grey
lines refer to single shot realizations, the black line refers to
the average over a hundred realizations. The black arrow indicates
the seeding region. } \label{biofh8B}
\end{figure}
The effect of the filtering process is illustrated, both in the time
and in the frequency domain, in Fig. \ref{biofh8B}.

\begin{figure}[tb]
\includegraphics[width=0.75\textwidth]{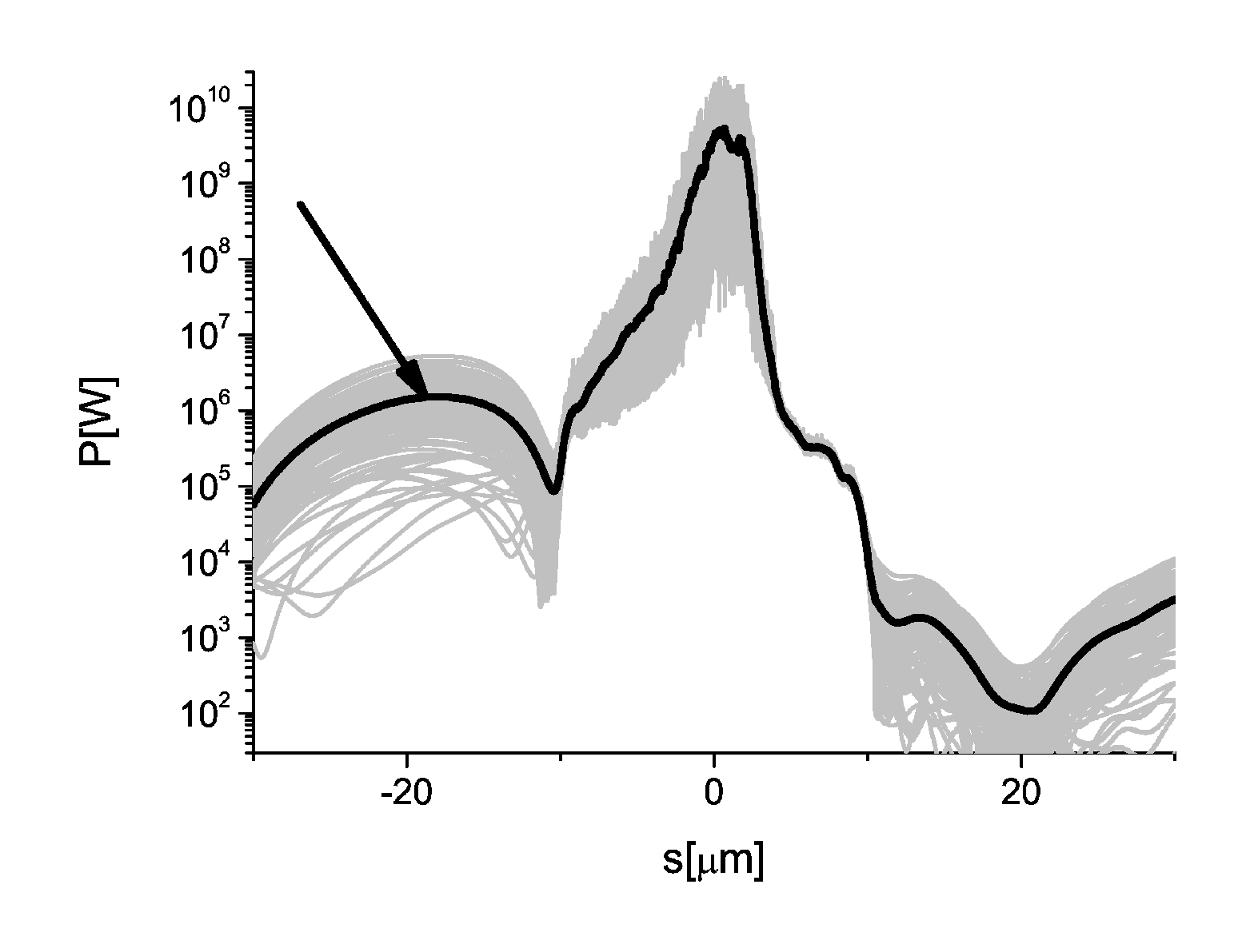}
\caption{Logarithmic plot of the power at $9$ keV after the single
crystal self-seeding X-ray monochromator.  The black arrow indicates
the seeding region. The region on the right hand side of the plot
(before the FEL pulse) is nominally zero because of causality
reasons. Differences with respect to zero give back the accuracy of
our calculations.} \label{9caus}
\end{figure}
The numerical accuracy with which causality is satisfied in our
simulations can be shown by a logarithmic plot of the FEL pulse
power after the crystal, Fig. \ref{9caus}. The peak in the center is
the main FEL pulse. On the left side one can identify the seed
pulse. One the right side, before the FEL pulse, one has nominally
zero power.

\begin{figure}[tb]
\begin{center}
\includegraphics[width=0.5\textwidth]{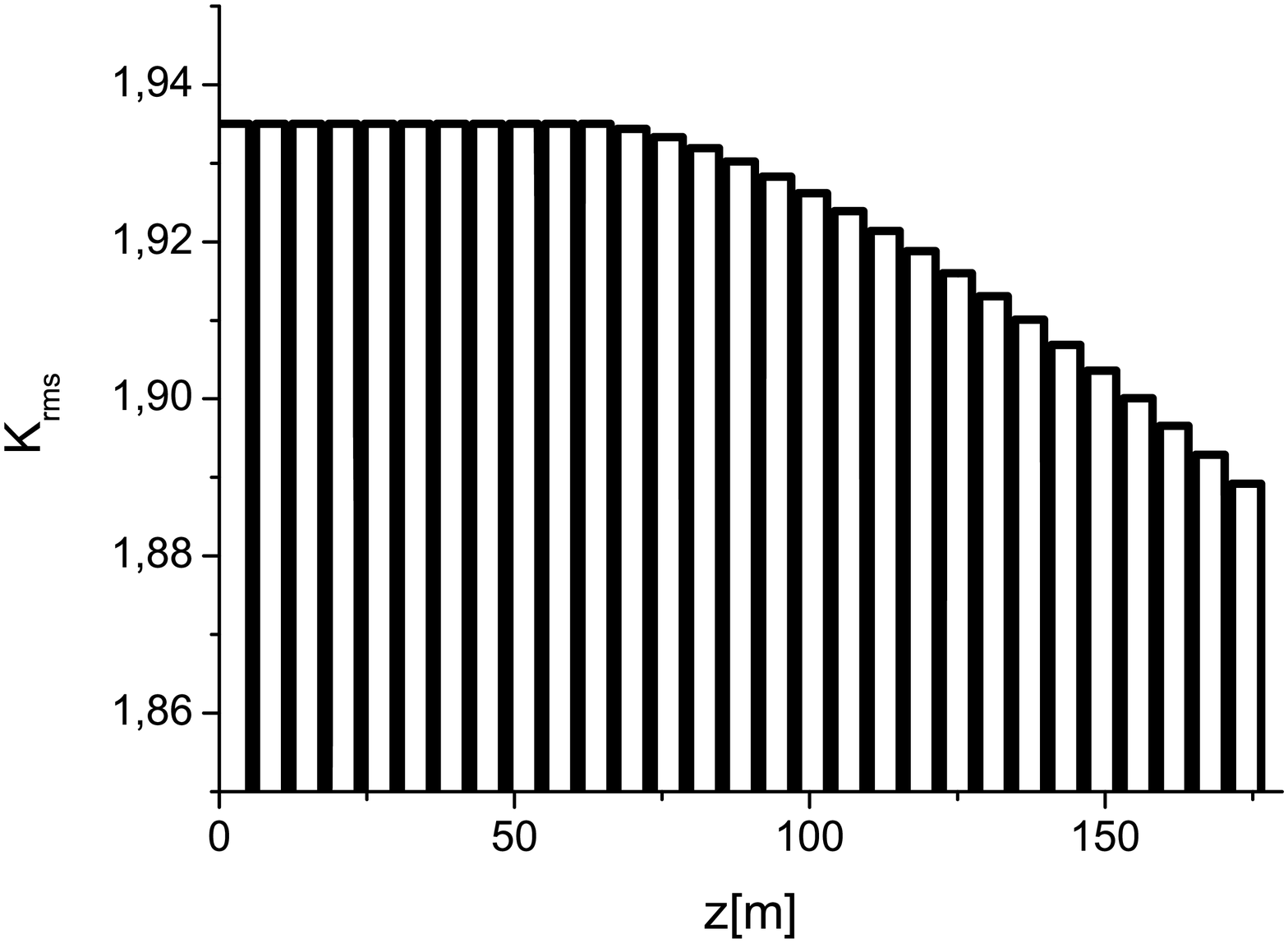}
\end{center}
\caption{Tapering law at $9$ keV.} \label{biofh3B}
\end{figure}
The seed is amplified up to saturation in the output undulator. As
already discussed, we can use post-saturation tapering to increase
the output power level. The tapering configuration in Fig.
\ref{biofh3B} is optimized for maximum output power level.

\begin{figure}[tb]
\includegraphics[width=0.5\textwidth]{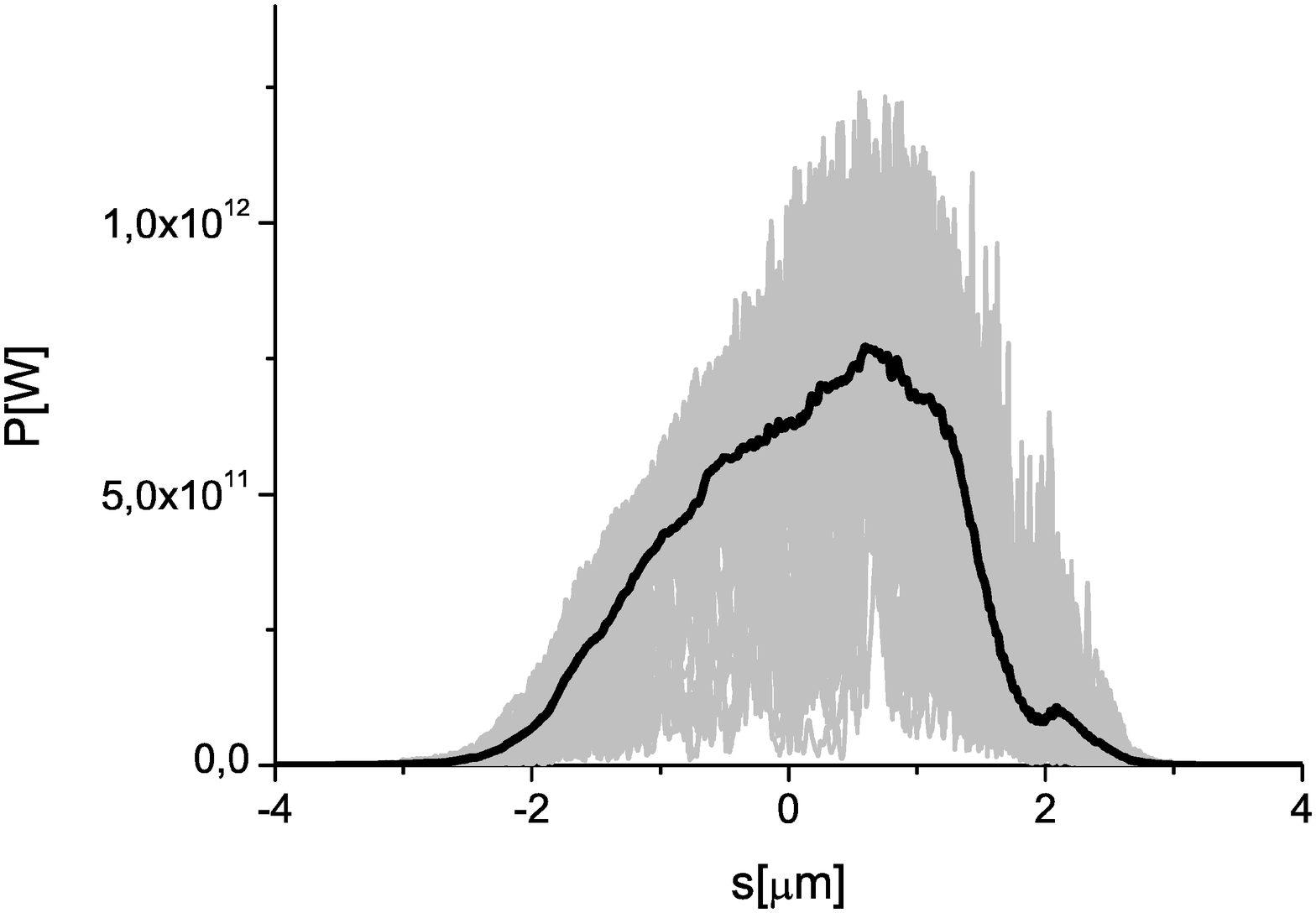}
\includegraphics[width=0.5\textwidth]{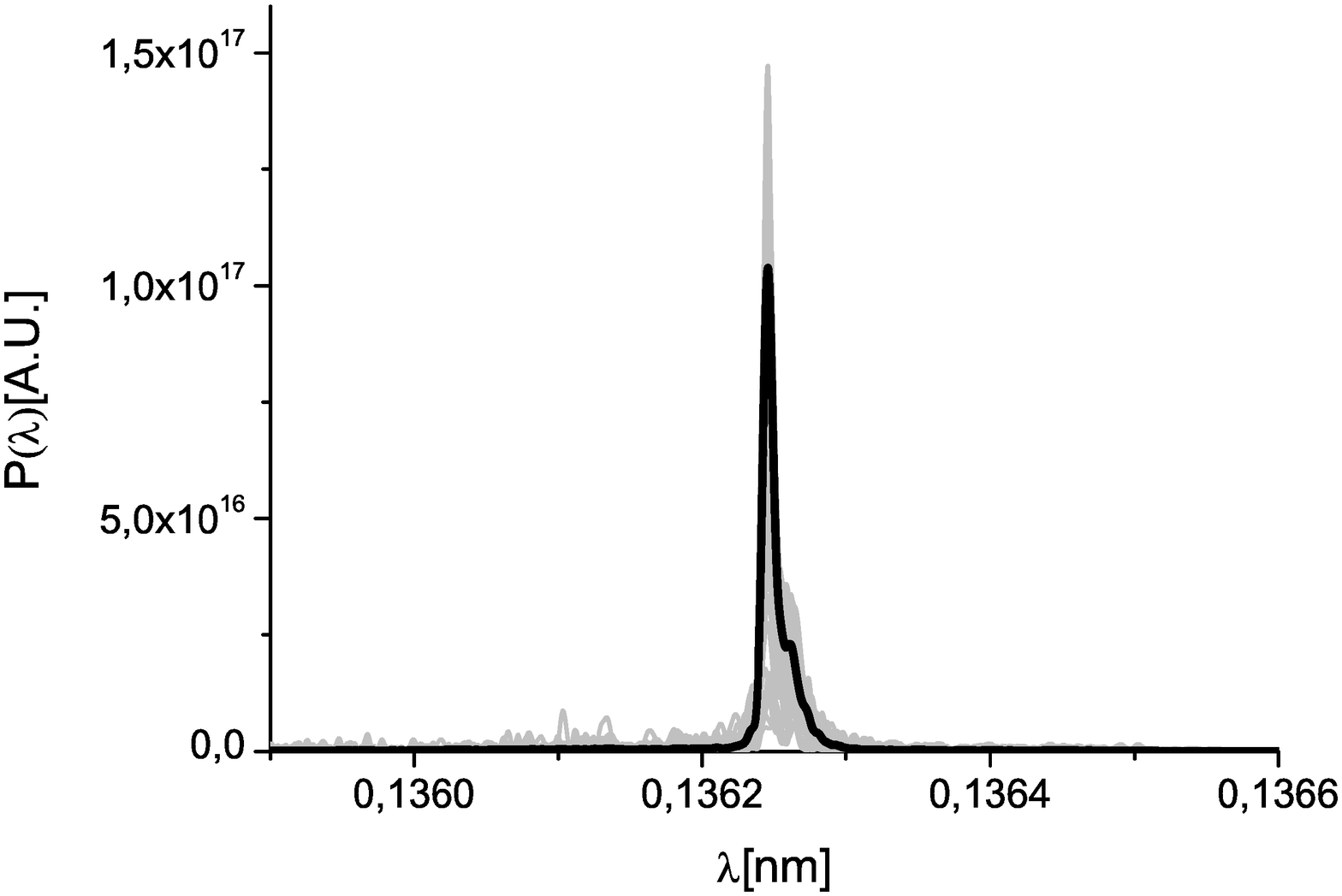}
\caption{Final output in the case of tapered output undulator at $9$
keV. Power and spectrum are shown. Grey lines refer to single shot
realizations, the black line refers to the average over a hundred
realizations.} \label{biofh6B}
\end{figure}

\begin{figure}[tb]
\includegraphics[width=0.5\textwidth]{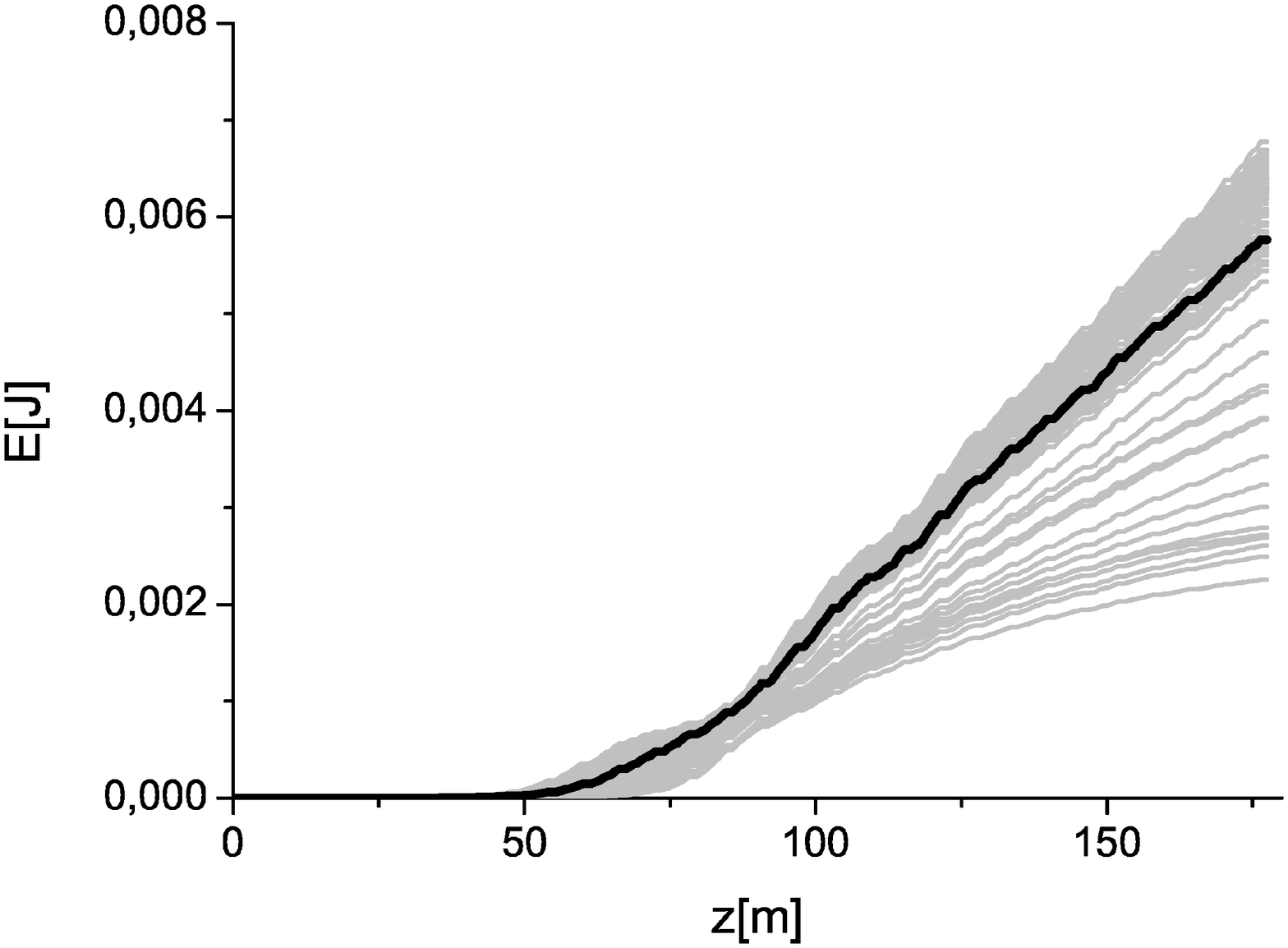}
\includegraphics[width=0.5\textwidth]{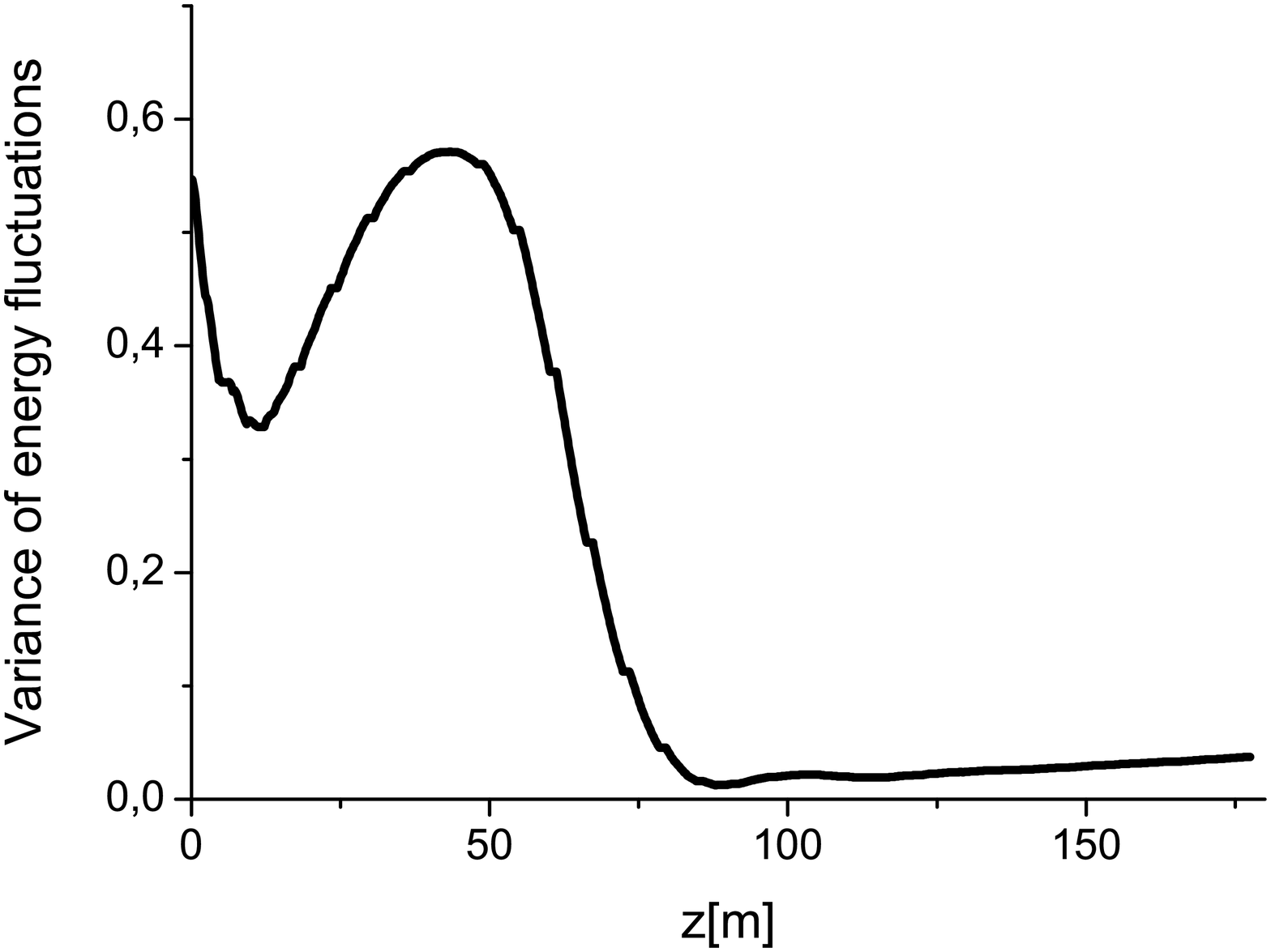}
\caption{Energy and energy variance of output pulses in the case of
tapered output undulator at $9$ keV. In the left plot, grey lines
refer to single shot realizations, the black line refers to the
average over a hundred realizations.} \label{biofh4B}
\end{figure}

\begin{figure}[tb]
\includegraphics[width=0.5\textwidth]{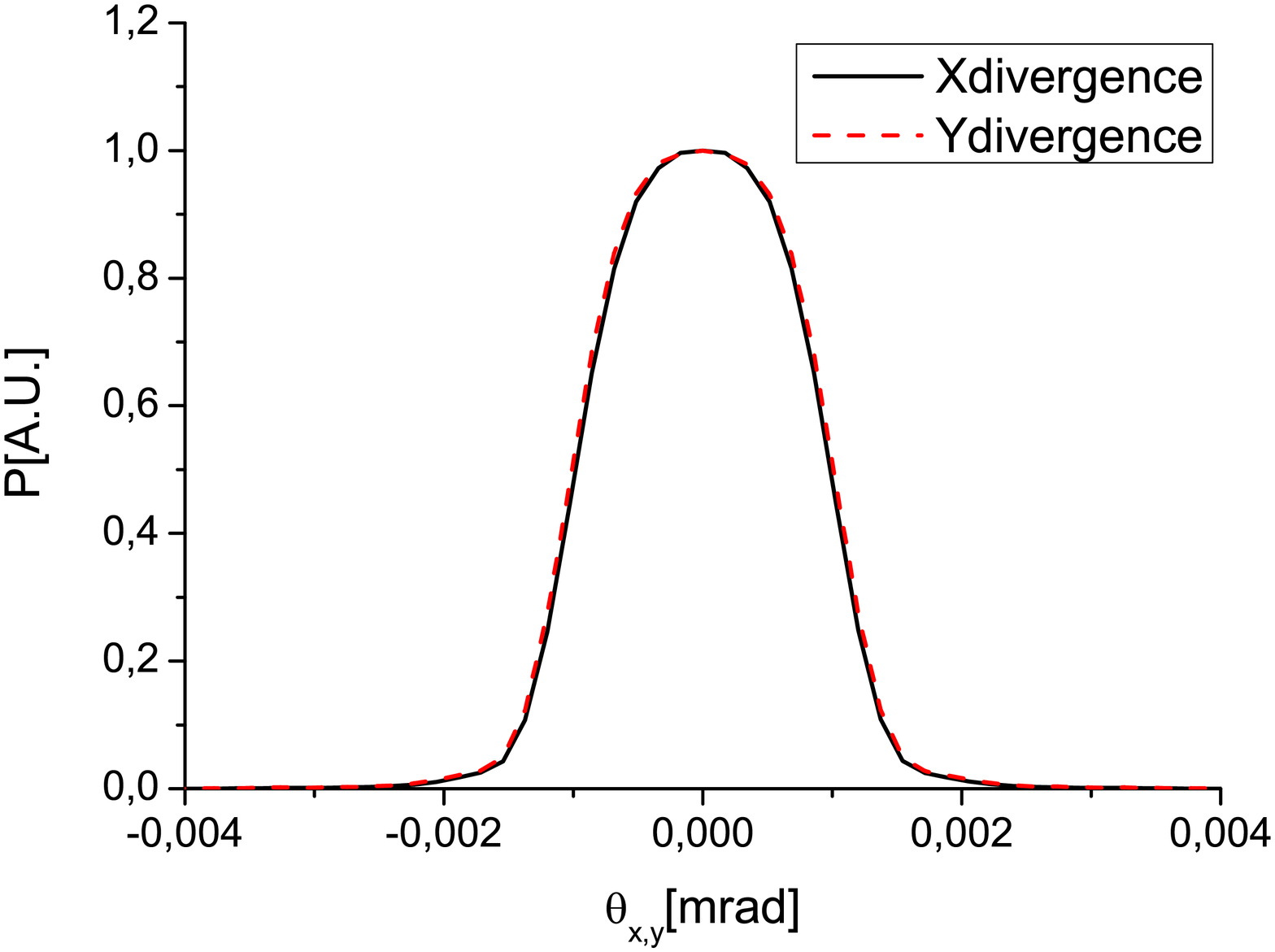}
\includegraphics[width=0.5\textwidth]{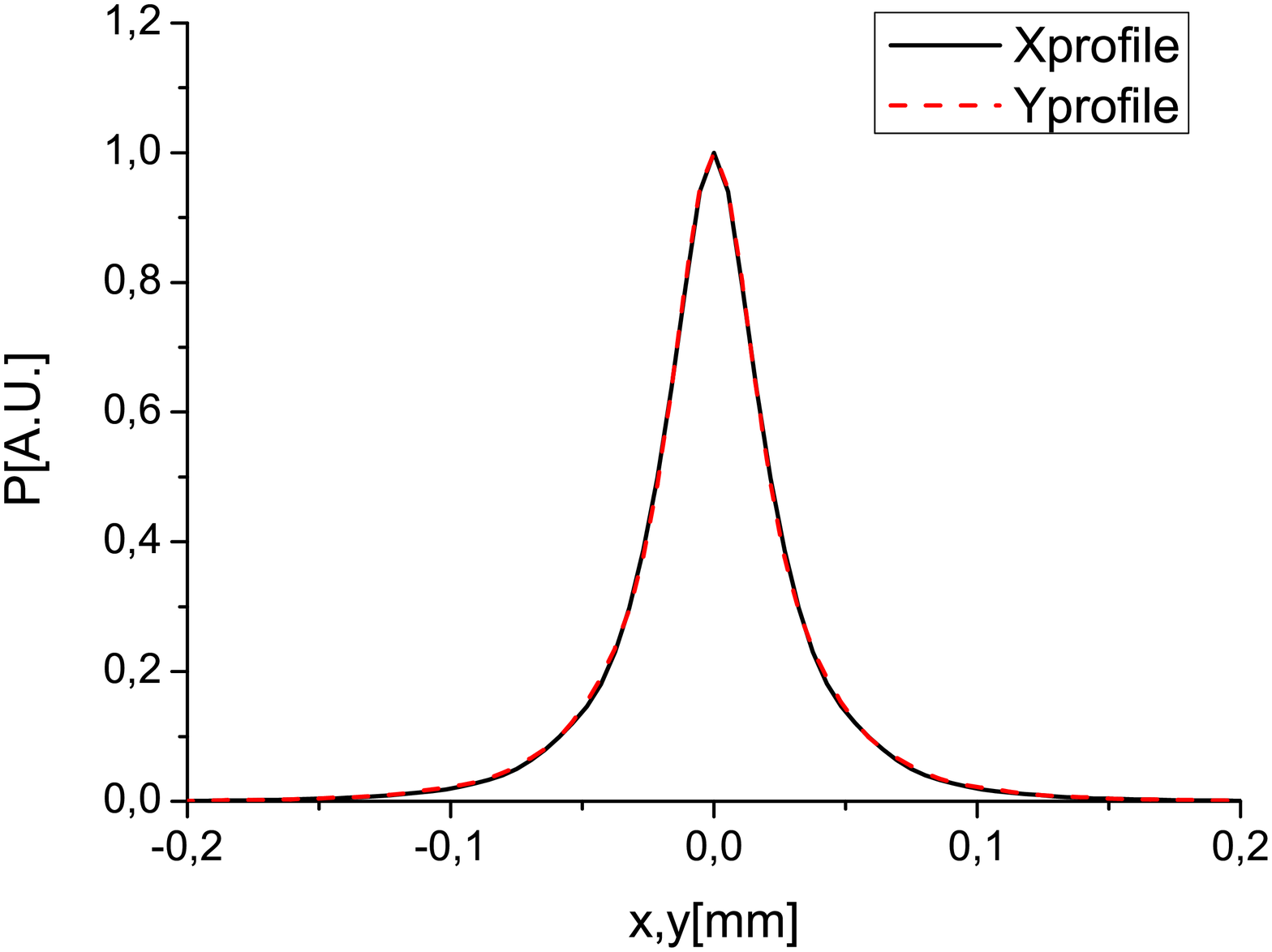}
\caption{Final output. X-ray radiation pulse energy distribution per
unit surface and angular distribution of the X-ray pulse energy at
$9$ keV at the exit of output undulator.} \label{biofh5B}
\end{figure}
The output characteristics, in terms of power and spectrum, are
shown in Fig. \ref{biofh6B}. The output power is increased of about
a factor $20$, allowing one to reach about one TW. The spectral
width remains almost unvaried. The output level has been optimized
by changing the tapering law, resulting in Fig. \ref{biofh3B}, and
by changing the electron beam transverse size along the undulator,
as suggested in \cite{LAST}. Optimization was performed empirically.
The evolution of the energy per pulse and of the energy fluctuations
as a function of the undulator length are shown in Fig.
\ref{biofh4B}. Finally, the transverse radiation distribution and
divergence at the exit of the output undulator are shown in Fig.
\ref{biofh5B}.

\section{\label{concl} Conclusions}

Self-seeding scheme with wake monochromators are routinely used in
generation of narrow bandwidth X-ray pulses at LCLS \cite{EMNAT}.
Recently, the photon energy range of self-seeding setup at LCLS was
extended from 8 keV to lower photon energies down to 5.5 keV by
exploiting different planes on the same diamond crystal \cite{FELC}.

In this article we propose a study of the performance of
self-seeding setups for the European XFEL based on wake
monochromators, exploiting different (symmetric and asymmetric,
Bragg and Laue) reflections of a single diamond crystal, similar to
that used into the LCLS setup. In our study we account the
spatiotemporal shift of the seed induced by the passage through the
crystal, and we study the dependence of the FEL input coupling
factor on this shift. We exemplify our results using the concept for
a dedicated bio-imaging beamline previously proposed by the authors.
We conclude that crystals similar to the one actually used at the
LCLS can be exploited to cover the energy range between $3$ keV and
$13$ keV with high performance, allowing one to reach TW-class
pulses.

\section{Acknowledgements}

We are grateful to Massimo Altarelli, Reinhard Brinkmann, Henry
Chapman, Janos Hajdu, Viktor Lamzin, Serguei Molodtsov and Edgar
Weckert for their support and their interest during the compilation
of this work.

\end{document}